\documentclass[twocolumn,times]{aastex631}

\newcommand{\beq}{\begin{equation}}
\newcommand{\eeq}{\end{equation}}

\usepackage{amsmath,amstext}
\usepackage[T1]{fontenc}
\usepackage{apjfonts} 
\usepackage[figure,figure*]{hypcap}
\usepackage[normalem]{ulem}             

\usepackage{graphicx}

\usepackage{xcolor}

\newcommand{\ionm}[2]{#1\,{\sc #2}}

\submitjournal{ApJ}

\shorttitle{EIS calibration}
\shortauthors{Del Zanna, Weberg, and Warren}

\begin{document}

\title{Hinode EIS: updated in-flight radiometric calibration}

\correspondingauthor{G. Del Zanna}
\email{gd232@cam.ac.uk}

\author[0000-0002-4125-0204]{G. Del Zanna}
\affiliation{DAMTP, Center for Mathematical Sciences, University of Cambridge, Wilberforce Road, Cambridge, CB3 0WA, UK}
\affiliation{School of Physics \& Astronomy, University of Leicester, Leicester  LE1 7RH, UK }
\author[0000-0002-4433-4841]{M. Weberg}
\author[0000-0001-6102-6851]{H. P. Warren}
\affiliation{Space Science Division, Naval Research Laboratory, Washington, DC 20375, USA}

\begin{abstract}
  We present an update to the in-flight radiometric calibration of the
  Hinode EUV Imaging Spectrometer (EIS), revising and extending our previous studies.
  We analyze full-spectral EIS observations of quiet Sun and active regions from 2007 until 2022. 
  Using CHIANTI version 10, we adjust the EIS relative effective areas for a selection of dates with emission measure analyses of off-limb quiet Sun. 
  We find generally good agreement (within typically $\pm$ 15\%) between measured and expected line intensities. We then consider selected intensity ratios for all the dates and apply an automatic fitting method to adjust the relative effective areas. To constrain the absolute values from 2010 and later
  we force agreement between  EIS and Solar Dynamics Observatory (SDO) Atmospheric Imaging Assembly
  (AIA) 193~\AA\ observations. The resulting calibration, with an uncertainty of about $\pm$ 20\%, is then validated in various ways, including flare line ratios from \ion{Fe}{24} and \ion{Fe}{17}, emission measure analyses of cool active region loops, and several density-dependent line ratios. 
\end{abstract}
  
\section{Introduction}
\label{sec:intro}

The Hinode
Extreme Ultraviolet Imaging Spectrometer (EIS) \citep{culhane_etal:2007}
 has observed
the solar corona in two wavelength bands
regularly since 2006.
These wavebands are commonly referred to as the shortwave (SW, 170--211~\AA) and longwave (LW, 242--292~\AA) channels.
An accurate EIS relative calibration  is necessary e.g. to measure electron densities \citep[see, e.g.][]{young_etal:2009_fe_13}, to obtain estimates of emission measures
\citep[see, e.g.][]{warren_etal:2011_arcore}, to measure electron temperatures and test for 
the presence of non-Maxwellian electron distributions via line ratios
\citep[see, e.g. the ][review]{delzanna_mason:2018}, and for estimates of the coronal magnetic fields \citep[see, e.g.][]{li_etal:2016}.
An accurate absolute calibration is necessary for e.g. a meaningful comparison
with other instruments such as IRIS \citep{delzanna_etal:2022}.

{ EUV spectrometers typically degrade in time \citep[cf][]{benmoussa_etal:2013}. 
One of the best ways to calibrate in-flight a spectrometer is with 
sounding rockets carrying similar instruments. 
An earlier check  on 11 EIS SW lines  was obtained in this way in 2007, see \cite{wang_etal:2011}.
A second method is to monitor radiances of low-temperature lines in the quiet Sun, as they are known to vary little over the solar cycle. 
Such method was used to apply a first correction for the 
calibration of the Coronal Diagnostic Spectrometer (CDS) on the Solar and Heliospheric Observatory (SOHO) \citep[see][and references therein]{delzanna_andretta:2015}, and was used by 
\cite{mariska:2013} to estimate the EIS degradation up to 2012 using 
two transition-region lines from Si VII and Fe VIII.  Unfortunately, 
such lines are somewhat affected by solar cycle variations, and cooler
ones in EIS are too weak to be used reliably (except a He II line which 
is blended with many transitions). 
A third way to monitor the relative in-flight degradation is to measure well-known
intensity ratios of lines that are insensitive to variations of densities
and temperatures and compare them to expected values, see e.g. 
\cite{neupert_kastner:83}.
 This  procedure has been used to obtain a relative calibration of 
 various spectrometers, including  the SOHO CDS \citep[see][and references therein]{delzanna_andretta:2015}. 
 }

One method to obtain the relative calibration is to use only spectral lines
from the same ion, while another one is to use lines
from ions formed at similar temperatures. 
Clearly, the first method is more robust as does not depend significantly on the
plasma conditions and on  uncertainties in the ionization fractions.
This was used by \cite{delzanna:13_eis_calib}, where
 a significant revision of the ground
calibration was presented.  In addition, a time-dependent decrease in the sensitivty
of the LW channel was introduced, increasing the calibrated
intensities of the LW lines by nearly a factor of two for data
after 2008. Only data from 2006 until Sept 2012 were
analysed. Still, deviations in several line ratios were present, indicating
the need for wavelength-dependent corrections within each EIS channel.

Such corrections were introduced by \cite{warren_etal:2014}, where
 observations of coronal lines during the 2006-2014 period
 were analysed. In this case, the second method was adopted:   off-limb quiet Sun (QS) observations were selected, and assuming isothermality, the emission measure loci method was used
 to enforce agreement between observed and predicted intensities using a range of ions. This was repeated for several  dates, to obtain a time- and wavelength-dependent revision of the ground calibration. 

 In the present paper, we extend our previous two studies in several ways,  as described below.
Section~2 outlines the methodology adopted,
while Section~3 describes the associated  
 selection of the observations and data analysis.
Section~4 describes a selection of the observations analysed
here, while Section~5 draws the conclusions.
An Appendix provides a few  more examples among the data analysed.

\section{Methods}

{ There are three general steps in the process we used to derive an updated radiometric calibration: (1) Careful DEM analysis of near-isothermal spectra for selected dates, (2) automated fitting of the EIS effective area curves using mostly density-insensitive line ratios, and (3) scaling the absolute, time-dependant calibration using Solar Dynamics Observatory (SDO)  Atmospheric Imaging Assembly
 (AIA, \citealt{lemen_etal:2012}) 193~\AA\ observations from 2010. 
Within (1), we relaxed the isothermal assumption and performed an emission measure
 analysis on seven off-limb QS  regions to obtain a relative calibration. 
A preliminary comparison between AIA and EIS data was carried out, providing
an absolute scale for the effective areas for the seven dates.
Step (1) was used to provide an input to step (2) which provided the 
final relative effective areas.
For step (2), we have analysed a wide range of observations covering the entire period until 2022, of the quiet Sun and active regions (AR), both on-disk and 
 off-limb. The line ratio analysis of \cite{delzanna:13_eis_calib} was extended 
 considering more cases. The relative calibration of step (2) was then scaled 
 using the AIA/EIS comparison after 2010. The calibration pre-2010 of the 
 SW channel was kept close to pre-flight values for the reasons described below.

Finally, we have run several checks that the calibration provides 
good agreement (within $\pm$20) for a sample of flare observations,
analysing \ionm{Fe}{xxiv}  and  \ionm{Fe}{xvii}  lines, for cool lines that are present in AR loops, and for the density-sensitive lines. 
 }

\subsection{DEM analysis on near-isothermal spectra}

{We first selected seven off-limb
QS observations that were located} far enough from the limb to reduce the
blending of cool lines, but close enough to retain enough counts.
 The advantage is that the
 plasma is nearly isothermal and isodensity in these cases,
 {see e.g. \cite{feldman_etal:1999a,warren_brooks:2009}. }

 This is an extension of the method used by \cite{warren_etal:2014}, as we 
 relax the isothermality assumption 
 and perform a differential  emission measure (DEM) analysis.
 This method is more reliable than the line ratio method, as several
 line ratios, especially those involving lines far apart in wavelength,
 are somewhat temperature-dependent. Indeed several
 ratios of lines from the same ion have now been identified
 and can be used to measure the electron temperature
 \citep[cf.][and references therein]{delzanna_mason:2018}.
 
 The strongest lines in these off-limb spectra are from \ion{Fe}{9}--\ion{Fe}{13}
 and \ion{Si}{10}.
 First, we estimate the average electron density from the  \ion{Si}{10} 258 vs. 261~\AA\  ratio.
 The  \ion{Fe}{13} 202.0, 203.8~\AA\ and other line ratios have also been
 considered, but the  \ion{Si}{10} was deemed the best case for these QS  observations as these lines are 
 close in wavelength and the uncertainty in the 
 relative calibration does not affect much the results.
 We  then calculate the $G(T)$ of all the lines possibly
 contributing to the  observed EIS lines, assuming a constant density.
 We then  perform a
 DEM analysis using  the CHIANTI v.10 DEM
 method, which defines the DEM  as a spline function, and uses a nonlinear, least squares levenberg—marquardt fitting algorithm (MPFIT, \citealt{Markwardt:2009}).
 
 We started with calibrated intensities using the ground calibration
 and aimed to obtain a correction to it,
 by defining  a set of spline nodes for the corrections
 across the EIS spectral
 range, most of which are  located where the strong  lines are.
{We used splines 
with nodes at 165, 171, 174.5, 177.2, 178.1, 180.4,
182.2, 184.5, 185.2, 186.9,  188.3, 190,  
   192.4, 192.8, 193.5, 194.7, 195.1, 196.6,
   197.4, 200,  201.1,  202.,   202.7, 204.9,
   208, 209.9, 211.3~\AA\ for the SW channel,
   and 245, 252, 255,  257, 259, 
        263 , 265, 268, 270, 272, 
        274, 277, 281, 286, 292~\AA\ for the LW channel. }
 
 The corrections were adjusted for each observation to obtain a good
 agreement between observed and predicted intensities.
 The DEM was obtained from a minimal list of lines from iron.
 
The absolute scale of the calibration was adjusted with a preliminary cross-calibration with AIA (described below).
These effective areas were used as initial constraint 
on the final effective areas, obtained by automatic fitting the line ratios as described below.

We then checked that the calibration obtained in this way produced
good agreement between predicted and measured intensities of a 
few cooler lines formed below 1 MK.  We selected a few AR loop regions, where the signal from the hotter lines was reduced, and performed a DEM analysis. 
Also in this case it is well known
 that the plasma distribution is nearly isothermal
 {\citep[cf.][]{delzanna_mason:2003,warren_etal:2008b}, } which simplifies the modelling.

 Clearly, we cannot expect for all lines agreement to say 10\%,
 as departures from the isodensity case, optical depth and other effects 
 can affect the results. Also, the atomic data for different
 ions do not have the same accuracy.
 However, what is important  is the fact that
 those lines we have selected, for which good agreement
 between observed and predicted radiances is present, should have the
 same agreement regardless of the time of the observation.

 \subsection{Fitting effective area curves} \label{fit_ratio_method}


The DEM analysis provides wavelength-dependent corrections to the ground 
calibration for a set of dates. 
{From the comparisons between observed and predicted 
radiances, it appears that this method can be very accurate,
down to $\pm$ 10\%, for the wavelengths where the key lines 
are observed. 
Unfortunately, however, there no sufficient suitable observations  to cover the entire mission, and the DEM analysis 
is time consuming, so it is not feasible to apply this method 
for the entire period.}
 To obtain a correction for any given date we fit the 
effective areas using a list of 39 line intensity ratios that are largely 
density-insensitive at typical AR and QS temperatures. 
The chosen ratios are listed in Table~\ref{tab:list_lines}.

\begin{table}[!htbp]
  \caption{Line ratios used for the EIS calibration. The second and third columns show the theoretical ratios R (photon units)  for suitable density ranges (log Ne [cm$^{-3}$] indicated in brackets) for the QS and AR. The last column indicates the channel.
    BR indicates a branching ratio, (w) weak lines. }
\centering
\footnotesize
\begin{tabular}{@{}lllllllllll@{}}
 \hline\hline \noalign{\smallskip}
Line ratio (\AA) & R (QS)         &   R (AR)   & Notes \\ 
                 & (8.2--8.8)      & (8.5--9.5)     &    &     \\
\hline \noalign{\smallskip}

\ionm{Fe}{ix} 189.94 /  197.85 & 1.0--1.2  & 1.1--1.5  & SW \\

\ionm{Fe}{ix} 188.49 /  197.85 & 2.0--2.1  & 2.0--2.5  & SW \\

\ionm{Fe}{x} 174.5 / 184.5 & 4.5--4.4  & 4.45--4.0  & SW \\
\ionm{Fe}{x} 177.2 / 184.5 & 2.6--2.5  & 2.58--2.35  & SW \\

\ionm{Fe}{x}  190.0 (bl) /184.5  &  0.338 &      & BR SW \\
\ionm{Fe}{x} 207.45 / 184.5   &  0.14--0.16    & 0.15--0.21   &  SW \\
\ionm{Fe}{x} 257.3 (sbl) / 184.5 & 1.5--1.2  & 1.35--0.8  &     LW/SW \\

\ionm{Fe}{xi} 178.1 / 182.17 &  0.274       &   &   BR SW (w) \\ 

\ionm{Fe}{xi} 188.2 / 192.8 (bl)& 4.8       &     & BR SW \\ 
\ionm{Fe}{xi} 202.7 / 188.3     & 0.1       &     & BR SW \\ 

\ionm{Fe}{xi} 180.4 / 188.2     & 2.0--1.98  & 1.99--1.88    &  SW \\

\ionm{Fe}{xi} 257.5 (sbl) / 188.2 & 0.17--0.19 & 0.175--0.24   & LW/SW \\

\ionm{Fe}{xii} 192.4 / 195.1 (sbl) &  0.315 &   &  SW \\

\ionm{Fe}{xii} 193.5 / 195.1 (sbl) &  0.67  &   &  SW \\

\ionm{Fe}{xii} 186.9 (sbl) / 196.6 (bl) & 3.45--3.6  & 3.5--4.1  & SW \\

\ionm{Fe}{xiii} 209.9 (bl) / 202.0      & 0.15   &   &  BR SW \\
\ionm{Fe}{xiii} 204.9 (bl) / 201.1 (bl) & 0.31   &   & BR SW \\
\ionm{Fe}{xiii} 201.1 (bl) /197.4       & 5.0    &   &  BR SW \\
\ionm{Fe}{xiii} 204.9 (bl) / 197.4      & 1.55   &   & BR SW \\

\ionm{Fe}{xiii} 209.6 / 200.            & 0.74-0.69 & 0.69--0.71 & SW \\
\ionm{Fe}{xiii} 246.2 (bl?) / 251.9 & 0.51        &  & BR LW  \\

\ionm{Fe}{xiii} 261.7 / 251.9 &  0.135-0.12     & 0.125--0.105  &  LW  \\
\ionm{Fe}{xiii} 261.7 / 201.1 (bl) & 0.185-0.17 & 0.17--0.18 &  LW/SW \\

\ionm{Fe}{xiii} 251.9 / 201.1 (bl) & 1.35--1.45 & 1.4--1.65    &   LW/SW \\

\ionm{Fe}{xiii} 251.9 / 204.9 & 4.6--4.9 & 4.7--5.5   &   LW/SW \\

\ionm{Fe}{xiv} 252.2 / 264.8 & 0.23  &     &  BR LW \\
\ionm{Fe}{xiv} 257.4 / 270.5 & 0.68  &     & BR  LW \\
\ionm{Fe}{xiv} 289.1 / 274.2 &  0.065 &  & BR LW (w) \\
\ionm{Fe}{xiv} 274.2 / 211.3 & 0.71   & 0.71--0.69 &   LW/SW \\
\ionm{Fe}{xiv} 270.5/(264.7+274.2) & 0.265  & 0.265-0.255  & LW \\

\ionm{Fe}{xvi} 251 / 263 & 0.57  & 0.57  &  LW \\
\ionm{Fe}{xvi} 265 / 263  & 0.098  &  0.098      & LW \\

\ionm{Fe}{xvii} { 204.7 / 254.9}  &    & 0.93 &  BR SW/LW \\
\ionm{Fe}{xxiv} { 192/ 255.1 }     &     &  1.85 &  SW/LW \\


\ionm{Si}{x} 253.8 / 258.4 & 0.194   &   & BR  LW \\
\ionm{Si}{x} 277.3 / 272.  & 0.84    &   & BR LW \\

\ionm{Si}{x}  277.3 / 261.1 & 0.68  & 0.68--0.70  &  LW \\

\ionm{S}{x} 257.1 (bl) / 264.2    & 0.348   & 0.348     &  LW \\
\ionm{S}{x} 259.5 / 264.2         & 0.68   &  0.68      &   LW \\

 \ionm{S}{xi} 285.6 (sbl) / 281.4 & 0.496  &    & BR LW \\

\ionm{Fe}{xii} 186.880 / 192.390 & N/A  & N/A  & AR/QS density \\
\ionm{Fe}{xii} 196.640 / 192.390 & N/A  & N/A  & AR/QS density \\
\ionm{Fe}{xiii} 196.525 / 202.044 & N/A  & N/A  & AR density \\
\ionm{Fe}{xiii} 203.826 / 202.044 & N/A  & N/A  & AR density \\
\ionm{Fe}{xiv} 264.787 / 270.524 & N/A  & N/A  & AR density \\
\ionm{Fe}{xiv} 264.787 / 274.203 & N/A  & N/A  & AR density \\
\ionm{Si}{xi} 258.372 / 261.056 & N/A  & N/A  & AR/QS density \\
\ionm{SI}{xi} 258.372 / 271.992 & N/A  & N/A  & AR/QS density \\

\noalign{\smallskip}\hline                                   
\end{tabular}
\normalsize
\label{tab:list_lines}
\end{table}

The general fitting methodology is as follows.
{
First, we select a dataset of  AR and QS observations that are 
relatively quiet and free of transient features and obtain a list of line ratios.
This dataset is an  extension of the observations considered in \cite{delzanna:13_eis_calib}. 
Next, we use the effective areas obtained from the DEM analysis as 
reference points and perform a linear 
interpolation to estimate the EIS effective area curve at each observation 
time and compute the line ratios.} 
Next, we produce a combined set of 
measured AR and QS line ratios by averaging, separately, all AR and QS 
observations within a six-month window and then concatenating the two 
lists together. Then, we fit the EIS effective area curve using a cubic spline in 
MPFIT such that the relative errors of the line ratios are minimized when compared 
to the theoretical values { and the values are close ($\pm$ 10\%) to the 
reference ones}. As an additional constraint, we also simultaneously 
minimize the relative ratios of select pairs of density-sensitive line ratios 
for \ion{Fe}{12}, \ion{Fe}{13}, \ion{Fe}{14}, \& \ion{Si}{10}.
{The inclusion 
of these pairs of line ratios helps ensure that density estimates computed using 
different lines of the same ion yields consistent, or at least comparable, values.
Since only the relative ratio of each pair of density-sensitive ratios is used 
in the minimization, the actual density value computed from each individual line 
ratio can be safely ignored during the fitting process. The pairs of density-sensitive line ratios used for the fitting are listed at the bottom of Table~\ref{tab:list_lines}.

The overall shape of the fit effective areas is sensitive to the number and 
spacing of the spline knots used. More knots allow for a better fit to each 
individual time period but also results in "wavy" profiles with more local 
maxima and minima and year-to-year variation of shapes. On the other hand, 
fewer spline knots results in smoother profiles but lower quality fits. 
Since we do not know the primary cause of the decline in EIS sensitivity, 
we do not have any reason to bias the fit towards wavy or smooth profiles. 
Therefore, after experimenting with a variety of spline knot spacings, we 
settled on using an average spacing of 5~\AA\  between spline knots in both the 
SW and LW bands, with one knot removed from 170 \AA, in the tail of the SW profile, 
and two extra knots added at 192.5 and 197.5 \AA, near the SW peak. This resulted 
in profiles that were moderately smooth, yet still fit the observed line ratios well.
Altogether, we used 20 spine knots to perform the fitting, which is a moderate 
increase over the 14 spline knots used for the fitting in \cite{warren_etal:2014}.
The spline knots used for the preliminary DEM analysis of the off-limb data
are slightly more irregular.
}

{ Lastly, we have reduced residual year-to-year variations by applying a 5-bin boxcar average along the time axis to the effective area results (i.e. effectively using a smoothing width of 2.5 years). This last step produces more long-term consistency, which is is particularly important for lines observed near the tails of the effective area profiles where small variations can result in large changes in calculated intensities.}

\subsection{Accuracy of line ratios and atomic data}

{
Most of the line ratios shown in Table~\ref{tab:list_lines} are the same 
used in \cite{delzanna:13_eis_calib}. 
The atomic data for most of them have been produced by Del Zanna 
with large-scale scattering calculations 
and have been benchmarked against  laboratory and astrophysical 
spectra, as described in a long series of papers, summarised in 
\cite{delzanna_mason:2018}. 
Such calculations and line identifications 
have dramatically improved the atomic data for the EIS spectral region,
and have been made available by Del Zanna via CHIANTI version 8 
\citep{delzanna_chianti_v8} and v.10 \citep{delzanna_etal:2020_chianti_v10}.

A detailed description of the atomic data and benchmark of the 
main line ratios has been given in \cite{delzanna:13_eis_calib} and 
is not repeated here. 
Unfortunately, within the EIS wavelengths, the only well-calibrated 
spectrum is that of \cite{malinovsky_heroux:73}. As shown in 
the benchmark studies and \cite{delzanna:13_eis_calib}, 
there is agreement between the line ratios of strong lines and the predicted ones within 10\% or so, which suggests that 
this is the typical accuracy of theoretical line ratios. The branching ratios also have a known accuracy 
of 10\% or better.  For a few ratios, the uncertainty in the absolute value
might be higher. However,  the CHIANTI atomic models are very accurate in 
identifying how a ratio varies with the electron density: in 
Table~\ref{tab:list_lines} we list the maximum expected variations 
in the ratios, for typical quiet Sun and active region observations.
For most of the ratios we considered in the appropriate set of 
observations, the expected variations are also within 10\%. 
And even if the absolute values of the ratios were offset from reality 
by more than 10\%, enforcing a ratio to be constant over time still gives 
a good constraint on the relative values of the effective areas at the 
wavelengths of the two lines.
}

\subsection{Absolute calibration}

Obtaining an absolute calibration is notoriously difficult. 
\cite{delzanna:13_eis_calib} noted that using the ground calibration, where
a few wavelengths were calibrated against laboratory standards,
the radiances of the SW lines in the QS were close to  previously
measured values, for observations in 2006-2007. This indicated that 
the ground calibration was relatively accurate for the SW channel. One problem with that
comparison was the paucity of previous radiance observations, 
and significant solar variability. Also, most previous measurements
were of the Sun as a star, and the irradiances had to be converted to
radiances.

Another indication that the SW ground calibration was correct in
2010 came with direct comparisons of EIS SW vs. SDO AIA 193~\AA\ observations,
as e.g. described in \cite{delzanna_etal:2011_aia,delzanna:2013_multithermal}.
As the SW observes the main wavelengths contributing that AIA band,
it is in principle straightforward to cross-calibrate the entire SW channel
against AIA, after taking into account the different
spatio-temporal resolutions, and carefully removing the EIS bias.
The AIA  193~\AA\ calibration was checked summing the signal from the
full Sun and comparing it against the predicted values from the SDO EVE irradiances.
{A similar study was carried out by 
\cite{2014SoPh..289.2377B}. }

In turn, the EVE calibration  relies on a comparison with
a few sounding rocket flights and adjustments using line ratios,
following the methods  adopted for EIS  \citep{delzanna:13_eis_calib}.
The prototype EVE flown on the sounding rocket flights
is regularly calibrated on the ground. The absolute calibration of the
EVE prototype is deemed accurate to within 20\%, although
detailed comparisons carried out on the first flight in 2008 showed
larger discrepancies (40\%) for some among the strongest lines
\citep{delzanna_andretta:2015,delzanna:2019_eve}.
Calibration flights were obtained in 2011, 2012, 2013, 2018, 2021 plus a recent
one in 2023.

The failure of EVE MEGS-A in 2014 meant that no direct AIA/EVE
cross-calibrations could be carried out after that date. 
After 2014, the only useful cross-calibration EVE sounding rockets
were flown in 2018, 2021, and 2023. The decrease in the sensitivity of the
AIA  193~\AA\ band has been linearly interpolated between 2014 and 2018,
which brings some additional uncertainty. The calibration of the recent
flights is still ongoing. 

Any cross-calibration is never straightforward, especially when
considering broad-band observations such as the AIA ones, as a
wavelength-dependent degradation would affect the results.
Such degradation is very common in space, and indeed has been
measured by EVE using different filters.
Fortunately, the overall count rates in the AIA  193~\AA\ band
 only show a decrease of about 10\% over the first
four years, when compared to EVE data. This indicates  moderate
degradation, and confirms that this band is usable for the EIS
calibration.

The absolute values for the EIS responsivity have been 
constrained by a direct comparison between the EIS SW and the 
AIA 193~\AA\ count rates using several observations after {June} 2010.
We gave preference to active region observations to have a better
signal in the EIS spectra. The background in the EIS spectra has been 
fitted pixel-by-pixel on the spectra in DN.
 This is a continuum dominated by the CCD bias, which is not measured
independently as in other instruments. 
To predict the AIA DN/s 
we have used the AIA effective areas obtained using the /evenorm 
and /time dependent keywords. We checked this AIA calibration 
against a few simultaneous SDO EVE observations, using the latest
EVE calibration, finding agreement to within a few percent. The 
AIA data have been resampled to the temporal and spatial resolution 
of EIS. On a pixel-by-pixel basis, there is generally a large scatter 
of values, partly due to the different point spread functions (PSF) 
of the instruments, and the lack of exact simultaneity in the 
observations. Nevertheless, averaging over spatial areas reduces the 
scatter, so the relative calibration can be estimated with only a 
few percent accuracy. More details for how the AIA data were processed 
are provided in  Appendix B.

 We obtained first the relative calibration between all the 
EIS wavelengths, then applied the SW calibration to the EIS data,
and from the EIS/AIA comparison obtained the absolute scale for the 
EIS effective area after June 2010. No scaling was applied 
before 2010.

\section{Data Selection and Analysis}

\subsection{Selected EIS observations}

{ We extended the analysis of  the dataset considered in
\cite{delzanna:13_eis_calib}, i.e. a set of on-disk QS regions, and a set of
AR observations. Whenever available, off-limb AR observations were used in order 
to reduce the contamination of the coronal lines from the cooler lines and opacity 
effects. We primarily selected EIS observations from the "ATLAS" series of studies, 
which consist of four data windows spanning the entire spectral range observed by 
EIS and taken at exposure times of 30, 60, or 120 s, depending on the target region. 
The ATLAS series, however, was not developed and run regularly until May of 2010. For 
time periods before this, we used other EIS observations that spanned the full spectral 
range, typically from the "SYNOP001" study which had exposure times of 90 s. For the 
DEM analysis we selected seven off-limb QS observations between 11 March 2007 and 
28 June 2019. Please see Tables \ref{tab:list_qs} and \ref{tab:list_ar} in the 
Appendix for a full list of the EIS observations used.}

\subsection{Data processing}
           {For each observation, suitable  areas (typically 100\arcsec$\times$100\arcsec\ wide
             were selected to obtain
averaged spectra and all known lines were fit.
The data were processed with custom-written 
software as described here and in  \cite{delzanna_etal:2011_aia}.
{
  The software (available on ZENODO\footnote{DOI: 10.5281/zenodo.10462448,
  see https://zenodo.org/records/10462448}) follows
very closely the standard EIS procedures available via SolarSoft.
}
The hot, warm and dusty pixels are flagged as missing,
following the database available via SolarSoft.
Unfortunately, the number of residual warm pixels
not flagged has increased significantly in data
during the past few years. The location of the residual
warm pixels is also variable 
{and the standard software fails to flag all the 
anomalous pixels. 
We have therefore avoided the most 
affected regions in each EIS raster, selecting areas with visual 
inspection.}
However, the weakest lines after 2018 are significantly
compromised by the presence of these warm pixels, which have the
effect of increasing their intensities.

{
The pixels  affected by particle events  are also flagged as `missing'.
We used the routine {\it  new\_spike}. 
All the exposures had to be visually inspected as often data 
dropouts at some wavelengths are present, as well as storms of 
particle hits, unusually strong cosmic rays, etc.
For visual inspection, the missing
pixels are then replaced with interpolated values. 
The full spectral 
atlases are then rotated to remove the slant in the spectra, 
effectively aligning the two SW and LW channels, using a modification
of the software written by B.Thompson for the CDS NIS. 
The spatial offset of 18 pixels between the SW and LW 
channels along the slit has been corrected, as well as the 
spatial offset along the dispersion direction for the early
datasets. 

As the dark frames (bias) are extremely variable and unknown, 
a base minimum value is removed from each exposure. 
Spatial regions for each date were carefully selected after visual inspection,
to obtain spatially averaged spectra before fitting the lines.
The spectra were normalised by the exposure times and the size of the slit,
to obtain an uniform set of spectra in DN/s per 1\arcsec\ per bin.
These averaged spectra are available as well as the results of the fitting.
}

Total data numbers (DN) in the lines are obtained with Gaussian fitting and 
a polynomial background. The background and line fitting is obtained 
over different spectral ranges. We note that the bias is very difficult 
to estimate in the central part of the SW channel. Different choices 
for the background can affect the weakest lines by a few percent. 
EIS saturation is near 16000 DN, however we found indications of
some non-linear effects for lower values \citep{delzanna_etal:2019_ntw},
so we have avoided areas near saturation.

\subsection{Selected line ratios and opacity in Fe XII lines}

{For fitting the EIS relative calibration curves, we selected several 
line ratios which are known to be rather constant. These ratios are listed 
in Table~\ref{tab:list_lines}. Most of the ratios are the same as used in 
\cite{delzanna:13_eis_calib}. Only a limited number of line ratios are well
measured in on-disk QS regions. Lines formed above 1.5 MK such as
\ion{Fe}{13} are only well observed in AR observations. The best ones are 
branching ratios, as they do not depend on electron density and have a 
small theoretical uncertainty, on the order of 10\%. The other ratios have 
a larger typical uncertainty, around 20\%. 
We then used different ratios for the QS and AR 
observations. Some ratios have some small, weak density 
sensitivity, which is indicated in the "Note" column of the Table. 
While most ratios are used to constrain the relative 
calibration within only the SW or LW channels, we also selected some 
some ratios to constrain the relative calibration of the two channels.}

The strongest EIS SW lines in the quiet Sun are three Fe XII decays to the ground state,
from $^4$P states, at 192.4, 193.5, and 195.1~\AA.
The latter line is a self-blend with a weak density-sensitive
transition \citep{delzanna_mason:05_fe_12}, which only contributes a
few percent for very high densities. 
The intensity ratios of the 192.4, 193.5, and 195.1~\AA\ lines 
are not sensitive to density or  temperature.

An analysis of hundreds  of EIS observations of different features, on-disk, off-limb,
with different exposures, slit combinations 
found several anomalies in the 192.4, 193.5, and 195.1~\AA\ lines \citep{delzanna_etal:2019_ntw}.
The main ones affect the instrumental widths of the 193.5, and 195.1~\AA\ lines, which are anomalous,
and reduced intensities of the same lines (compared to the weaker 192.4~\AA),
in all on-disk active region and many off-limb observations.
The 195~\AA\ line is typically weaker than expected by 30--40 \%.
Note that the intensity of the 195.1~\AA\ line should
actually increase in active region observations, due to the presence
of the above-mentioned self-blend, and not decrease.
The only explanation found for the anomalous ratios of these lines was
that optical depth effects are  ubiquitous.

The intensity ratios of the 192.4, 193.5, and 195.1~\AA\ lines
in on-disk QS observations are stable and have been used to constrain
the relative calibration between these wavelengths.
There are a few other strong decays to the ground states of other
ions, with large oscillator strengths, however establishing
opacity in these other cases is non-trivial.
For this reasons, we rely more on the line ratios of the
on-disk QS observations to assess the validity of our revised calibration.

We have also preferred active region observations off-limb,
and selected areas away from the bright moss regions, to reduce 
possible effects in some line ratios due to non-Maxwellian electron 
distributions (NMED). In fact, some evidence based on EIS and IRIS 
observations have indicated that NMED are present in active regions
\citep{2020ApJ...893...34L, delzanna_etal:2022}.

\section{Results}

\subsection{QS off-limb DEM analysis}

Table~\ref{tab:lines_11_mar_2007} lists the selection of coronal lines
and the results of the DEM for the QS off-limb
observation on 11 Mar 2007. Only a few lines from iron, as listed in the
Table, were used
for the DEM, so the results are independent of chemical abundance issues.
The resulting DEM is shown in  Fig.~\ref{fig:eis_dem_11_mar_2007}.  
The points for all the coronal lines are plotted at the temperature $T_{\rm max}$
where the line contribution function $G(T)$ has a maximum. In the ordinate, we plot the ratio of the  theoretical vs. the observed
 intensity multiplied by the DEM value at that temperature, to provide a visual indication of which main lines constrain the DEM, and how close the predicted 
 radiances are to the observed ones. 
The peak of the DEM is close to log $T$ [K] = 6.15, typical of the QS near the limb,
but the distribution is not strictly isothermal.
Nevertheless, most of the strongest lines are formed at this
temperature, which makes the comparison between lines from different
ions much easier.
This can be seen in Table~\ref{tab:lines_11_mar_2007} where 
$T_{\rm max}$, as well as the effective temperature $T_{\rm eff}$:
\beq
T_{\rm eff} = \int G{\left({T}\right)}~
DEM{\left({T}\right)} ~T~dT / 
{\int G{\left({T}\right)}~DEM{\left({T}\right)}~dT} \quad ,
\eeq
are listed. $T_{\rm eff}$ is  an average temperature more indicative of 
where a line is formed.
The Table also shows the total counts in each line and the
calibrated radiance obtained with the present calibration.
$R$ is the ratio between the predicted and observed radiances.
 The program calculates the emissivities of all the lines in the
  CHIANTI database that are within each observed line width. All the
  line emissivities are then summed up; $r$ in  Table~\ref{tab:lines_11_mar_2007}
gives the fractional contribution to the total of the main lines.

We can see that generally all lines are well reproduced within
$\pm$15\%, but examining lines from the same ion, relative
differences are generally within $\pm$5\%.
Also note that the list includes lines that are strongly density
dependent, indicating that the isodensity approximation is
relatively accurate.
To calculate the line emissivities we used
a constant electron density of 2$\times$10$^8$  cm$^{-3}$.
The few \ion{Si}{10} lines in the LW channel are also very useful
to constrain the LW sensitivity. Typically, the silicon
abundance variations are in line with those of iron, so again
the results are virtually independent of the chemical abundances.

A few of the extremely weak lines, with total data numbers below 100
are not well reproduced. That is of no concern. 
The intensity ratios of the 192.4, 193.5, and 195.1~\AA\ lines
indicate some opacity effects.

\begin{figure}[!htbp]
\centering
\includegraphics[clip,width=5cm,angle=-90]{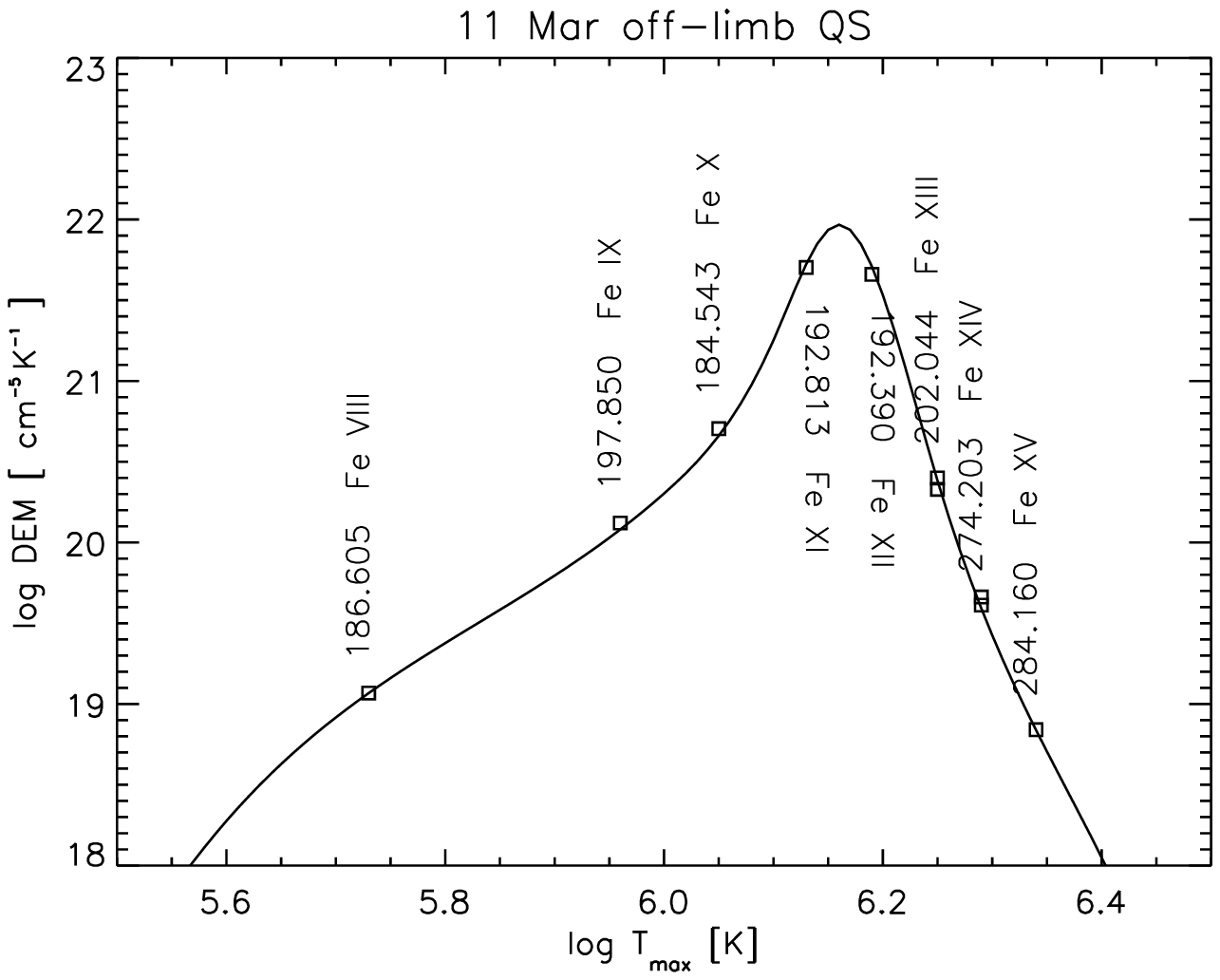}
  \caption{DEM for the QS off-limb on 11 Mar 2007.
The points are plotted at the temperature $T_{\rm max}$, 
 and at the theoretical vs. the observed
 intensity ratio multiplied by the DEM value.  }
\label{fig:eis_dem_11_mar_2007}
\end{figure}

\begin{table}[!htbp]
  \caption{QS off-limb  observation on 11 Mar 2007 (90 s, 1\arcsec\ slit). 
    $\lambda_{\rm obs}$ and  $\lambda_{\rm exp}$ (\AA) are the observed and CHIANTI wavelengths,
DN are total counts, $I_{\rm obs}$
is the  radiance (phot cm$^{-2}$ s$^{-1}$ 
arc-second$^{-2}$).
$T_{\rm max}$ and $T_{\rm eff}$ are the maximum and effective temperature 
(log values, in K),
$R$ the ratio between predicted and observed radiances,  and 
$r$ the fractional contribution to the blend. 
The lines used for the DEM have a * while those density-dependent have a dd.} 
\centering
  \scriptsize
  \setlength\tabcolsep{3.pt}
\begin{tabular}{@{}rrrccrlrr@{}}
\hline\hline \noalign{\smallskip}
 $\lambda_{\rm obs}$  & DN & $I_{\rm obs}$   & log $T_{\rm max}$ & log $T_{\rm eff}$  & $R$ & Ion & $\lambda_{\rm exp}$   &  $r$ \\
\noalign{\smallskip}\hline\noalign{\smallskip} 
 194.66 &       90 & 1.2 &  5.72 &  6.08 &  0.72 &  \ionm{Fe}{viii} &  194.661 & 0.92 \\ 
 
 185.22 &       61 & 4.0 &  5.73 &  6.08 &  1.02 &  \ionm{Fe}{viii} &  185.213 & 0.94 \\ 
 
 * 186.60 &       70 & 3.3 &  5.73 &  6.08 &  1.00 &  \ionm{Fe}{viii} &  186.598 & 0.83 \\ 
 
 189.94 &      155 & 3.9 &  5.94 &  6.12 &  0.91 &  \ionm{Fe}{ix} &  189.935 & 0.95 \\ 
 
 188.49 &      247 & 8.2 &  5.95 &  6.12 &  0.93 &  \ionm{Fe}{ix} &  188.493 & 0.89 \\ 
 
 * 197.85 &      268 & 4.6 &  5.96 &  6.13 &  0.91 &  \ionm{Fe}{ix} &  197.854 & 0.93 \\ 
 
 257.26 &      464 & 54.7 &  6.04 &  6.14 &  0.89 &  \ionm{Fe}{x} &  257.259 & 0.21 \\ 
                             &    &   &  &  &  &  \ionm{Fe}{x} &  257.263 & 0.76 \\ 
 
 177.24 &       87 & 101.1 &  6.05 &  6.14 &  1.05 &  \ionm{Fe}{x} &  177.240 & 0.97 \\ 
 
 174.53 &       53 & 179.2 &  6.05 &  6.14 &  1.03 &  \ionm{Fe}{x} &  174.531 & 0.98 \\ 
 
 * 184.54 &      580 & 45.4 &  6.05 &  6.14 &  0.90 &  \ionm{Fe}{x} &  184.537 & 0.96 \\ 
 
 dd 190.04 &      728 & 18.2 &  6.06 &  6.15 &  1.03 &  \ionm{Fe}{xii} &  190.040 & 0.17 \\ 
                           &      &   &  &  &  &  \ionm{Fe}{x} &  190.037 & 0.71 \\ 
 
 257.55 &      142 & 16.4 &  6.12 &  6.15 &  1.15 &  \ionm{Fe}{xi} &  257.554 & 0.70 \\ 
                           &      &   &  &  &  &  \ionm{Fe}{xi} &  257.547 & 0.26 \\ 
 
 256.92 &      233 & 28.2 &  6.12 &  6.15 &  0.86 &  \ionm{Fe}{xi} &  256.919 & 0.91 \\ 
 
 dd 182.17 &      178 & 30.0 &  6.13 &  6.16 &  1.02 &  \ionm{Fe}{xi} &  182.167 & 0.97 \\ 
 
 188.30 &     1920 & 66.1 &  6.13 &  6.16 &  1.07 &  \ionm{Fe}{xi} &  188.299 & 0.97 \\ 
 
 188.22 &     3110 & 108.8 &  6.13 &  6.16 &  1.06 &  \ionm{Fe}{xi} &  188.216 & 0.97 \\ 
 
 * 192.81 &     1440 & 23.7 &  6.13 &  6.16 &  1.05 &  \ionm{Fe}{xi} &  192.813 & 0.96 \\ 
 
 180.40 &      608 & 194.2 &  6.13 &  6.16 &  1.19 &  \ionm{Fe}{xi} &  180.401 & 0.97 \\ 
 
dd 202.71 &      141 & 9.3 &  6.13 &  6.16 &  1.14 &  \ionm{Fe}{xi} &  202.705 & 0.94 \\ 
 
 253.79 &       72 & 11.3 &  6.15 &  6.16 &  0.90 &  \ionm{Si}{x} &  253.790 & 0.97 \\ 
 
dd 258.37 &      536 & 58.7 &  6.15 &  6.16 &  0.90 &  \ionm{Si}{x} &  258.374 & 0.97 \\ 
 
dd 271.99 &      475 & 30.4 &  6.15 &  6.16 &  0.89 &  \ionm{Si}{x} &  271.992 & 0.97 \\ 
 
dd 261.06 &      375 & 34.6 &  6.15 &  6.16 &  0.90 &  \ionm{Si}{x} &  261.056 & 0.98 \\ 
 
dd 277.27 &      273 & 23.5 &  6.15 &  6.16 &  0.89 &  \ionm{Si}{x} &  277.264 & 0.98 \\ 
 
dd 256.40 &      435 & 54.8 &  6.16 &  6.16 &  1.07 &  \ionm{Si}{x} &  256.377 & 0.63 \\ 
                               &  &   &  &  &  &  \ionm{Fe}{xii} &  256.410 & 0.18 \\ 
 
dd 196.64 &      491 & 7.4 &  6.19 &  6.16 &  1.17 &  \ionm{Fe}{xii} &  196.640 & 0.93 \\ 
 
dd 186.88 &      610 & 27.3 &  6.19 &  6.16 &  1.09 &  \ionm{Fe}{xii} &  186.854 & 0.16 \\ 
                               &  &   &  &  &  &  \ionm{Fe}{xii} &  186.887 & 0.77 \\ 
 
dd 203.73 &      132 & 12.3 &  6.19 &  6.17 &  0.79 &  \ionm{Fe}{xii} &  203.728 & 0.96 \\ 
 
 193.51 &     6640 & 99.5 &  6.19 &  6.17 &  1.23 &  \ionm{Fe}{xii} &  193.509 & 0.95 \\ 
 
 * 192.39 &     2860 & 50.5 &  6.19 &  6.17 &  1.12 &  \ionm{Fe}{xii} &  192.394 & 0.96 \\ 
 
 195.12 &    10300 & 136.0 &  6.19 &  6.17 &  1.31 &  \ionm{Fe}{xii} &  195.119 & 0.97 \\ 
 
dd 196.52 &      170 & 2.5 &  6.24 &  6.17 &  0.99 &  \ionm{Fe}{xiii} &  196.525 & 0.89 \\ 
 
dd 200.02 &      303 & 7.6 &  6.24 &  6.17 &  0.77 &  \ionm{Fe}{xiii} &  200.021 & 0.94 \\ 
 
dd 261.73 &       40 & 3.6 &  6.25 &  6.18 &  1.10 &  \ionm{Fe}{xiii} &  261.743 & 0.97 \\ 
 
dd 203.83 &      243 & 23.3 &  6.25 &  6.18 &  1.12 &  \ionm{Fe}{xiii} &  203.795 & 0.28 \\ 
                               &  &   &  &  &  &  \ionm{Fe}{xiii} &  203.826 & 0.62 \\ 
 
dd 246.21 &       50 & 15.6 &  6.25 &  6.18 &  1.03 &  \ionm{Fe}{xiii} &  246.209 & 0.95 \\ 
 
 * 251.95 &      144 & 26.7 &  6.25 &  6.18 &  1.13 &  \ionm{Fe}{xiii} &  251.952 & 0.97 \\ 
 
dd 204.94 &       57 & 7.4 &  6.25 &  6.18 &  0.88 &  \ionm{Fe}{xiii} &  204.942 & 0.96 \\ 
 
 209.92 &       51 & 18.6 &  6.25 &  6.18 &  0.97 &  \ionm{Fe}{xiii} &  209.916 & 0.96 \\ 
 
 * 202.04 &     2040 & 103.0 &  6.25 &  6.18 &  0.96 &  \ionm{Fe}{xiii} &  202.044 & 0.97 \\ 
 
 dd 264.79 &      296 & 24.2 &  6.29 &  6.18 &  0.89 &  \ionm{Fe}{xiv} &  264.789 & 0.67 \\ 
                               &  &   &  &  &  &  \ionm{Fe}{xi} &  264.772 & 0.30 \\ 
 
 * 274.20 &      349 & 24.2 &  6.29 &  6.19 &  0.96 &  \ionm{Fe}{xiv} &  274.204 & 0.95 \\ 
 
 252.20 &       19 & 3.5 &  6.29 &  6.19 &  1.01 &  \ionm{Fe}{xiv} &  252.200 & 0.94 \\ 
 
 dd 257.40 &       55 & 6.5 &  6.29 &  6.19 &  1.05 &  \ionm{Fe}{xiv} &  257.394 & 0.96 \\ 
 
 dd 270.52 &      128 & 9.4 &  6.29 &  6.19 &  1.06 &  \ionm{Fe}{xiv} &  270.521 & 0.97 \\ 
 
 * 211.32 &       80 & 37.1 &  6.29 &  6.19 &  0.85 &  \ionm{Fe}{xiv} &  211.317 & 0.96 \\ 
 
* 284.16 &      169 & 31.3 &  6.34 &  6.20 &  1.00 &  \ionm{Fe}{xv} &  284.163 & 0.96 \\ 
\end{tabular}
\normalsize
 \label{tab:lines_11_mar_2007} 
\end{table}

The other observations were processed
in a similar way, and the results are provided in the Appendix.
Generally, the consistency between observed and predicted radiances
is the same, indicating that for those observations, the 
accuracy of the relative sensitivity at specific wavelengths is
within $\pm$15\%. 
{Note that the calibrated intensities listed in the 
Tables are those obtained with the provisional 
calibration, adjusted to take into account the EIS/AIA
relative calibration.}
We used the set of effective areas obtained 
in this way as a baseline when using the line ratios as shown below.

\subsection{On-disk AR loop DEM analysis}

Several low-$T$ lines become prominent in AR loops, and they are
useful to cross-check the relative calibration at wavelengths
different than those of the coronal lines.
AR loop legs are nearly isodensity and isothermal
\citep[cf][and references therein]{delzanna_mason:2018} so they
are also good sources for the cross-calibration.

As we are not interested in the coronal lines
in this case, only the cooler lines are considered to check
the relative calibration.
As multiple isothermal  structures are present in AR cores
\citep[cf][]{delzanna:2013_multithermal}, obtaining a DEM
which reproduces well all lines is challenging,
compared to the QS off-limb case.
Nevertheless, as we are primarily concerned to check the
relative agreement between lines from the same ion or
lines with similar effective temperatures, an accurate estimate of the DEM is not necessary. 

We have performed the analysis on a few cool loops at different dates,
and found general consistency using the effective areas
obtained with the previous method. 
One example is provided in the Appendix.

\subsection{Line radiances}

\begin{figure}[!htbp]
\centering
\includegraphics[clip,width=5cm,angle=90]{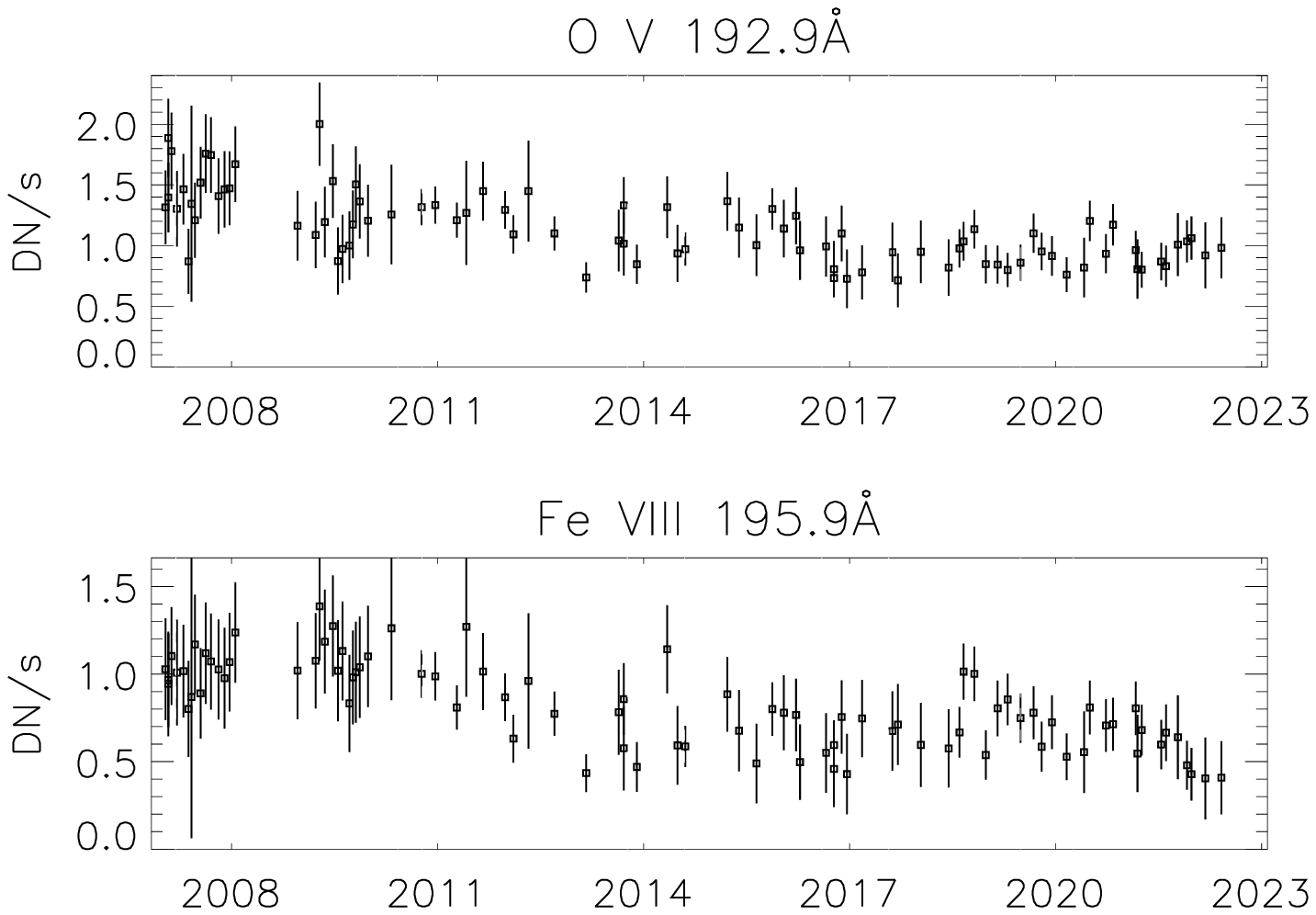}
\smallskip
\includegraphics[clip,width=5cm,angle=90]{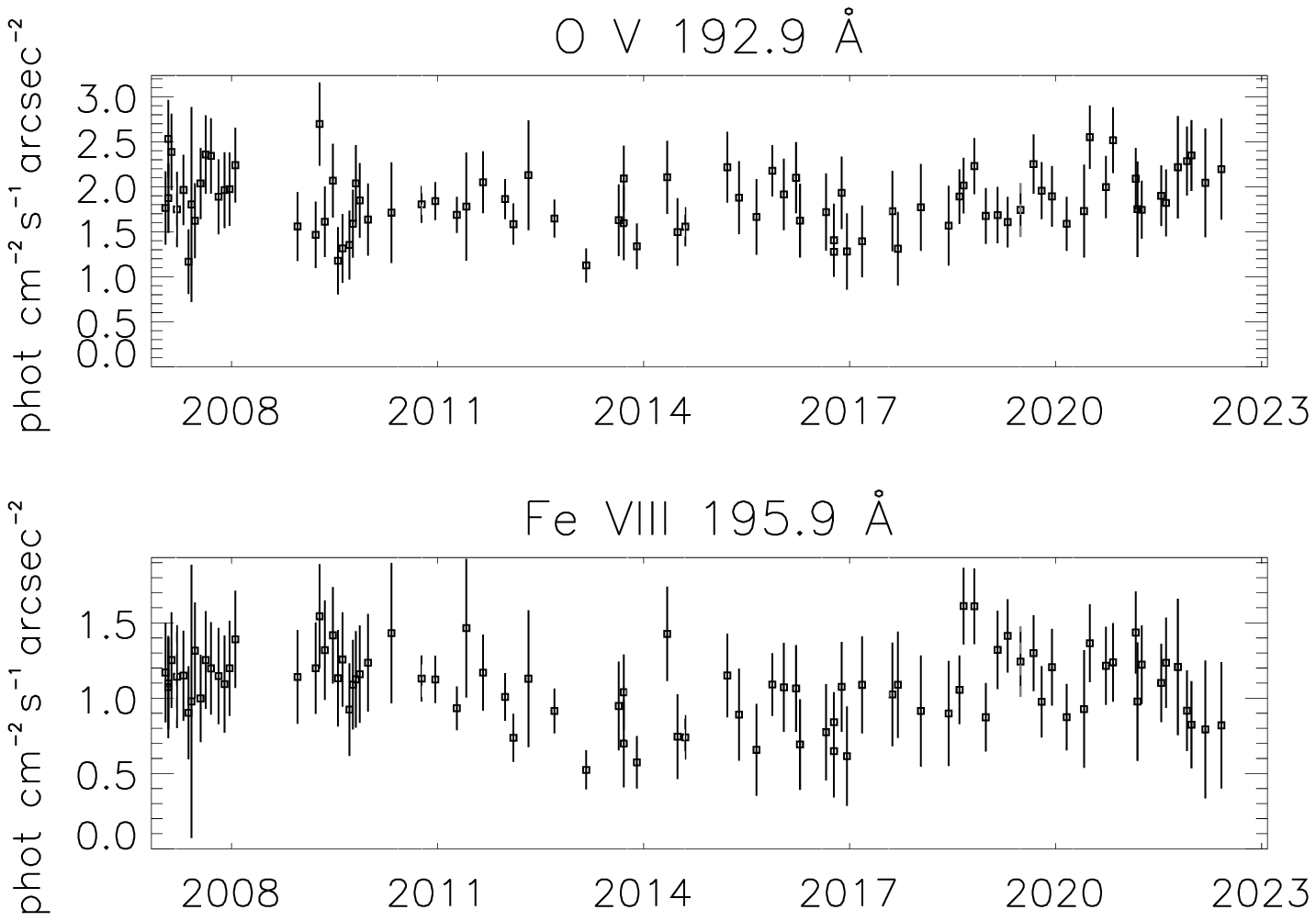}
\caption{Top: Hinode EIS quiet Sun count rates (DN/s) in two cooler SW lines.
Bottom: radiances with the present calibration. }
\label{fig:qs_sw_int}
\end{figure}

The large solar variability precludes the use of QS radiances for the
EIS calibration, as carried out for the SOHO CDS, as only two low-temperature lines
are available, the \ion{He}{2} 256.3~\AA\ and the \ion{O}{5} 192.9~\AA.
The former is blended  with several coronal lines, and the latter is weak
and hard to measure. 
However, several transition-region lines are expected to show a small
variation with the solar cycle, and should have similar radiances during the
2008 and 2019 minima. 
Fig.~\ref{fig:qs_sw_int} (top) shows the count rates in two cool SW lines.
It is clear that there is a decrease in the signal, especially after 2010.
Fig.~\ref{fig:qs_sw_int} (bottom) shows the corrected radiances with the
present calibration.

\begin{figure}[!htbp]
\centering
\includegraphics[clip,width=5cm,angle=90]{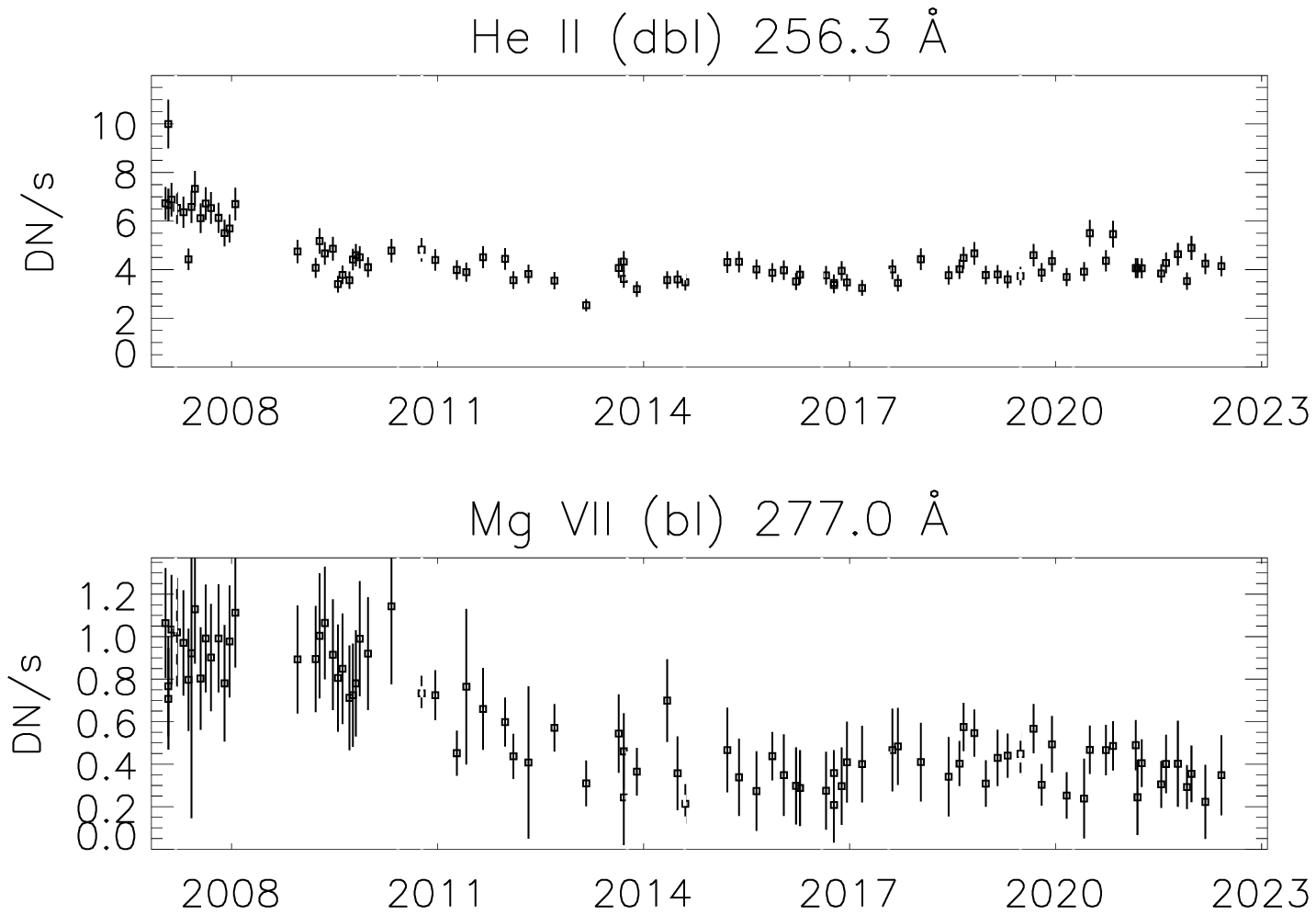}
\smallskip\smallskip
\includegraphics[clip,width=5cm,angle=90]{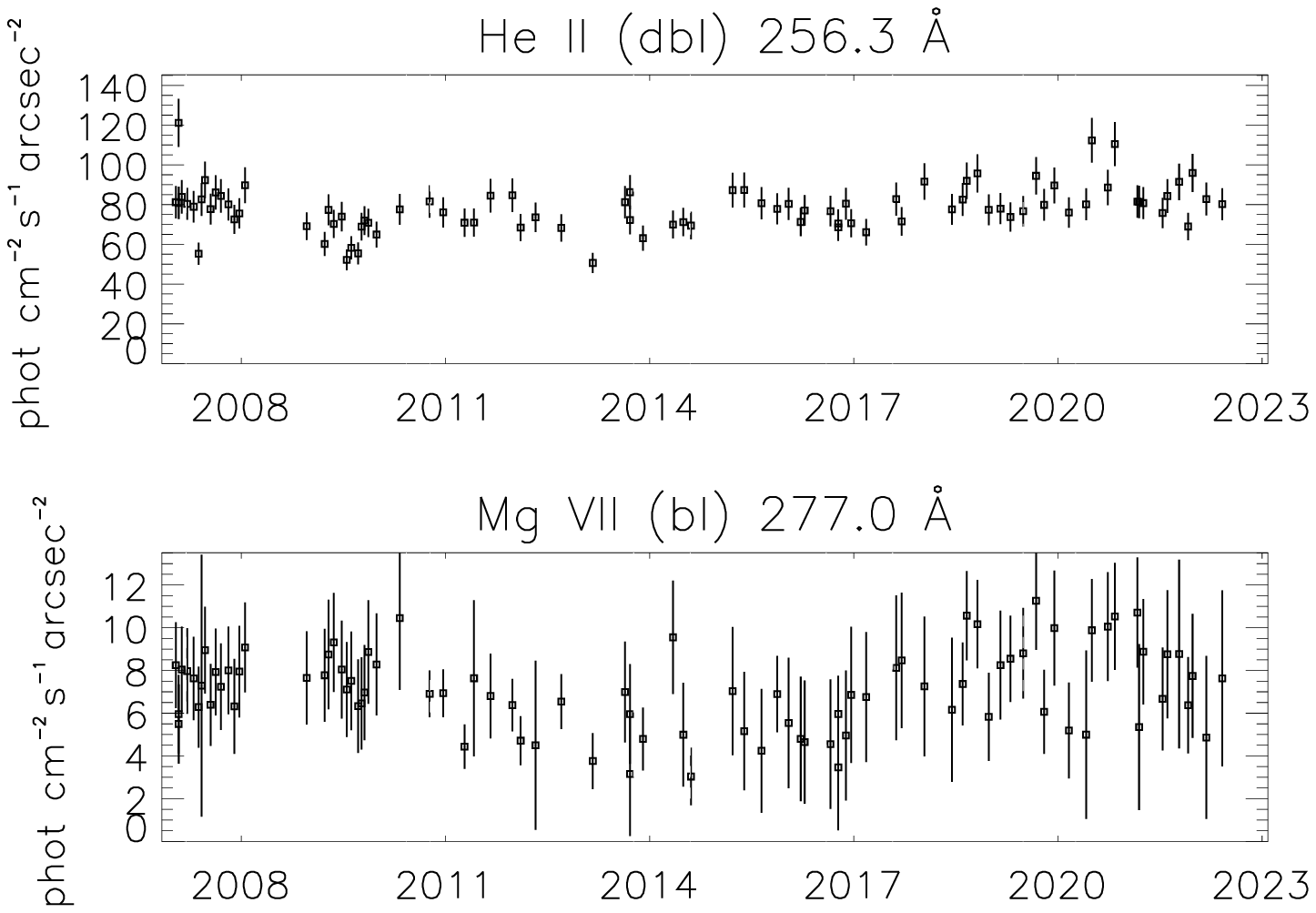}
\caption{Top: Hinode EIS quiet Sun count rates in two cooler LW lines.
Bottom: radiances with the present calibration. }
\label{fig:qs_lw_int}
\end{figure}

Fig.~\ref{fig:qs_lw_int} (top) shows the count rates in
the  \ion{He}{2} 256.3~\AA,
which has been deblended from its coronal contributions,
and a weak \ion{Mg}{7} line. 
The signal in the \ion{He}{2} line during 2007 was relatively constant, but then had a significant
drop in 2010, with a subsequent constant behaviour.  On the other hand,
all the lines at longer wavelengths show a very different behaviour:
a nearly constant one until 2010, followed by a drop until 2016.
This is a clear indication of a very different decrease in sensitivity
at different wavelengths, which is difficult to explain and account for.

\subsection{Line ratios}

\begin{figure*}
\includegraphics[width=17cm]{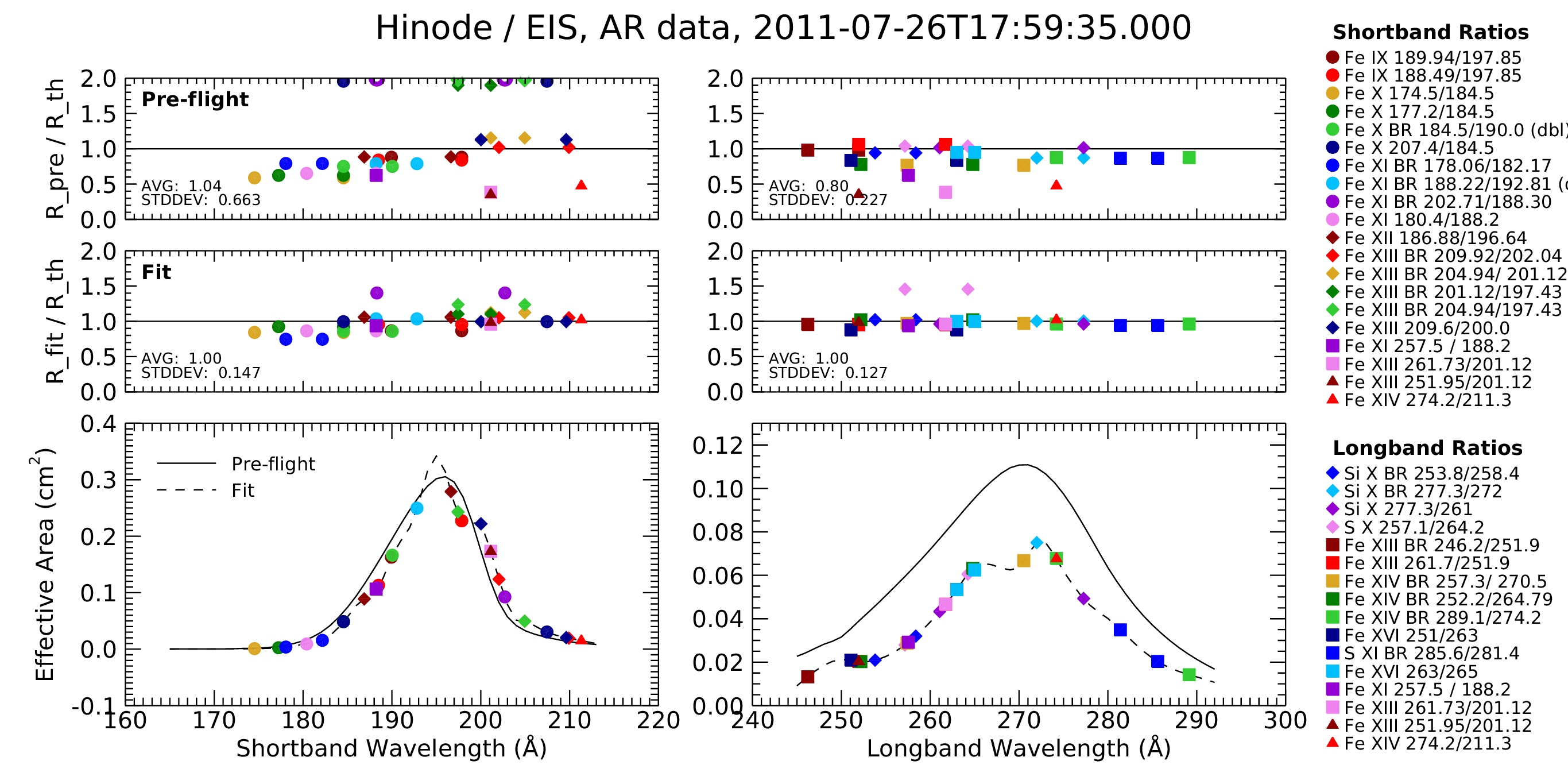}
  \caption{Example fit effective area curves for an active region observed on 2011-07-26.}
\label{fig:fit_ar_2011}
\end{figure*}

\begin{figure*}
\includegraphics[width=17cm]{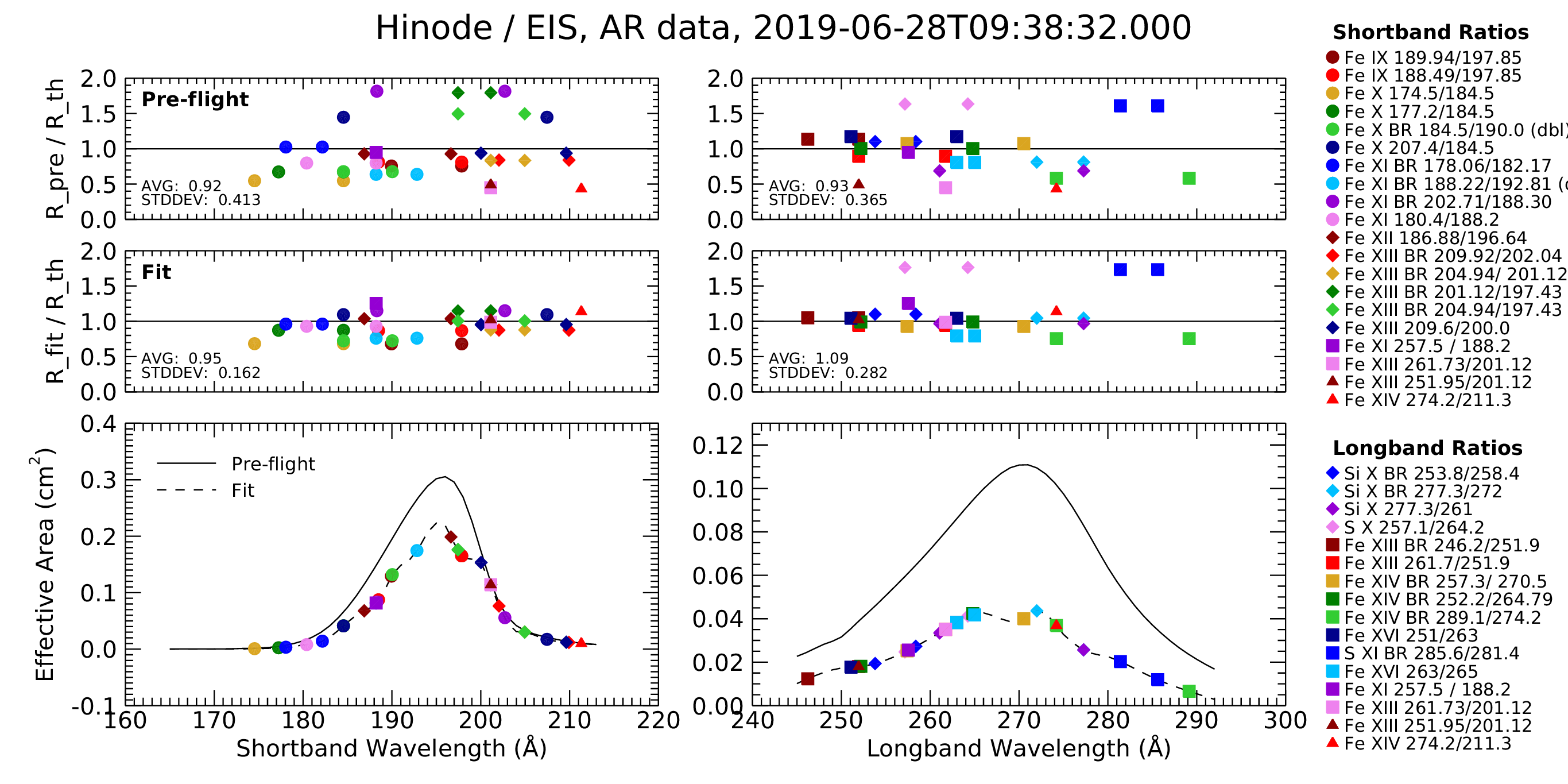}
  \caption{Example fit effective area curves for an active region observed later in the mission on 2019-06-28.}
\label{fig:fit_ar_2019}
\end{figure*}

\begin{figure*}
\includegraphics[width=17cm]{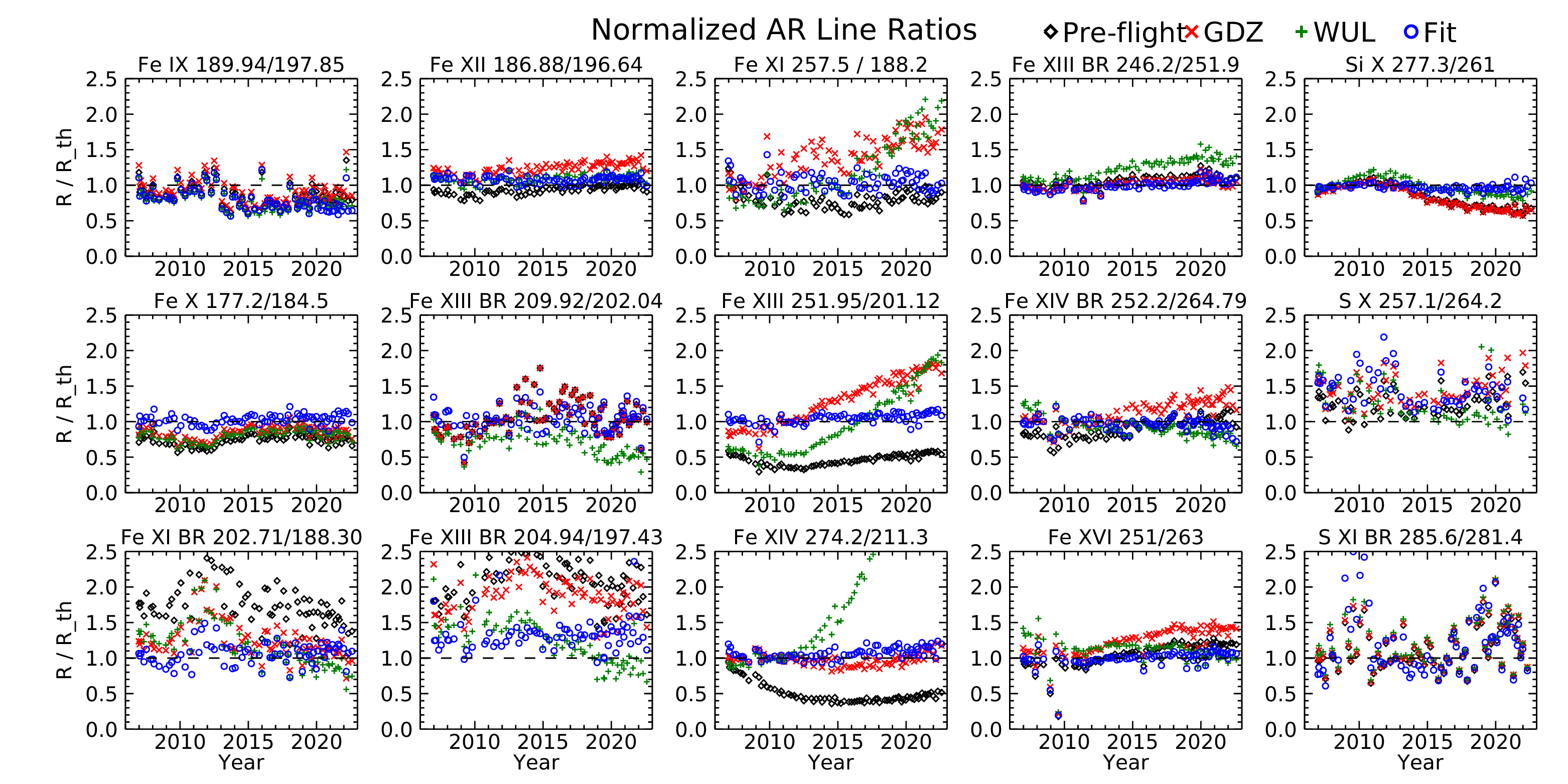}
  \caption{Relative line ratios observed in AR over time. Ratios computed using the pre-flight (black diamonds), \cite{delzanna:13_eis_calib} (GDZ; red X's), \cite{warren_etal:2014} (WUL; green crosses), and new fit (blue circles) calibrations are shown for comparison.}
\label{fig:ar_rel_ratios}
\end{figure*}

\begin{figure*}
\includegraphics[width=17cm]{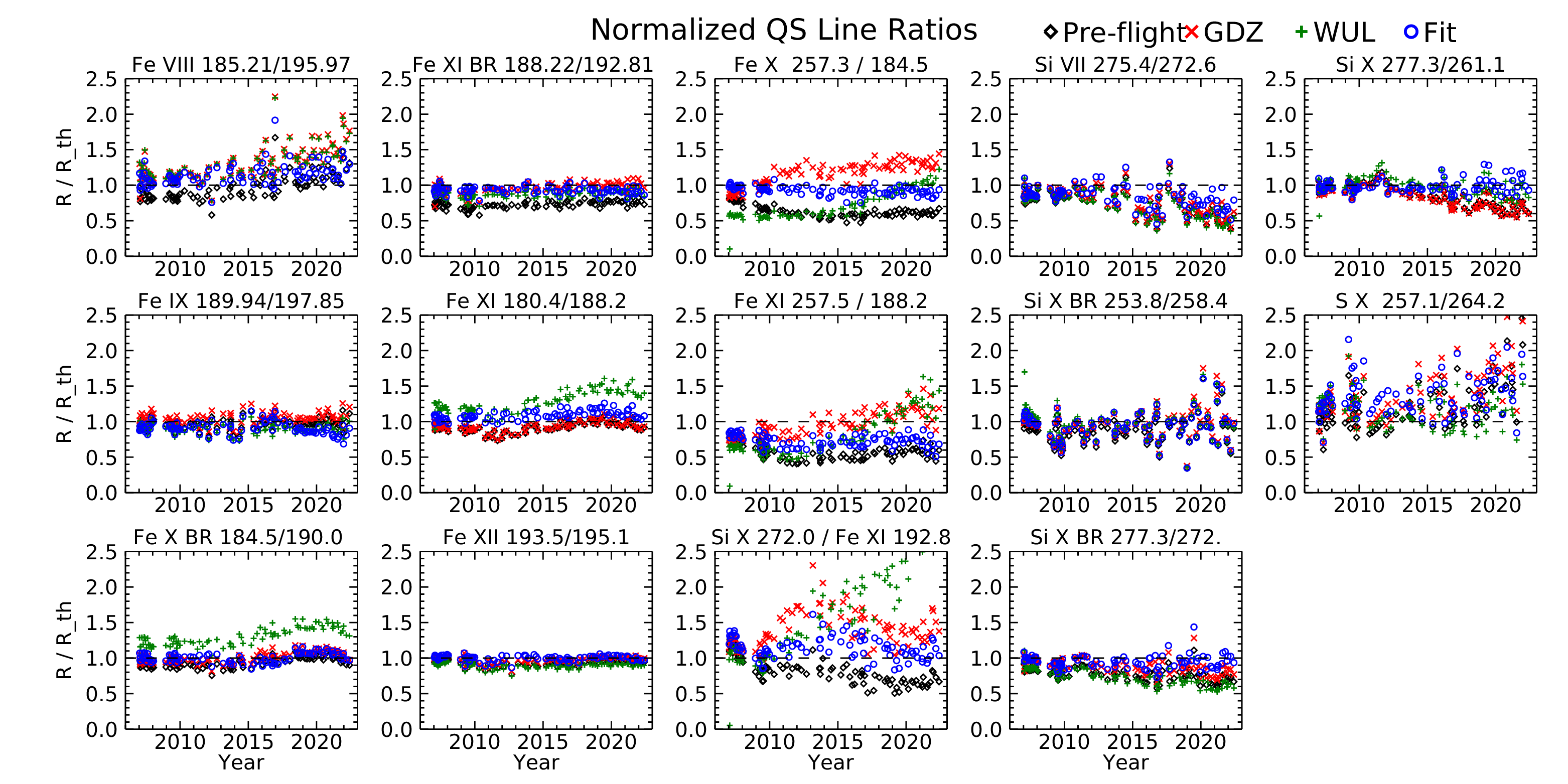}
  \caption{Relative line ratios observed in QS over time. Ratios computed using the pre-flight (black diamonds), \cite{delzanna:13_eis_calib} (GDZ; red X's), \cite{warren_etal:2014} (WUL; green crosses), and new fit (blue circles) calibrations are shown for comparison.}
\label{fig:qs_rel_ratios}
\end{figure*}

Figures ~\ref{fig:fit_ar_2011} and ~\ref{fig:fit_ar_2019} show example fit effective 
areas for two different ARs observed, respectively, in 2011 and 2019.
The left column of each figure contain the SW effective areas while the 
right column contain the LW. The top two panels in both columns show the 
relative line ratios of the pre-flight (top) and fit (middle) effective area 
profiles. A relative ratio of 1 indicates a perfect match with the theoretical
line ratio. In both AR, the fit profiles greatly improve the relative ratios. 
Compared to the pre-flight profiles, the fit effective areas exhibit finer-scaled 
variations and have notable "shoulders" in both the SW (around 197 -- 200 \AA) 
and LW (~265 -- 269 \AA) bands. The LW profiles shows more significant degradation
over time than the SW profiles.

Figures ~\ref{fig:ar_rel_ratios} and ~\ref{fig:qs_rel_ratios} compare 
selected line ratios computed over time using the pre-flight (black diamonds) 
and fit (blue crosses) effective areas. The ratios were  normalized by their 
respective theoretical values. 
{We also show for comparison the results obtained with the earlier calibrations, by \cite{delzanna:13_eis_calib} (GDZ; red X's) and \cite{warren_etal:2014} (WUL; green crosses).
}

In both figures, the left two columns contain 
lines in the SW channel, the right two columns show ratios in the LW 
channel, and the middle column contains cross-channel ratios. Most of the fit line 
ratios show dramatic improvement over the pre-flight ratios, with new values 
falling within 20\% of the theoretical values. However, a few ratios show little 
to no improvement.

\subsection{Flare observations: \ion{Fe}{24} and \ion{Fe}{17}}

\begin{figure}
\centerline{\includegraphics[width=0.45\textwidth]{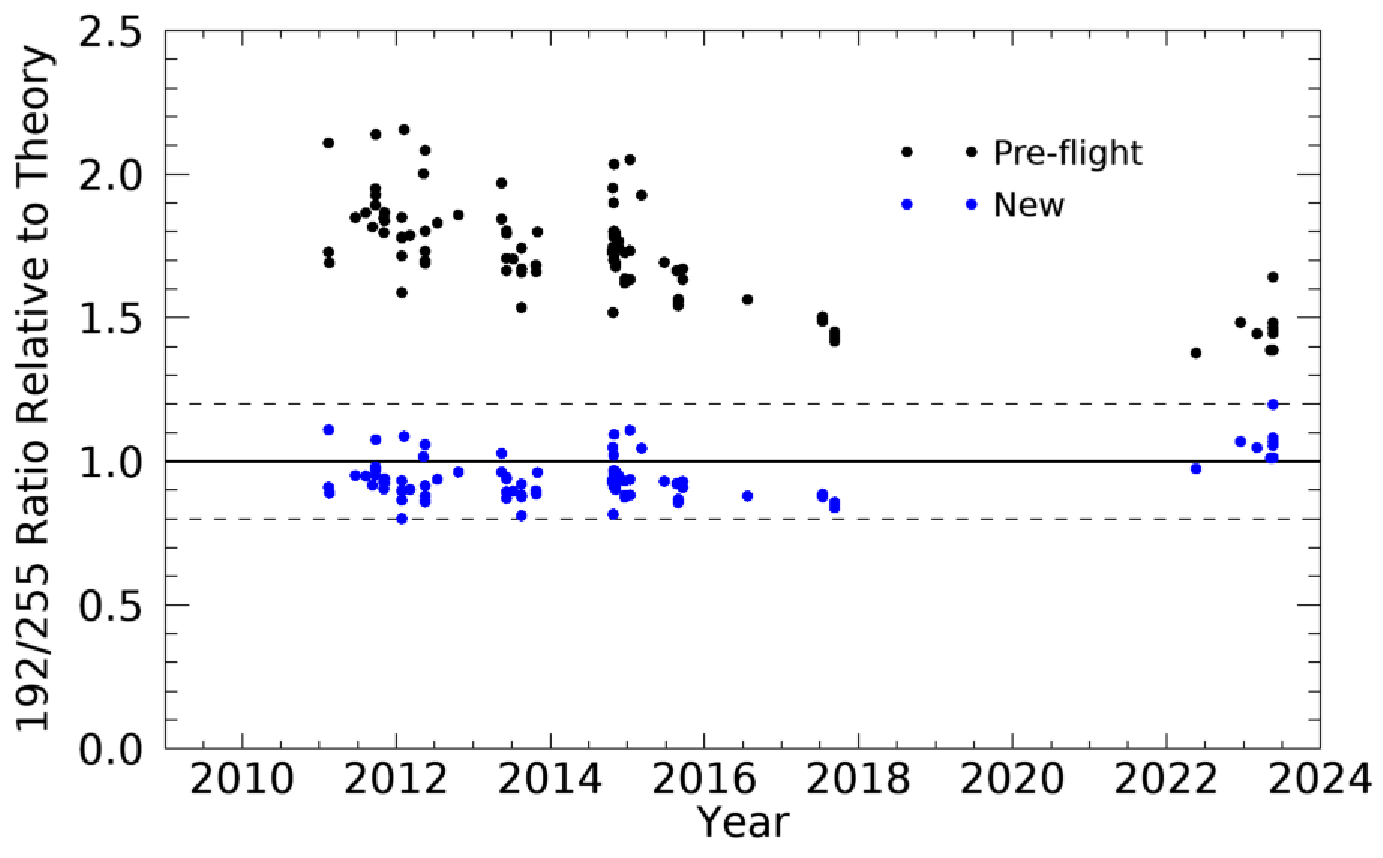}}
  \caption{The \ion{Fe}{24} 192/255~\AA\ ratio relative to theory as a function of time using the pre-flight calibration and the present time-dependant effective areas.}
\label{fig:eis_fe24_ratio}
\end{figure}

\begin{figure}[!htbp]
\centering
\includegraphics[clip,width=6.5cm,angle=90]{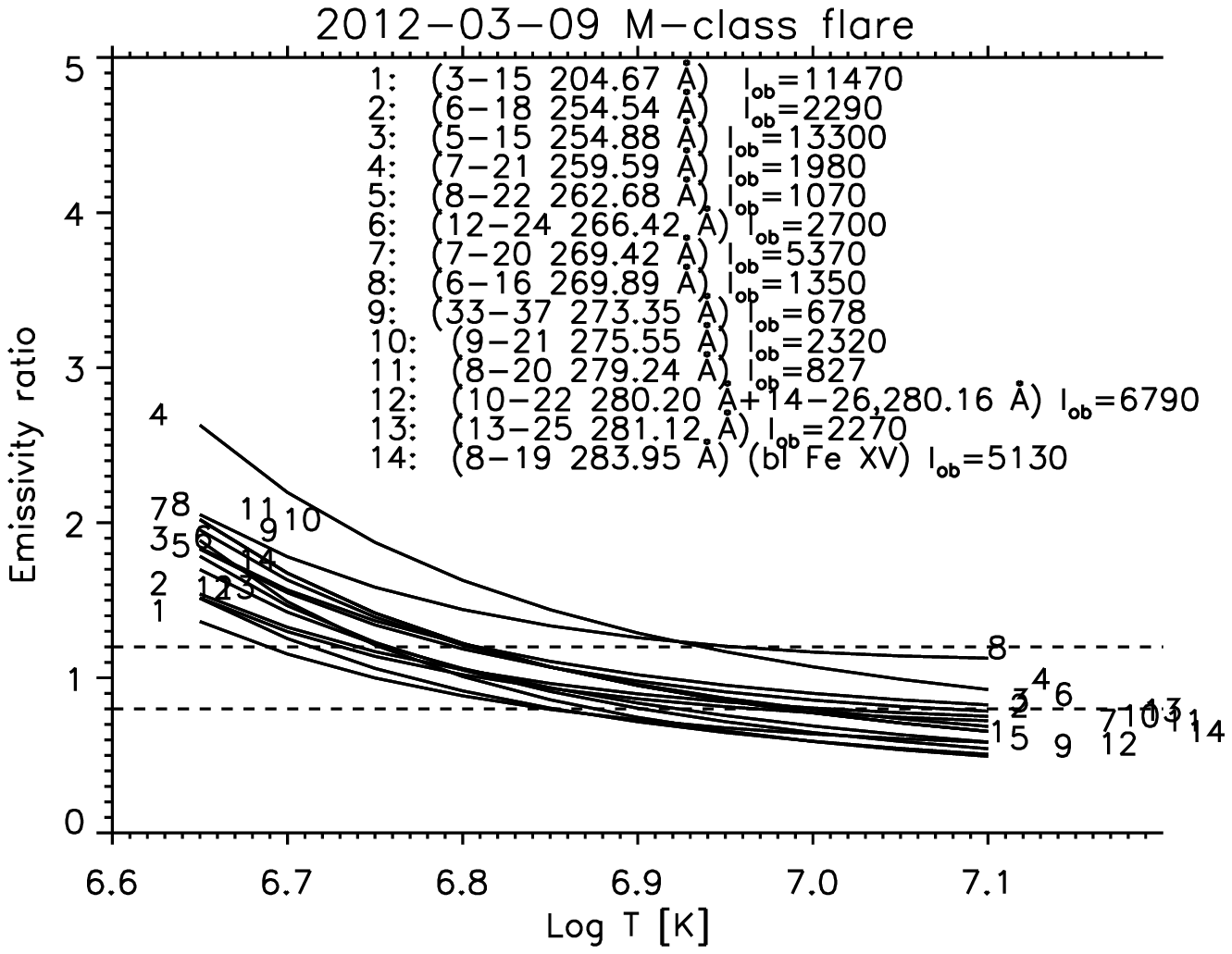}
\includegraphics[clip,width=6.5cm,angle=90]{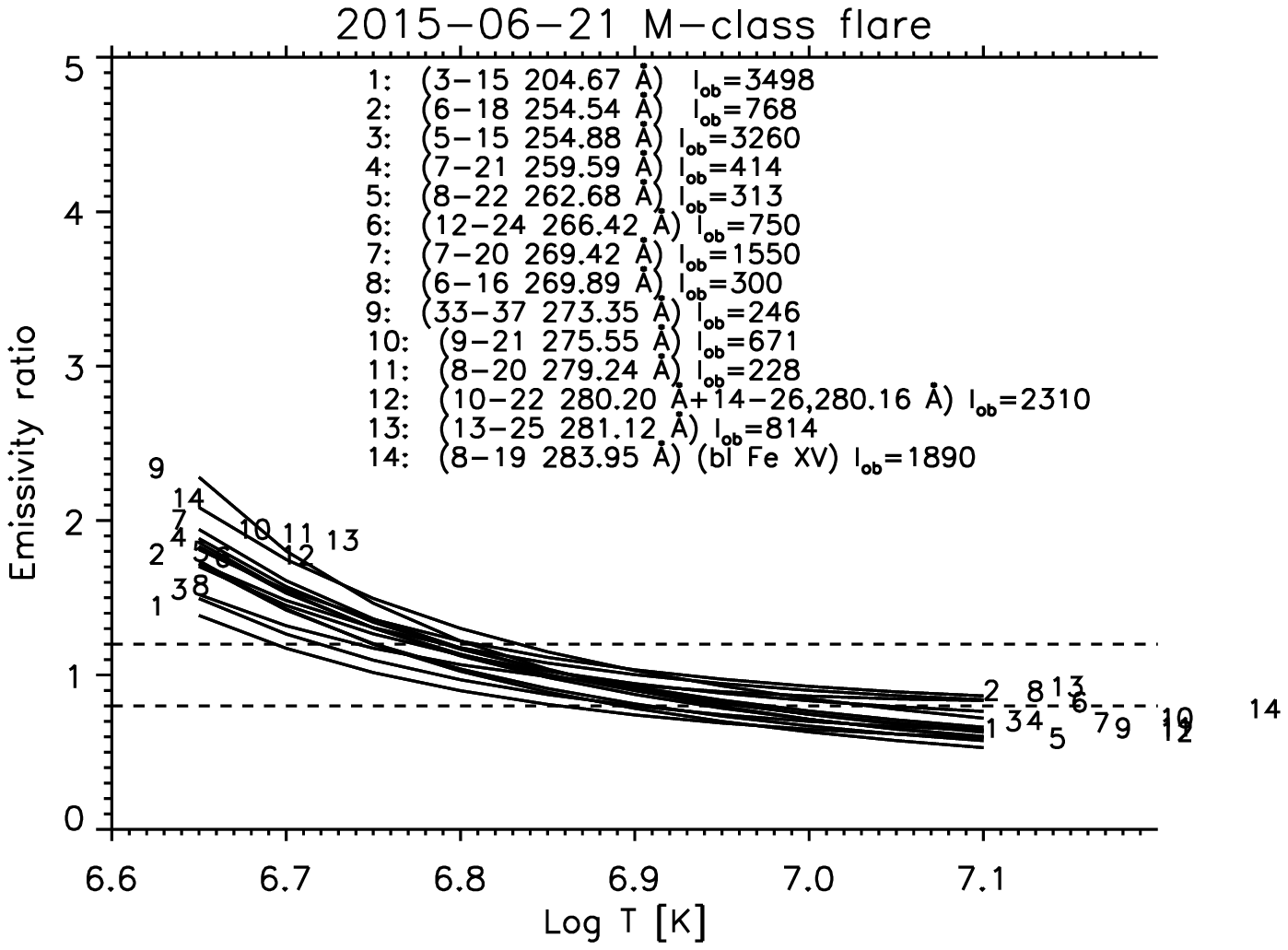}
  \caption{Emissivity ratio  curves obtained with the present calibration for the 
    main \ionm{Fe}{xvii} EUV transitions observed  by Hinode  EIS during two M-class flares,
    on 2012 March 9 and 2015 June 21. 
I$_{\rm ob}$ is the calibrated measured intensity in erg cm$^{-2}$ s$^{-1}$ sr$^{-1}$.}
  \label{fig:fe_17_flare}
\end{figure}

A few lines become prominent in flare observations and offer
the possibility to check the relative calibration at different wavelengths.
As only a few flares have been observed by EIS with a sufficient number
of lines, the check is rather limited. 
The best ions are  \ion{Fe}{24} and \ion{Fe}{17}.

The \ion{Fe}{24} 192/255~\AA\ is a good ratio to validate the SW/LW cross-calibration, as it does not depend on density or temperature, and could only be affected by H, He absorption or opacity effects. The lines are usually blended with several transitions \citep[ e.g.][]{delzanna:08_bflare}, but for strong flares the \ion{Fe}{24} is dominant. We have computed the \ion{Fe}{24} 192 and 255~\AA\ intensities for all available EIS \verb+FlareResponse+ and \verb+HH_Flare+ rasters and manually selected observations where the flare produced more than 50 pixels with statistically significant intensities, but were not dominated by pixels with saturation or diffracted signal. As is seen in Figure~\ref{fig:eis_fe24_ratio}, when using the ground calibration the ratio departs by more than 50\% from the expected value. The few data points between 2015 and 2018 also indicate a trend for the ratio to decrease. Using the updated calibration described in this paper brings the ratios to agreement with theory to within $\pm$20\%. 

As reviewed by \cite{delzanna_ishikawa:09},  \ion{Fe}{17} produces several 
lines in the EIS channels. The 204.6 and 254.9~\AA\ form a branching ratio,
for which atomic data  are very accurate, and therefore are useful to 
validate the SW/LW cross-calibration.  Many other \ion{Fe}{17} lines are 
available across the LW channel, although they become usable and unblended 
only in larger flares. The line ratios do not have any density dependence and 
negligible temperature dependence, hence are excellent for the calibration.

Figure~\ref{fig:fe_17_flare} shows the emissivity ratio  curves  relative to  the 
main \ionm{Fe}{xvii} lines observed  by Hinode  EIS on two 
M2-class flares.
  With two exceptions,   
  all the  \ion{Fe}{17} lines (several
  of which were  identified by \cite{delzanna_ishikawa:09})
have   emissivities that agree with theory within $\pm$ 20\%.
The first exception is the weak line No.4, which actually has a wavelength
of 259.71~\AA, and is likely blended. The second is a very weak line 
observed by EIS at  288.95~\AA, and not shown in the Figures.

\subsection{Absolute calibration before 2010}

Establishing the absolute calibration before 2010 is difficult.
We would normally expect the sensitivity at all wavelengths to decrease with time,
hence the effective areas at the beginning of the mission should be higher  than
in 2010. However, many published  studies indicated that the ground calibration
in the SW channel was reasonable for the 2007 data, except for 
the shorter wavelengths, as shown in \cite{delzanna:13_eis_calib}.
\cite{wang_etal:2011} performed a direct comparison between EIS SW 
and EUNIS quiet Sun (QS) observations in 2007, finding a small (20\%)
decrease compared to the ground calibration, a variation within 
the  EUNIS calibration uncertainties (the EUNIS SW channel did not have a direct
 radiometric calibration, but a line ratio method was used).

Another way to assess the calibration is to check that the quiet Sun calibrated 
radiances in low-temperature lines agree with literature values. 
There is always a scatter in the radiances of the quiet Sun, but on average 
they are relatively constant and largely independent of the solar cycle.
In fact, they have been used for the in-flight calibration of the SOHO CDS 
instrument \citep[cf][]{delzanna_andretta:2015,andretta_delzanna:2014}. 
As in \cite{delzanna:13_eis_calib}, we have checked that the calibrated radiances
are within 20--30\% the literature values, for the few lines 
for which measurements were found. Furthermore, we have carried out an emission measure analysis
using averaged quiet Sun CDS and EIS radiances, to check that we have consistency
within 20\%. Results will be discussed in a separate paper. 

\subsection{Absolute calibration after 2010: EIS vs. AIA}

\begin{figure}
\centerline{\includegraphics[width=0.45\textwidth]{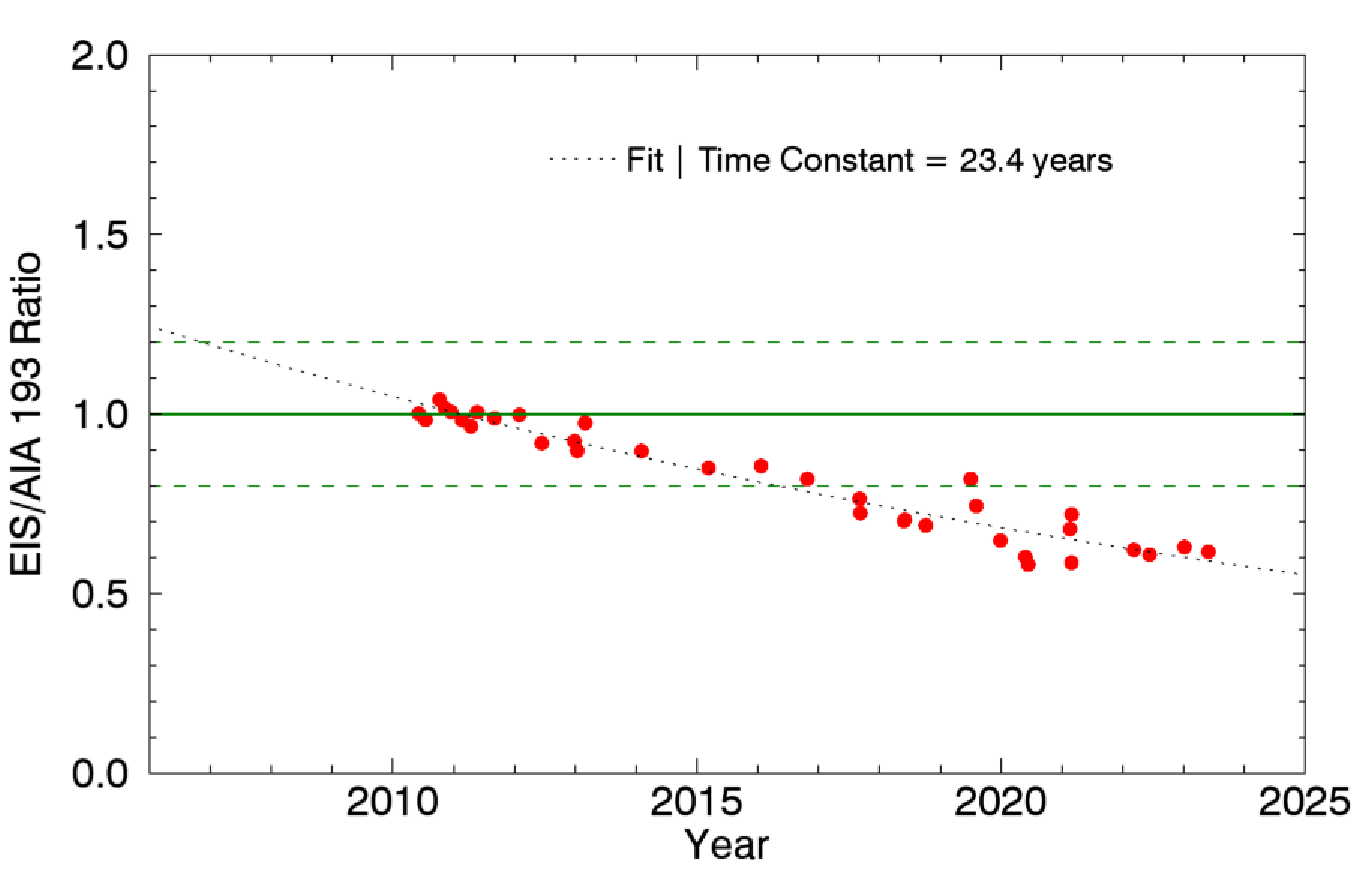}}
\centerline{\includegraphics[width=0.45\textwidth]{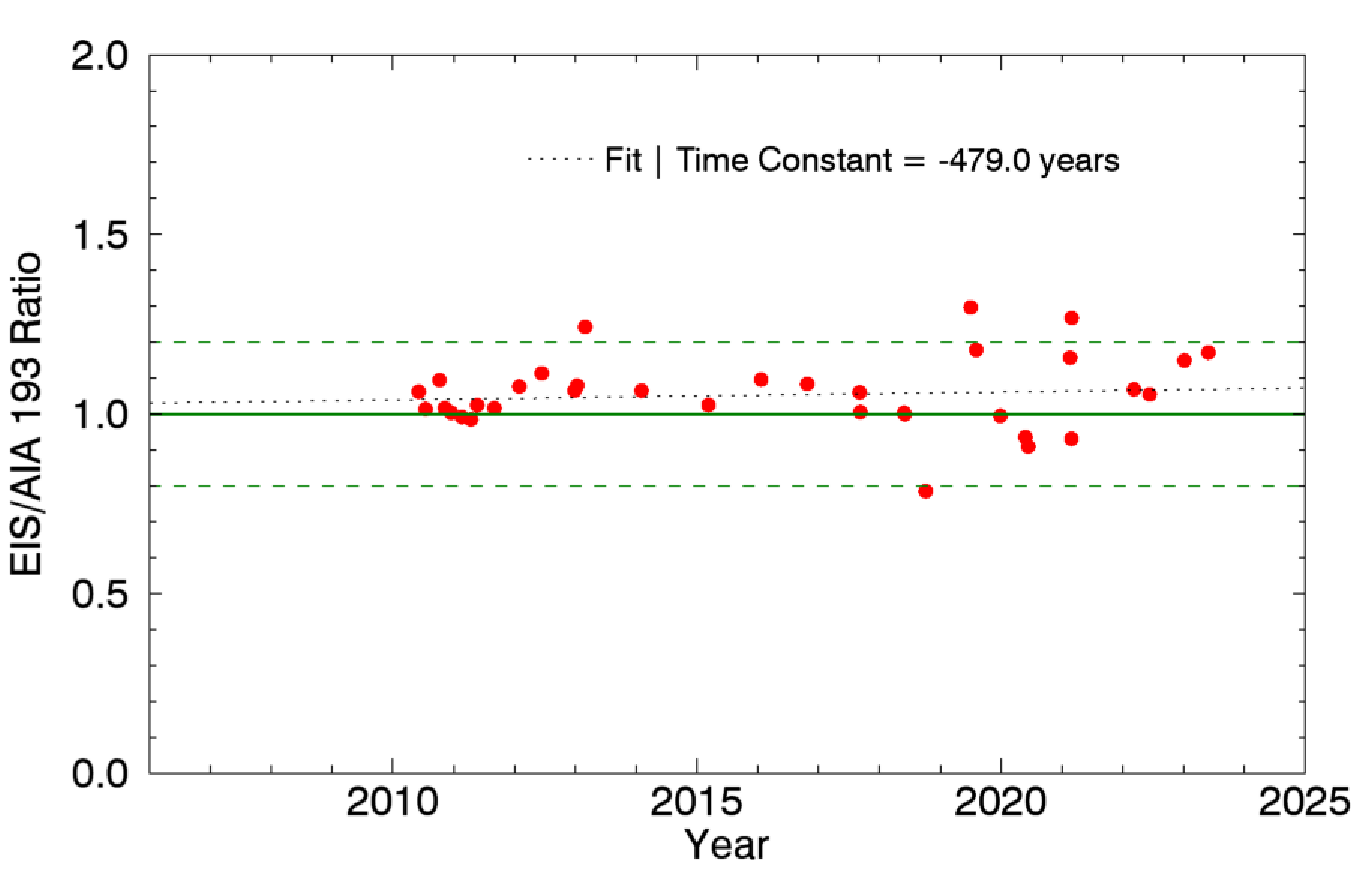}}
  \caption{The ratio of EIS 193 to AIA 193~\AA\ as a function of time using the pre-flight calibration (top panel) and the new, time-dependant effective areas (bottom panel). The EIS intensities have been convolved with the AIA 193 effective area to compute the expected AIA 193 count rates. }
\label{fig:eis_aia_ratio}
\end{figure}

The averaged ratio between predicted and measured AIA DN
obtained using the EIS ground calibration indicates a degradation of nearly a factor of two, as shown in Fig.~\ref{fig:eis_aia_ratio} (top). This is clearly due to
a decrease in the sensitivity of the EIS SW channel with time.

To obtain the EIS absolute calibration, we have first used the 
 revised EIS SW relative calibration
 obtained with the previous methods  to obtain calibrated units,
 to then convert them to  predicted AIA counts.
 We have then applied corrections factors to the grid of selected dates,
 to obtain a better match with AIA. The results are shown in
 Fig.~\ref{fig:eis_aia_ratio} (bottom).

\begin{figure*}
 \centerline{\includegraphics[clip,width=0.65\linewidth]{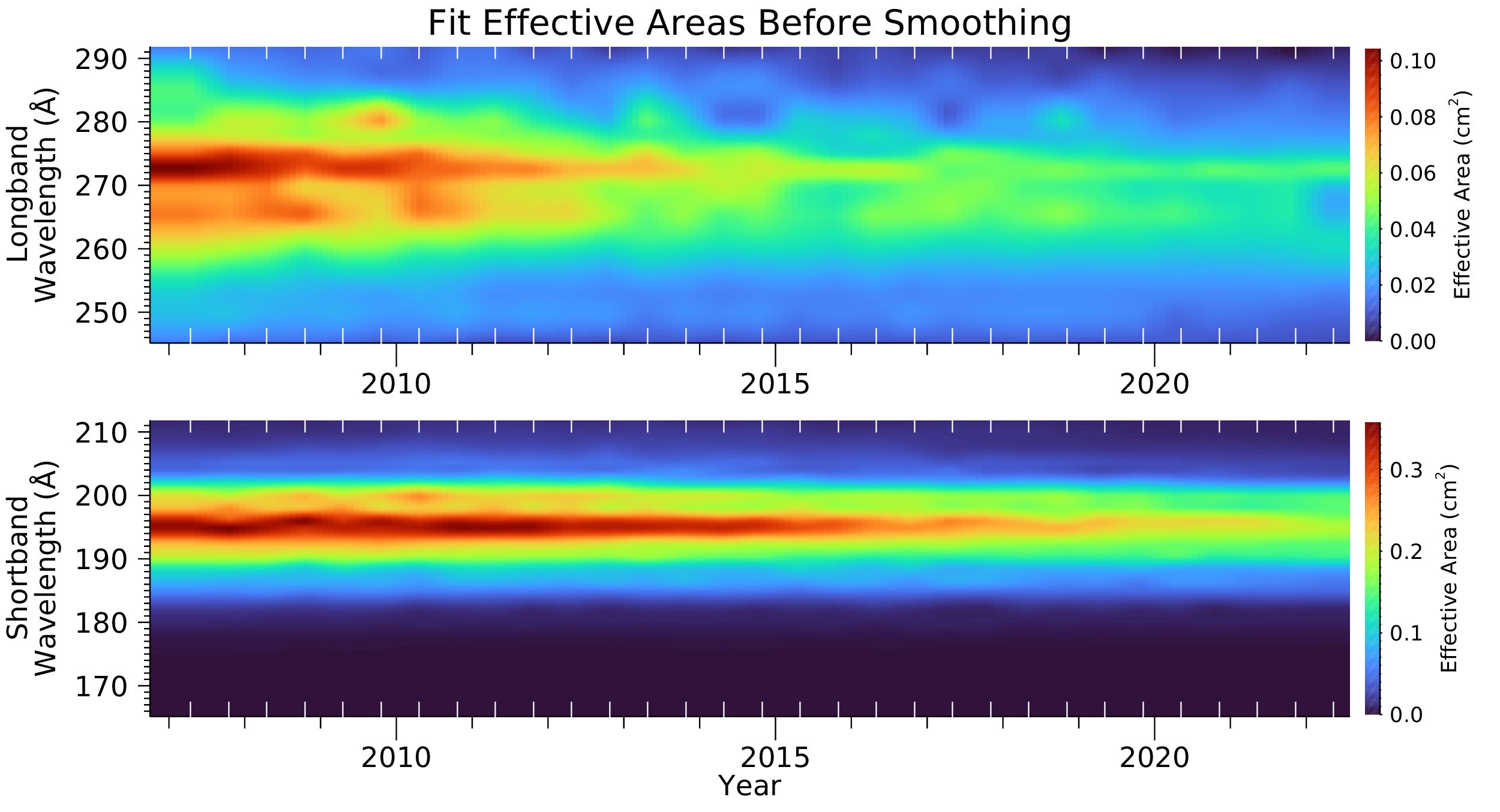}}
 \centerline{\includegraphics[clip,width=0.65\linewidth]{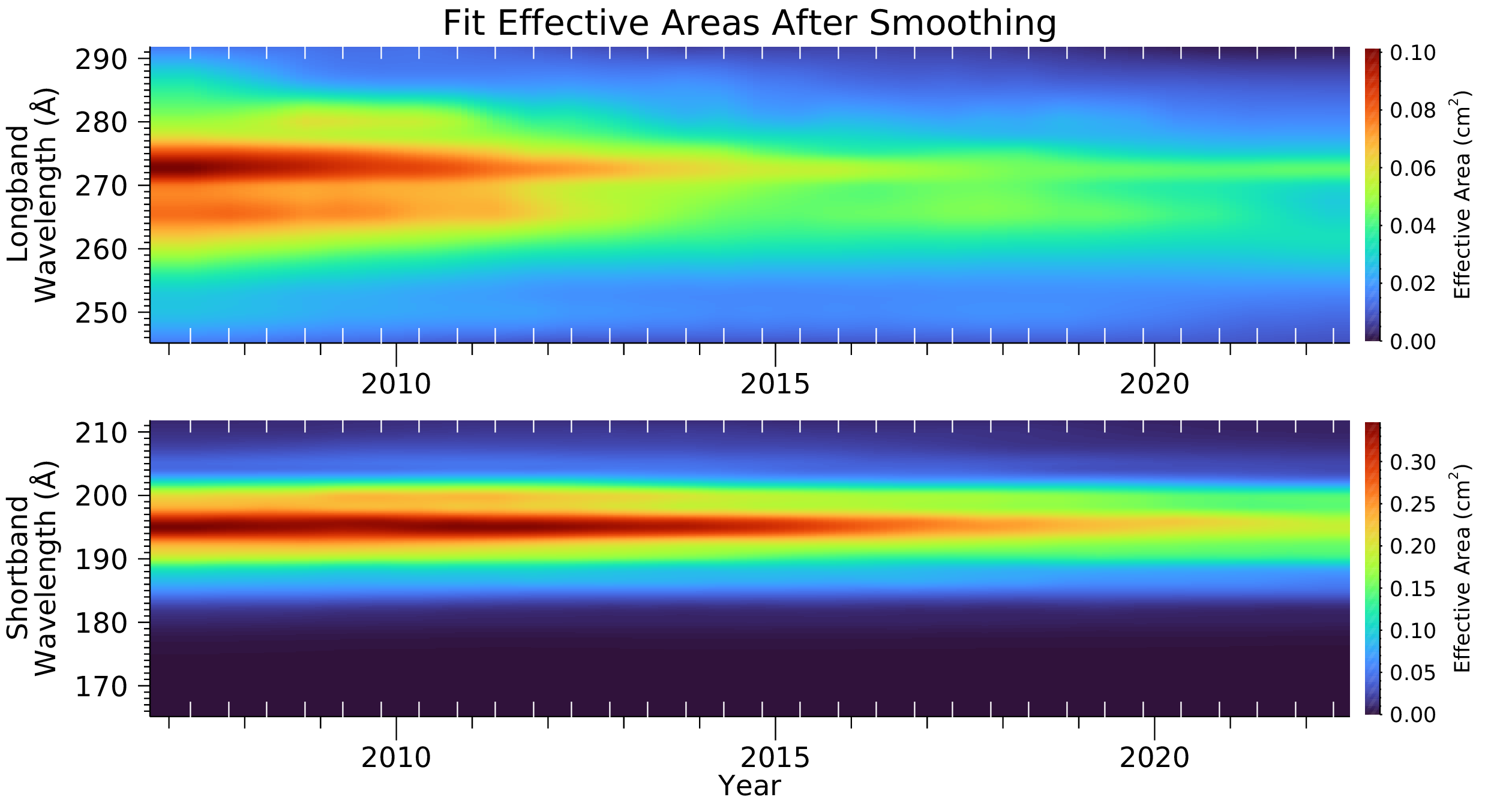}}
 \centerline{\includegraphics[clip,width=0.65\linewidth]{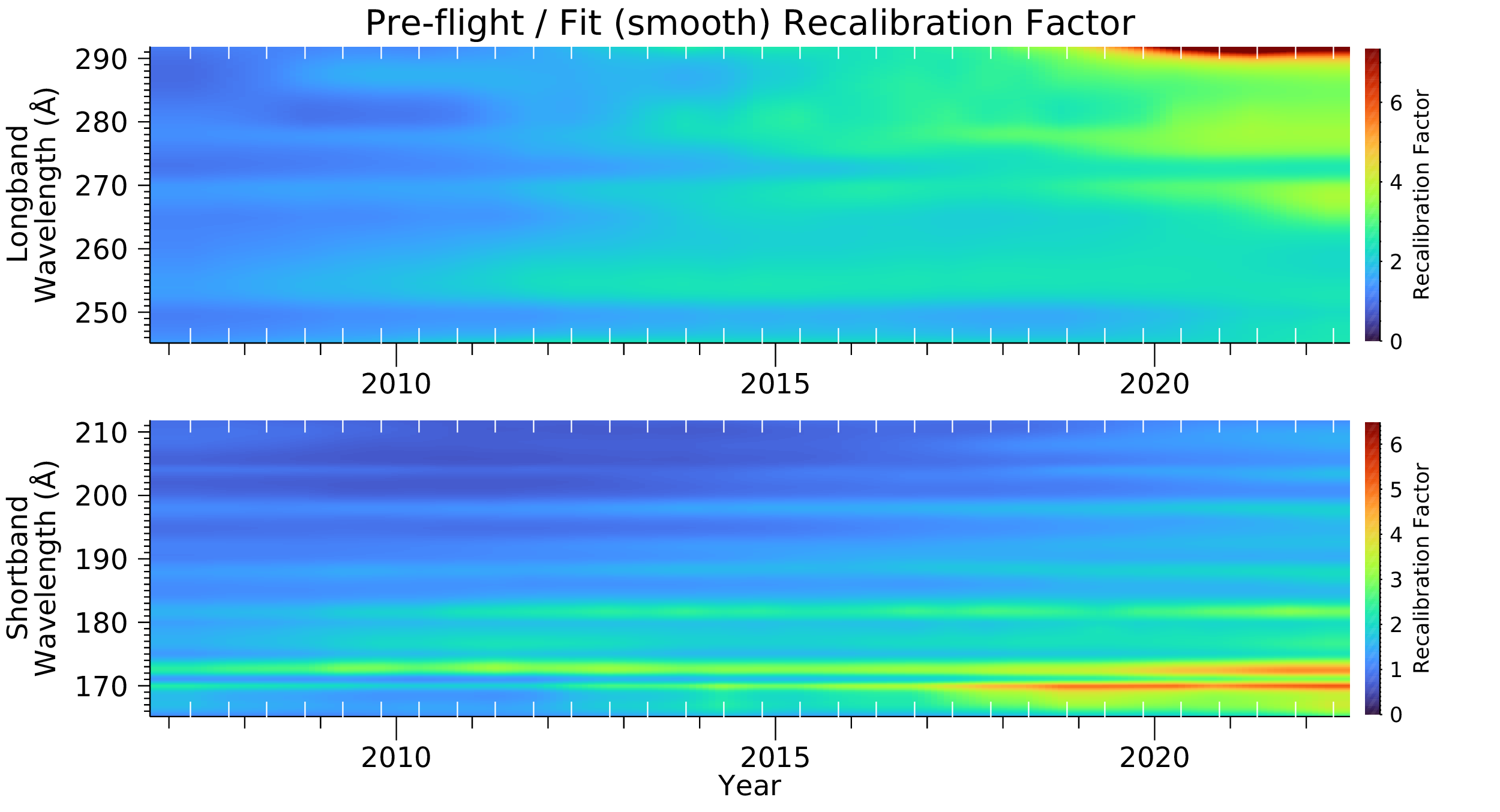}}
 \caption{2D histograms of the fit effective areas before smoothing (top two panels) and after smoothing (middle two panels). The bottom two panels show the the "recalibration factor" computed as the ratio of pre-flight / fit (smooth) effective area. The white tick marks along the edge of the plots indicate the center of each six-month fitting window.}
\label{fig:ea_over_time}
\end{figure*}

 \begin{figure}[!ht]
\centerline{\includegraphics[clip,width=1.1\linewidth, angle=0]{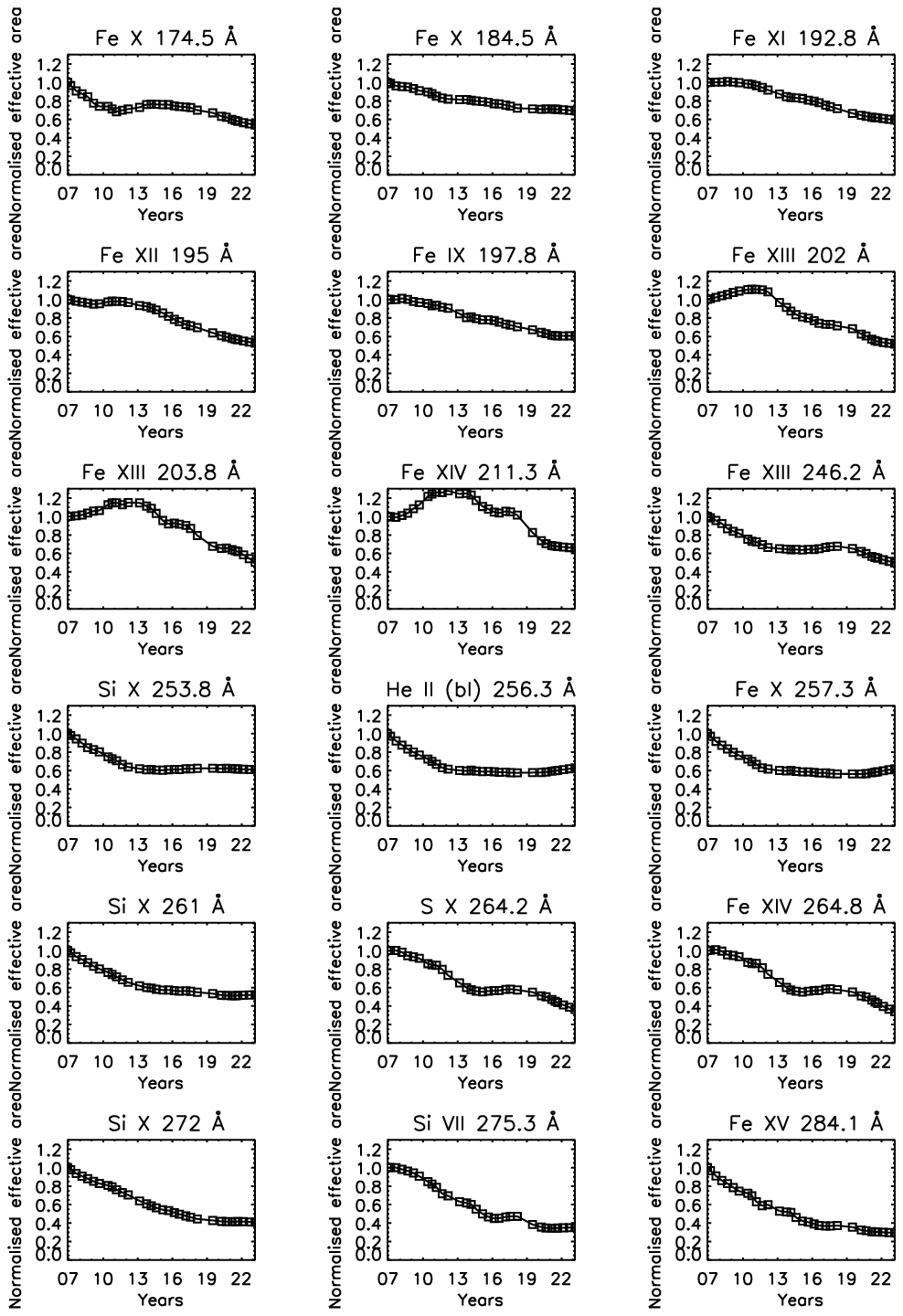}}
 \caption{Normalized effective areas at selected wavelengths in both the shortwave and longwave channels. For each line, the trend has been normalized by the initial value in order to show the relative change over time.}
\label{fig:norm_ea}
\end{figure}

\subsection{Effective areas}

Figure ~\ref{fig:ea_over_time} shows the changes in the fit effective areas over time. The top two panels show the full 2D histograms for all wavelengths in each wave channel before the final smoothing process and the middle two panels show the effective areas after smoothing. The bottom two panels give the "recalibration factor" calculated as the ratio of pre-flight / fit (smooth) effective area. This ratio illustrates which wavelengths are most affected by the updated calibration. Figure ~\ref{fig:norm_ea} plots the effective areas over time at select wavelengths and normalized by their initial values.
These figures indicate that the SW band exhibits relatively little change until late 2012 and then a gradual decline in sensitivity after. In contrast, the sensitivity in the LW band declines much faster at first, but then appears to plateau after 2017. Additionally, the LW band is more variable, which may indicate that the fit in that channel is more susceptible to outliers. This complex, channel-dependent decay in sensitivity is similar to the trends noted by \cite{warren_etal:2014} for the years 2007 -- 2013.
{In the Appendix, we provide a few comparisons 
between the present effective areas, the ground calibration and the previous two.}

\subsection{The problems with observations after 2021}
{

As already mentioned, observations after 2021 turn out to be 
difficult to calibrate, for a series of reasons.
First, the signal shortward of 174~\AA\ and longward of 
284~\AA\ is very low - even exposures of 120s in the 
quiet Sun are not enough to measure any lines. 
Second, the number and behaviour of warm/hot pixels 
further reduces the signal.  Third, we found no suitable 
off-limb quiet Sun observations to assess the calibration.
The last useful observation we have analysed was taken on 
2021 June 16 with the EIS study 
``dhb\_atlas\_120m\_30", with 120s exposures. 
We selected a region, applied our final calibration
and performed a DEM analysis, using the
photospheric abundances of \cite{asplund2021}. 
The results are shown in Table~\ref{tab:ol_last} in the 
Appendix and indicate that at least until that date 
our calibration still produces consistently good results, i.e. almost
all the stronger lines are well represented within 20\%,
including the density-sensitive ones, which are all included in the Table.
There are however few exceptions. The most notable one is the 
\ionm{S}{x} 264.23~\AA, which has been widely used to 
measure the First Ionization Potential (FIP) effect, in conjunction
with one of the nearby \ionm{Si}{x} lines, as in remote-sensing 
observations the sulphur abundance generally varies in line 
with those of the high-FIP elements such as Ar, Ne, O. 
Now there is an abundant body of literature which has 
indicated that the quiet Sun corona at the temperature of 
formation of \ionm{S}{x} and \ionm{S}{xi} has photospheric 
abundances, see for example the reviews in 
\cite{delzanna_mason:2018} and in  Appendix A of \cite{delzanna_etal:2023}.
Indeed all the quiet Sun  observations 
we have analysed do confirm that. 

However, for observations after 
2021 the \ionm{S}{x} 264.23~\AA\ presents an anomalous behavior as it is also clear from Figure~\ref{fig:si_s_ratio} in the Appendix.
For the 2021 June 16 observation the \ionm{S}{x} 264.23~\AA\
line is about 2.5 weaker than 
expected. The main problem is only in this line, as the other  \ionm{S}{x} 
lines (and also lines from other sulphur and argon ions) are relatively
close to the expected values. 
The effective area at 264.23~\AA\ as shown in Fig.~12 has a 
smooth behaviour, close to the nearby \ionm{Fe}{xiv} 264.8~\AA\
which does not show such anomalous behaviour. However it is clear
that such wavelengths suffered a stronger degradation than 
others. 
}

\section{Conclusions}

After trying several approaches we have presented what we believe 
is the best combination of methods, to provide a reasonable 
in-flight radiometric calibration  for EIS. Overall, it provides 
calibrated intensities which should be accurate to within $\pm$20 \%.
The overall degradation over time in the central parts of the 
two channels has been about a factor of two over the 2006-2022 
period, which is an excellent performance, one of the best 
of any EUV instrument in space 
(SOHO CDS degraded by a factor of about 2 in 13 years, see \citealt{delzanna_andretta:2015}). 
{
Significant degradation is however 
present in various spectral regions, 
in particular in the long wavelength part of the LW channel, with factors of more than five, 
see Fig.12. Within this region, as well as 
shortward of 174~\AA, the degradation actually becomes un-measurable as there is nearly no signal, see Fig. 13.}

Nearly all  previous EUV/UV  space instruments have shown 
significant degradation \citep[see e.g.][]{benmoussa_etal:2013},
which was usually thought to be associated with polimerization
of chemical compounds (mostly including carbon) onto 
the surfaces of the front filters, caused 
by exposure to the strong solar flux. Such degradation is typically 
wavelength-dependent and the simple aluminium filters, used for the 
EUV (as for Hinode EIS), are particularly affected.

Recently, { after over a decade of studies,} it has been found that 
oxidation, caused by outgassing of water molecules, has also 
been causing
major degradation for SDO instruments, see e.g. 
\cite{2023SoPh..298...32T}.
{ The wavelength-dependent degradation was studied with the 
SDO/EVE spectrometer, alternating different front filters. Those 
exposed regularly to EUV radiation showed most degradation. 
EIS has a single Al front filter, so a similar study could not be performed.
 As we have shown, there is no clear evidence for a systematic wavelength-dependent degradation caused by e.g. polimerization of carbon. 
It could well be that various chemical compounds have been present near the 
EIS Al entrance filter, and have caused the unusual evolution of the effective areas we have observed, where nearby wavelengths have shown
a very different behaviour over time. }

It seems unlikely that there is contamination on the EIS detector itself.  Three EIS detector bakeouts were performed (2016 February 23--26, 2017 August 8--10, and 2018 January 17--21) and no significant changes in response have been observed (see Figure~\ref{fig:eis_aia_ratio}). 

{Another possibility is the degradation of the EIS
Mo/Si multilayers (MLs) on the primary mirror and grating. 
As we have mentioned, degradation of EUV instruments is common. However,
 wavelength-dependent effects on MLs have not been studied before,
 as they would require a spectrometer with MLs.

 There is relatively little experimental work carried out on the
stability of multilayers (MLs). Some multilayers such as Mg/SiC have shown degradation
over time with anomalous wavelength-dependent changes in the reflectivities
\citep[see, e.g.][]{pelizzo_etal:2012}. The Mo/Si ML, adopted for the EIS mirror and
grating, is known to be stable if left on the ground.
Space is a harsh environment with significant thermal excursions
and large photon and particle fluxes.
Degradation of  Mo/Si ML could be caused by polimerization of
compounds caused by UV and EUV photons. For example,
\cite{benoit_etal:2006} found significant degradation of an Mo/Si at
135~\AA\ due to oxidation caused by EUV photons.
The polimerization of carbon compounds due to EUV/UV light is
also a well-known source of degradation. 
Such process could have occurred, if there were any
contaminants inside the closed EIS instrument.
Polimerization of compounds caused by UV and EUV photons was
a possible cause of the degradation of the CDS NIS instrument,
as the instrument did not have a front filter nor a ML. 

A few experiments have been carried out to estimate the effects
of particle hits. They have all found wavelength-dependent changes
in the reflectivities, with large shifts in the location of the peak.
Such changes are thought to be mainly caused by low energy protons, as the
high-energy ones tend to pass through the thin ML structure.
\cite{lv_etal:2012} irradiated a Mo/Si ML
with 100 keV protons and a dose of 6$\times$10$^9$ cm$^{-2}$.
\cite{pelizzo_etal:2011} irradiated samples of Mo/Si with 1 keV
protons, with doses similar to those expected in space
by Solar Orbiter within one year, and also found significant
changes in the shape of the reflectivity. The peak shifted by about 5--10~\AA\
towards longer wavelengths.
In a similar study, \cite{nardello_etal:2013} irradiated a Mo/Si ML
with 4 keV alpha particles, also finding significant changes,
with the peak this time shifted by about 5~\AA\ towards shorter wavelengths.
The experiments were carried out with high doses of particles, much higher than
those likely experienced by EIS. However, EIS would have experienced
protons with a whole range of energies.  To provide quantitative estimates,
the effects of a spectrum of proton energies would have to be measured in the
laboratory, combined with rough estimates of proton fluxes that could have entered the
instrument. An alternative  for the future would be to leave ML exposed in
space and then retrieve them for laboratory analysis.

 }

{
We have biased the effective areas to keep a relatively smooth behaviour over wavelength and over time, associated with an overall  degradation of the multilayers. 
However, it is clear that nearby wavelength regions suffered 
very different degradations, and in some cases our smoothed
calibration  produces  departures from 
the stated overall 20\% uncertainty. 
The extreme case is the degradation at  264.2~\AA\ after 2021 
where the important  \ionm{S}{x} line is. 
We caution the use of the present calibration for data taken after 2021.
A more accurate relative calibration can be obtained by analysing line ratios observed close in time to a specific observation. 

Continuous monitoring of the degradation is recommended, by 
running EIS atlases in quiet Sun on-disk and off-limb regions,  
with exposure times of at least 120s, and off-limb active regions,
with exposures of at least 60s}. 
{The procedures to analyse the data, the processed
data, and the results of the fitting, with a program to 
calculate the effective area is availabe on ZENODO
(DOI: 10.5281/zenodo.10462448)}

{Acknowledgments}

\noindent
 GDZ  acknowledges partial 
 support from STFC (UK) via the consolidated grants 
 to the atomic astrophysics group (AAG) at DAMTP, University of Cambridge (ST/P000665/1. and ST/T000481/1)
 during the few years of work on this long-term project.
 HPW and MJW were supported by NASA's Hinode program. 
 We thank the reviewer for useful comments which led
 to a significant improvement of the paper.
 We also thank David Brooks for pointing out the 
 problem with the recent observations of the \ionm{S}{x} 264.2~\AA\ line.
 

\bibliography{eis_cal}{}
\bibliographystyle{aasjournal}



 \clearpage

 \appendix
\section{Tables of QS and AR observations}

\begin{table}[!htbp]
\caption{List of the seven quiet Sun off-limb observations used for the DEM analysis, showing the date and start time, the EIS study name, the slit size and exposure time.}
\centering
\begin{tabular}{@{}rccccc@{}}
\hline\hline \noalign{\smallskip}
Date-Obs &    STUDY  & (\arcsec) & Exp (s)  \\
\noalign{\smallskip}\hline\noalign{\smallskip}
2007-03-11T02:32:12 &  	HPW001\_FULLCCD\_RAST & 1 & 90 \\
 2010-10-16T10:29:41 &  Atlas\_120 & 2 & 120 \\
 2013-10-07T19:02:03 & Atlas\_120 & 2 & 120 \\
 2014-08-07T04:56:39 &  Atlas\_60 & 2 & 60 \\
 2016-08-06T11:00:13 & Atlas\_60 &  2 & 60 \\
 2017-07-28T06:55:42 & Atlas\_60 &  2 & 60 \\
 2019-06-28T09:38:32 & Atlas\_60 &  2 & 60 \\
\noalign{\smallskip}\hline 
\end{tabular}
\label{tab:list_qs1} 
\end{table}

\begin{table}[!htbp]
\caption{List of the quiet Sun EIS observations analysed. }
\centering
\scriptsize
\begin{minipage}[t]{0.496\textwidth}
\begin{tabular}{@{}rccccc@{}}
\hline\hline \noalign{\smallskip}
No & Date-Obs & STUDY  & Region &  (\arcsec) & Exp (s)  \\
\noalign{\smallskip}\hline\noalign{\smallskip}
1 &  2007-01-14T22:08:19   & SYNOP001 & QS & 1 & 90 \\
2 &  2007-01-30T11:19:12   & HPW001\_FULLCCD\_RAST & QS & 1 & 90 \\
3 &  2007-02-16T11:23:50   & SYNOP001 & QS & 1 & 90 \\
4 &  2007-03-16T18:01:27   & SYNOP001 & QS & 1 & 90 \\
5 &  2007-04-21T03:06:43   & SYNOP002 & QS & 1 & 90 \\
6 &  2007-05-17T00:04:50   & SYNOP001 & QS & 1 & 90 \\
7 &  2007-06-02T13:15:20   & HPW008\_FULLCCD\_RAST & QS & 1 & 25 \\
8 &  2007-06-20T18:08:35   & SYNOP001 & QS & 1 & 90 \\
9 &  2007-07-20T11:08:22   & SYNOP001 & QS & 1 & 90 \\
10 &  2007-08-17T06:29:35   & SYNOP001 & QS & 1 & 90 \\
11 &  2007-09-13T17:58:36   & SYNOP001 & QS & 1 & 90 \\
12 &  2007-10-24T06:18:35   & SYNOP001 & QS & 1 & 90 \\
13 &  2007-11-25T10:46:56   & SYNOP001 & QS & 1 & 90 \\
14 &  2007-12-22T11:12:05   & SYNOP001 & QS & 1 & 90 \\
15 &  2008-01-21T16:02:13   & SYNOP002 & QS & 1 & 90 \\
16 &  2008-12-17T11:05:19   & HPW001\_FULLCCD\_RAST & QS & 1 & 90 \\
17 &  2009-03-23T17:42:30   & SYNOP001 & QS & 1 & 90 \\
18 &  2009-04-13T17:50:41   & HPW001\_FULLCCD\_RAST & QS & 1 & 90 \\
19 &  2009-05-11T18:09:29   & SYNOP001 & QS & 1 & 90 \\
20 &  2009-06-23T18:28:12   & SYNOP001 & QS & 1 & 90 \\
21 &  2009-07-20T06:00:35   & SYNOP001 & QS & 1 & 90 \\
22 &  2009-08-13T18:04:29   & SYNOP001 & QS & 1 & 90 \\
23 &  2009-09-19T18:05:57   & SYNOP001 & QS & 1 & 90 \\
24 &  2009-10-07T12:02:19   & HPW001\_FULLCCD\_RAST & QS & 1 & 90 \\
25 &  2009-10-23T06:05:50   & SYNOP001 & QS & 1 & 90 \\
26 &  2009-11-13T18:05:29   & SYNOP001 & QS & 1 & 90 \\
27 &  2009-12-27T06:35:35   & SYNOP001 & QS & 1 & 90 \\
28 &  2010-05-01T05:40:13   & Atlas\_30 & QS & 2 & 30 \\
29 &  2010-10-08T10:15:26   & Atlas\_120 & QS & 2 & 120 \\
30 &  2010-12-20T05:05:26   & Atlas\_120 & QS & 2 & 120 \\
31 &  2011-04-13T13:20:33   & Atlas\_120 & QS & 2 & 120 \\
32 &  2011-06-03T11:30:20   & Atlas\_060x512\_60s & QS & 1 & 60 \\
33 &  2011-08-31T05:45:34   & Atlas\_60 & QS & 2 & 60 \\
34 &  2011-12-26T18:19:40   & Atlas\_120 & QS & 2 & 120 \\
35 & 2012-02-08T18:33:43  & Atlas\_120 & QS & 2 & 120 \\
36 &  2012-04-28T15:13:19   & Atlas\_060x512\_60s & QS & 1 & 60 \\
37 &  2012-09-13T18:25:34   & Atlas\_120 & QS & 2 & 120 \\
38 &  2013-03-01T22:39:02   & Atlas\_120 & QS & 2 & 120 \\
39 & 2013-08-22T06:40:33  & Atlas\_60 & QS & 2 & 60 \\
40 &  2013-09-17T09:28:49   & Atlas\_60 & QS & 2 & 60 \\
41 & 2013-11-26T13:47:34  & Atlas\_120 & QS & 2 & 120 \\
42 &  2013-09-17T09:28:49   & Atlas\_60 & QS & 2 & 60 \\
43 & 2014-05-06T15:08:54  & Atlas\_60 & QS & 2 & 60 \\
44 & 2014-07-01T01:34:41  & Atlas\_60 & QS & 2 & 60 \\
\noalign{\smallskip}\hline
\end{tabular}
\end{minipage} \hfill
\begin{minipage}[t]{0.496\textwidth}
\begin{tabular}{@{}rccccc@{}}
\hline\hline \noalign{\smallskip}
No & Date-Obs & STUDY & Region & (\arcsec) & Exp(s)  \\
\noalign{\smallskip}\hline\noalign{\smallskip}
45 &  2014-08-11T03:22:51   & Atlas\_120 & QS & 2 & 120 \\
46 &  2015-03-22T10:01:48   & Atlas\_60 & QS & 2 & 60 \\
47 & 2015-05-23T19:15:34  & Atlas\_60 & QS & 2 & 60 \\
48 & 2015-08-24T09:07:30  & Atlas\_60 & QS & 2 & 60 \\
49 & 2015-11-17T03:05:48  & Atlas\_120 & QS & 2 & 120 \\
50 &  2016-01-17T09:04:47   & Atlas\_60 & QS & 2 & 60 \\
51 &  2016-03-22T03:25:03   & Atlas\_60 & QS & 2 & 60 \\
52 &  2016-04-11T22:17:23   & Atlas\_60 & QS & 2 & 60 \\
53 & 2016-08-28T10:48:53  & Atlas\_60 & QS & 2 & 60 \\
54 & 2016-10-09T16:31:55  & Atlas\_60 & QS & 2 & 60 \\
55 &  2016-11-19T11:19:34   & Atlas\_60 & QS & 2 & 60 \\
56 &  2016-12-17T11:11:22   & Atlas\_60 & QS & 2 & 60 \\
57 &  2017-06-04T11:08:54   & Atlas\_60 & QS & 2 & 60 \\
58 & 2017-08-16T11:30:50  & Atlas\_60 & QS & 2 & 60 \\
59 & 2017-09-14T15:59:35  & Atlas\_60 & QS & 2 & 60 \\
60 & 2016-10-09T16:31:55  & Atlas\_60 & QS & 2 & 120 \\
61 & 2018-01-14T13:06:41  & Atlas\_60 & QS & 2 & 60 \\
62 & 2018-06-12T00:04:14  & Atlas\_60 & QS & 2 & 60 \\
63 & 2018-08-08T18:55:33  & dhb\_atlas\_120m\_30" & QS & 2 & 120 \\
64 & 2018-08-30T04:17:26  & dhb\_atlas\_120m\_30" & QS & 2 & 120 \\
65 & 2018-10-26T13:36:49  & dhb\_atlas\_120m\_30" & QS & 2 & 120 \\
66 & 2018-12-25T20:43:42  & dhb\_atlas\_120m\_30" & QS & 2 & 120 \\
67 & 2019-02-27T04:17:30  & dhb\_atlas\_120m\_30" & QS & 2 & 120 \\
68 & 2019-04-21T16:09:05  & dhb\_atlas\_120m\_30" & QS & 2 & 120 \\
69 & 2019-06-29T03:42:13  & dhb\_atlas\_120m\_30" & QS & 2 & 120 \\
70 & 2019-09-06T00:20:26  & dhb\_atlas\_120m\_30" & QS & 2 & 120 \\
71 & 2019-10-19T02:11:12  & dhb\_atlas\_120m\_30" & QS & 2 & 120 \\
72 & 2019-12-12T22:44:36  & dhb\_atlas\_120m\_30" & QS & 2 & 120 \\
73 & 2020-02-29T15:06:06  & dhb\_atlas\_120m\_30" & QS & 2 & 120 \\
74 & 2020-05-31T14:54:40  & Atlas\_60 & QS & 2 & 60 \\
75 & 2020-07-01T13:22:13  & dhb\_atlas\_120m\_30" & QS & 2 & 120 \\
76 & 2020-09-25T02:59:43  & Atlas\_120 & QS & 2 & 120 \\
77 & 2020-11-02T07:12:19  & dhb\_atlas\_120m\_30" & QS & 2 & 120 \\
78 & 2021-03-02T22:16:41  & dhb\_atlas\_120m\_30" & QS & 2 & 120 \\
79 & 2021-03-12T04:44:43  & Atlas\_60 & QS & 2 & 60 \\
80 & 2021-04-03T18:20:40  & Atlas\_120 & QS & 2 & 120 \\
81 & 2021-07-16T02:49:02  & Atlas\_120 & QS & 2 & 120 \\
82 & 2021-08-10T16:49:20  & Atlas\_120 & QS & 2 & 120 \\
83 & 2021-10-13T03:12:42  & Atlas\_60 & QS & 2 & 60 \\
84 & 2021-11-30T14:02:30  & Atlas\_120 & QS & 2 & 120 \\
85 & 2021-12-24T02:15:32  & Atlas\_120 & QS & 2 & 120 \\
86 & 2022-03-07T16:03:13  & Atlas\_60 & QS & 2 & 60 \\
87 & 2022-06-01T04:25:13  & Atlas\_60 & QS & 2 & 60 \\
& &  &  &  \\ 
\noalign{\smallskip}\hline 
\end{tabular}
\end{minipage}
\normalsize
\label{tab:list_qs} 
\end{table}

\begin{table}[!htbp]
\caption{List of the active region EIS observations analysed. }
\centering
\scriptsize
\begin{minipage}[t]{0.496\textwidth}
\begin{tabular}{@{}rccccc@{}}
\hline\hline \noalign{\smallskip}
No & Date-Obs & STUDY  & Region &  (\arcsec) & Exp (s)  \\
\noalign{\smallskip}\hline\noalign{\smallskip}
1 &  2006-12-25T22:50:13   & HPW001\_FULLCCD\_RAST & AR & 1 & 90 \\
2 &  2007-01-18T12:04:35   & SYNOP001 & AR & 1 & 90 \\
3 &  2007-02-20T05:40:28   & SYNOP001 & AR & 1 & 90 \\
4 &  2007-05-11T10:55:44   & SYNOP001 & AR & 1 & 90 \\
5 &  2007-07-14T00:09:49   & SYNOP001 & AR & 1 & 90 \\
6 &  2007-11-14T00:07:07   & SYNOP001 & AR & 1 & 90 \\
7 &  2008-01-07T10:14:48   & SYNOP001 & AR & 1 & 90 \\
8 &  2008-02-04T10:47:00   & SYNOP001 & AR & 1 & 90 \\
9 &  2008-06-20T23:03:39   & HPW001\_FULLCCD\_RAST & AR & 1 & 90 \\
10 &  2008-12-17T11:05:19   & HPW001\_FULLCCD\_RAST & AR & 1 & 90 \\
11 &  2009-03-22T06:06:30   & SYNOP001 & AR & 1 & 90 \\
12 &  2009-05-21T18:05:29   & SYNOP001 & AR & 1 & 90 \\
13 &  2009-07-25T05:50:49   & SYNOP001 & AR & 1 & 90 \\
14 &  2009-10-25T21:46:25   & HPW008\_FULLCCD\_RAST & AR & 1 & 25 \\
15 &  2010-01-23T17:15:32   & FILL001 & AR & 1 & 30 \\
16 &  2010-05-17T13:57:41   & Atlas\_60 & AR & 2 & 60 \\
17 &  2010-09-22T11:26:33   & Atlas\_60 & AR & 2 & 60 \\
18 & 2010-11-09T10:07:33  & Atlas\_30 & AR & 2 & 30 \\
19 &  2011-01-21T12:37:57   & Atlas\_60 & AR & 2 & 60 \\
20 &  2011-05-22T10:33:54   & Atlas\_60 & AR & 2 & 60 \\
21 &  2011-07-26T17:59:35   & Atlas\_60 & AR & 2 & 60 \\
22 &  2011-10-22T10:05:43   & Atlas\_60 & AR & 2 & 60 \\
23 &  2011-12-17T12:58:56   & Atlas\_60 & AR & 2 & 60 \\
24 &  2012-04-16T12:40:33   & Atlas\_60 & AR & 2 & 60 \\
25 &  2012-07-04T22:36:57   & Atlas\_60 & AR & 2 & 60 \\
26 &  2012-08-30T23:40:44   & Atlas\_60 & AR & 2 & 60 \\
27 & 2013-01-27T08:16:49  & Atlas\_60 & AR & 2 & 60 \\
28 & 2013-03-27T02:17:09  & Atlas\_60 & AR & 2 & 60 \\
29 & 2013-06-01T11:58:57  & Atlas\_60 & AR & 2 & 60 \\
30 & 2013-09-15T15:13:40  & Atlas\_60 & AR & 2 & 60 \\
31 & 2013-11-04T07:51:10  & Atlas\_60 & AR & 2 & 60 \\
32 & 2014-01-14T12:24:52  & Atlas\_30 & AR & 2 & 30 \\
33 & 2014-05-08T18:20:49  & Atlas\_30 & AR & 2 & 30 \\
34 & 2014-07-18T13:25:30  & Atlas\_60 & AR & 2 & 60 \\
35 & 2014-10-13T11:13:16  & Atlas\_30 & AR & 2 & 30 \\
36 & 2014-12-23T10:21:42  & Atlas\_60 & AR & 2 & 60 \\
37 & 2015-03-22T21:55:24  & Atlas\_60 & AR & 2 & 60 \\
38 & 2015-06-15T19:57:49  & Atlas\_60 & AR & 2 & 60 \\
39 & 2015-10-22T01:41:09  & Atlas\_60 & AR & 2 & 60 \\
40 & 2015-12-30T18:27:26  & Atlas\_60 & AR & 2 & 60 \\
\noalign{\smallskip}\hline
\end{tabular}
\end{minipage} \hfill
\begin{minipage}[t]{0.496\textwidth}
\begin{tabular}{@{}rccccc@{}}
\hline\hline \noalign{\smallskip}
No & Date-Obs & STUDY & Region & (\arcsec) & Exp(s)  \\
\noalign{\smallskip}\hline\noalign{\smallskip}
41 & 2016-03-17T20:30:54  & Atlas\_60 & AR & 2 & 60 \\
42 & 2016-05-30T21:35:25  & Atlas\_60 & AR & 2 & 60 \\
43 & 2016-08-04T08:13:48  & Atlas\_60 & AR & 2 & 60 \\
44 & 2016-10-01T18:29:41  & Atlas\_60 & AR & 2 & 60 \\
45 & 2016-12-10T20:20:27  & Atlas\_60 & AR & 2 & 60 \\
46 & 2017-02-08T11:17:33  & Atlas\_60 & AR & 2 & 60 \\
47 & 2017-05-03T11:53:41  & Atlas\_60 & AR & 2 & 60 \\
48 & 2017-08-02T13:06:40  & Atlas\_60 & AR & 2 & 60 \\
49 & 2017-10-08T01:48:41  & Atlas\_60 & AR & 2 & 60 \\
50 & 2017-12-02T00:03:40  & Atlas\_60 & AR & 2 & 60 \\
51 & 2018-01-17T08:51:40  & Atlas\_60 & AR & 2 & 60 \\
52 & 2018-06-12T03:21:14  & Atlas\_60 & AR & 2 & 60 \\
53 & 2018-08-10T08:32:43  & Atlas\_60 & AR & 2 & 60 \\
54 & 2018-10-06T14:12:12  & Atlas\_30 & AR & 2 & 30 \\
55 & 2018-12-16T10:13:41  & Atlas\_60 & AR & 2 & 60 \\
56 & 2019-01-30T23:47:41  & Atlas\_60 & AR & 2 & 60 \\
57 & 2019-03-26T13:00:41  & Atlas\_60 & AR & 2 & 60 \\
58 & 2019-06-15T18:27:48  & Atlas\_60 & AR & 2 & 60 \\
59 & 2019-06-28T09:38:32  & Atlas\_60 & AR & 2 & 60 \\
60 & 2019-09-03T11:18:22  & Atlas\_60 & AR & 2 & 60 \\
61 & 2019-10-11T23:58:43  & Atlas\_60 & AR & 2 & 60 \\
62 & 2019-12-24T21:38:13  & Atlas\_60 & AR & 2 & 60 \\
63 & 2020-02-02T13:33:41  & Atlas\_60 & AR & 2 & 60 \\
64 & 2020-05-03T11:20:00  & Atlas\_60 & AR & 2 & 60 \\
65 & 2020-06-15T00:47:41  & Atlas\_60 & AR & 2 & 60 \\
66 & 2020-08-06T00:10:10  & Atlas\_60 & AR & 2 & 60 \\
67 & 2020-10-06T11:15:49  & Atlas\_60 & AR & 2 & 60 \\
68 & 2020-11-15T21:48:10  & Atlas\_60 & AR & 2 & 60 \\
69 & 2020-12-02T11:29:40  & Atlas\_60 & AR & 2 & 60 \\
70 & 2021-02-28T15:59:13  & Atlas\_60 & AR & 2 & 60 \\
71 & 2021-04-20T20:00:14  & Atlas\_60 & AR & 2 & 60 \\
72 & 2021-05-29T15:18:12  & Atlas\_60 & AR & 2 & 60 \\
73 & 2021-08-02T22:46:11  & Atlas\_60 & AR & 2 & 60 \\
74 & 2021-09-03T08:48:40  & Atlas\_60 & AR & 2 & 60 \\
75 & 2021-11-04T03:08:41  & Atlas\_60 & AR & 2 & 60 \\
76 & 2021-12-26T12:00:39  & Atlas\_60 & AR & 2 & 60 \\
77 & 2022-03-06T23:26:33  & Atlas\_30 & AR & 2 & 60 \\
78 & 2022-04-28T23:41:08  & Atlas\_60 & AR & 2 & 60 \\
79 & 2022-08-07T14:23:40  & Atlas\_60 & AR & 2 & 60 \\
& &  &  &  \\ 
\noalign{\smallskip}\hline 
\end{tabular}
\end{minipage}
\normalsize
\label{tab:list_ar} 
\end{table}

Fig.~\ref{fig:qs_spectra} gives as an example  three
averaged quiet Sun EIS  spectra, taken near the beginning of the mission in 2008,
in 2012, and 2017. Some of the hotter lines have a strong
variability with the solar cycle and location, but those formed
at 1 MK and below are not expected to show strong variability \citep[see, e.g. the discussion in][ related to the SOHO/CDS calibration]{delzanna_andretta:2015}.
The figure shows the clear decrease in the signal with time,
and the fact that the shorter and longer wavelengths of the LW
channel degraded significantly already by 2017.

\begin{figure*}[!htbp]
\centering
\includegraphics[clip,width=12cm,angle=0]{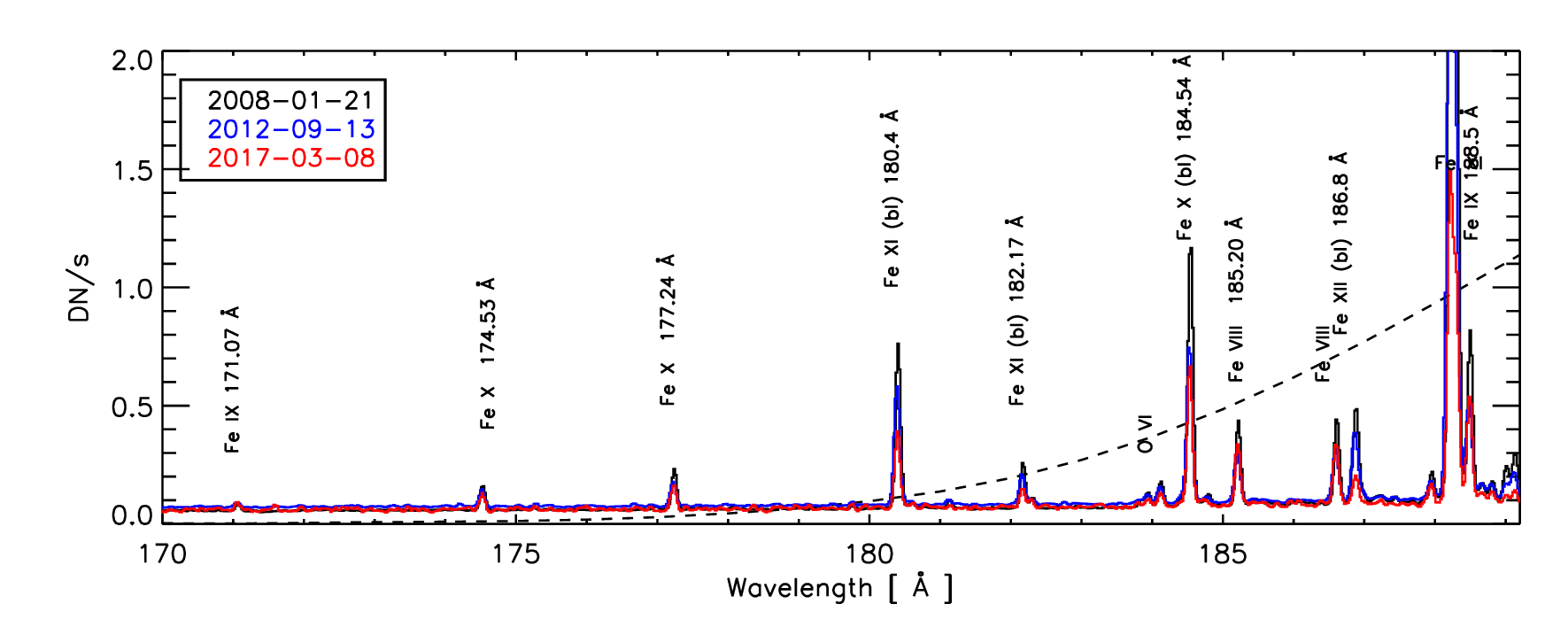}
\includegraphics[clip,width=12cm,angle=0]{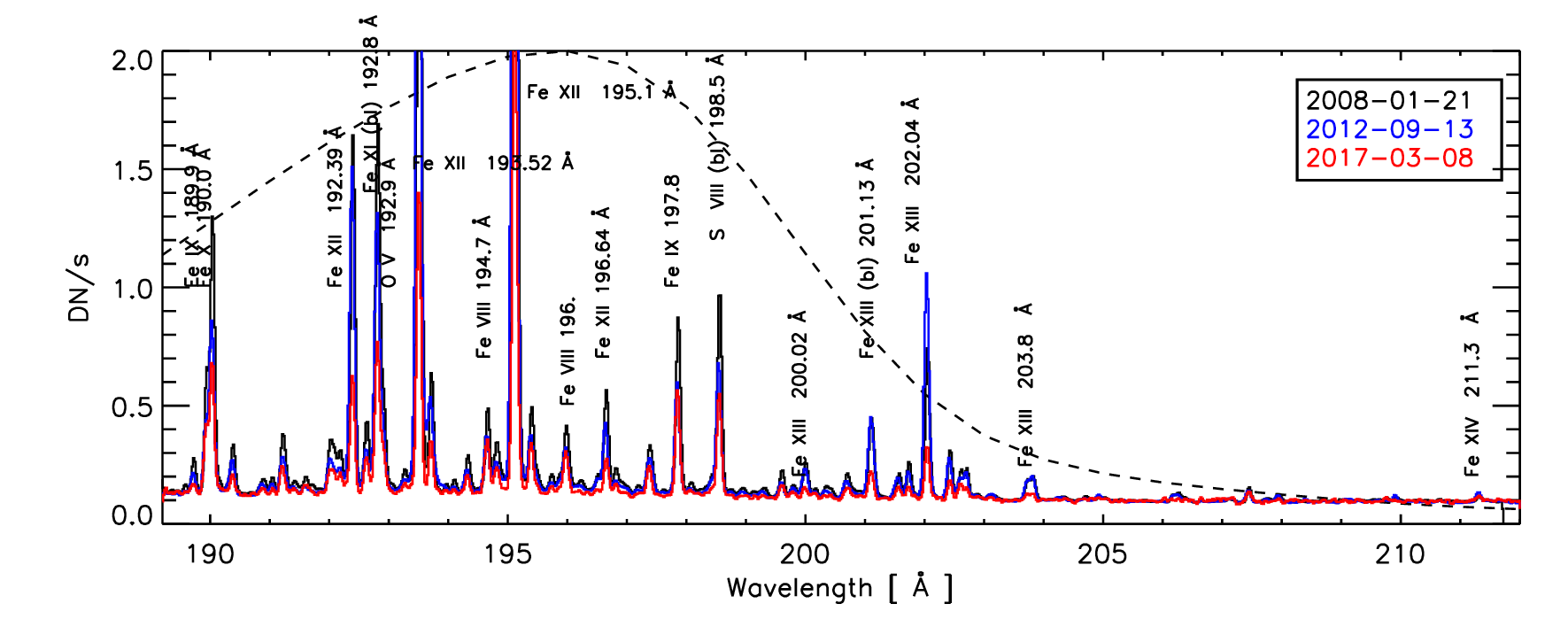}
\includegraphics[clip,width=12cm,angle=0]{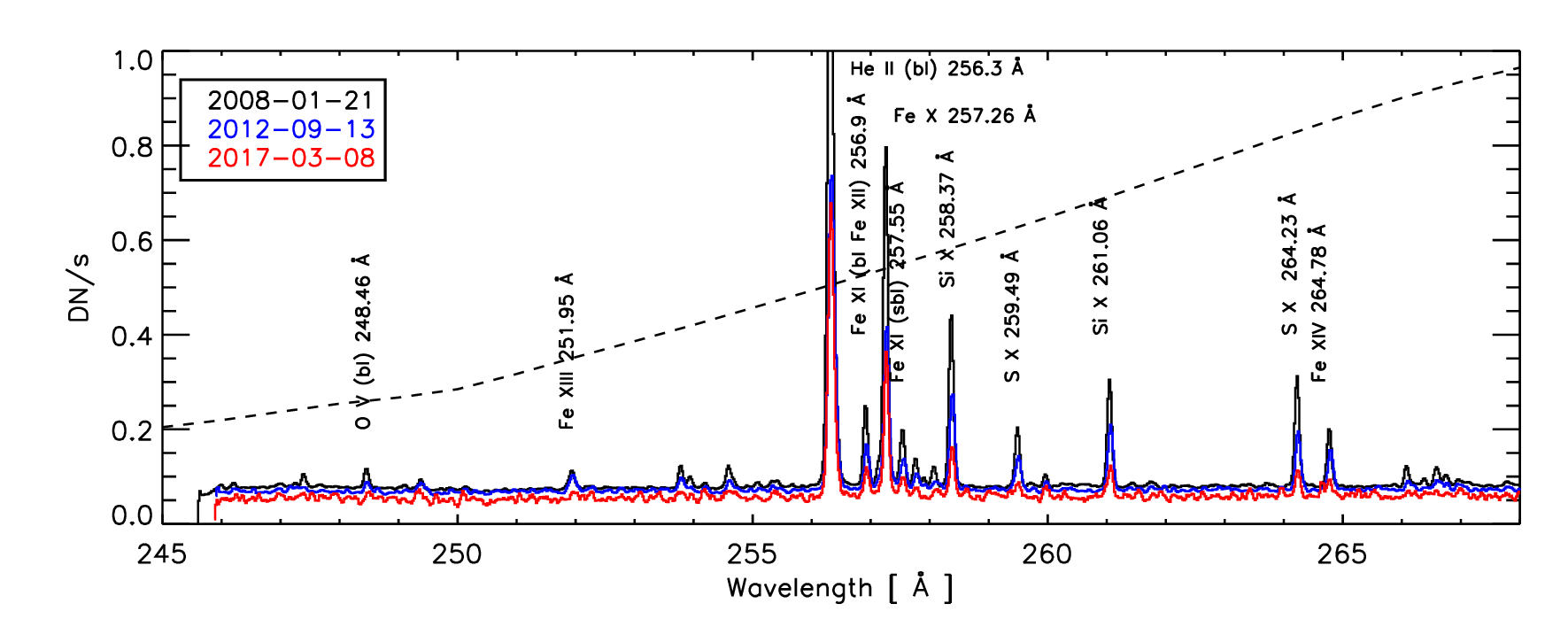}
\includegraphics[clip,width=12cm,angle=0]{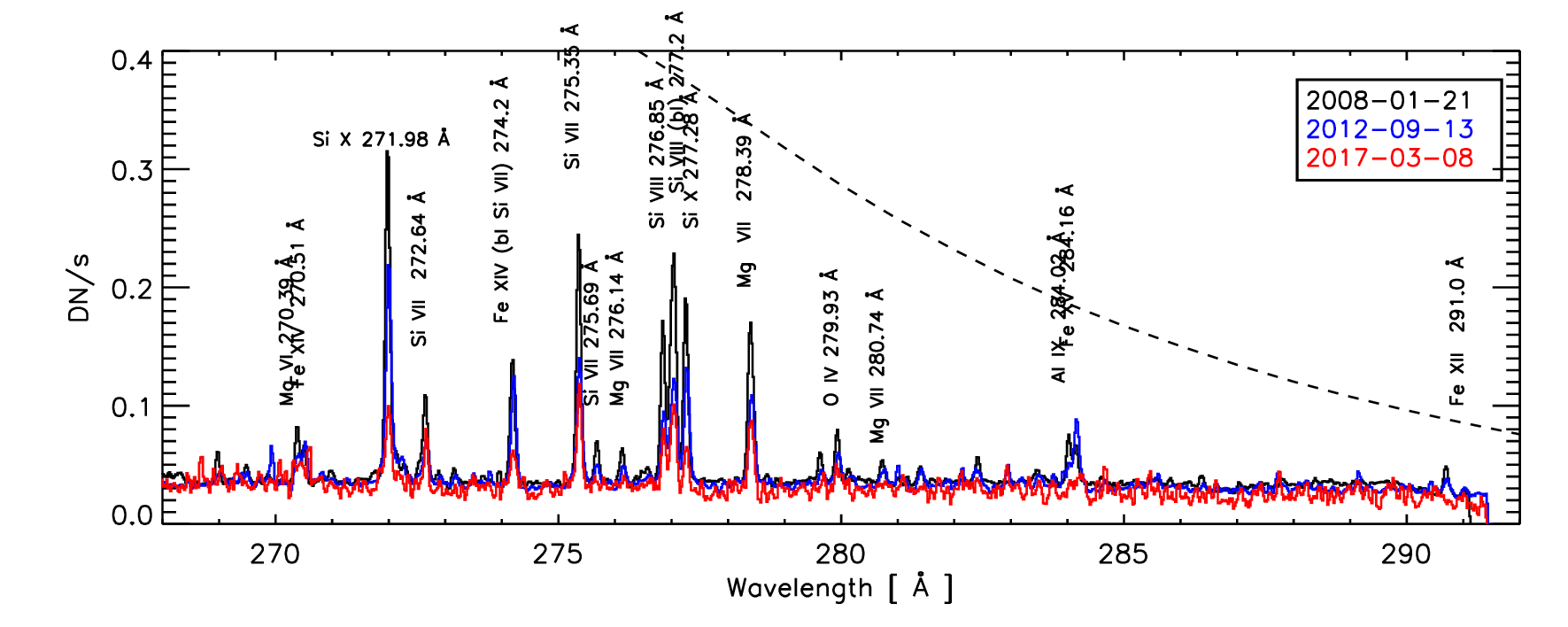}
\caption{Hinode EIS spectra (DN/s/1'' slit) for the  quiet Sun on-disk,
  for three sample dates. The dashed line is the scaled ground-calibration
  effective area.
}
 \label{fig:qs_spectra}
\end{figure*}

\section{Absolute calibration after 2010: EIS vs. AIA}

As a first step, the EIS data were processed with
eis\_prep and the pre-flight calibration.
The AIA 193~\AA\ data were cutouts downloaded from the
JSOC with an expanded FOV relative to EIS and a
cadence of 36 s.
These data are processed with aia\_prep. 

 We considered full EIS spectral atlases and took the SW channel,
  which covers very nicely the entire spectral range where the
AIA 193~\AA\ band is sensitive. We processed 35 dates. 
For each EIS exposure, the backgrond was estimated,
and the spectrum convolved with the AIA effective area (EA)
calculated using the  evenorm and timedepend\_date
keywords, to compute the expected AIA counts.
The nearest 3 AIA images were found and averaged.
The averaged was then smoothed over 5 pixels (3\arcsec)
and rebinned to the EIS pixel size of 1\arcsec.
The predicted and observed AIA images were then
cross-correlated to find the optimal EIS pointing, i.e.
finding  (aia\_x, aia\_y), the position of the bottom of the eis slit
derived from AIA.
Since the cross correlation is done exposure by exposure, it can be noisy,
particularly if there
is little structure along the slit. We expect that the aia\_y values should all be very similar
and the aia\_x values to increase linearly with time due to solar rotation.
A second pass is then computed, and a
mean aia\_y and a linear fit to aia\_x obtained. 

\section{Normalized Emissivity Curves}

Density-sensitive line ratios are often used in the literature 
to measure the electron density. Such measurements rely on 
accurate atomic data and relative radiometric calibration. 
As the EIS effective area has a strong wavelength variation, 
any small change in the relative effective area can affect the 
density measurements significantly. We have checked for 
a sample of dates that the electron densities obtained from the 
main line ratios and ions are reasonable, using the emissivity 
ratio method, described in the main part of the paper. 
Fig.~\ref{fig:norm_emiss} shows one example for the main 
coronal lines.
We expected the ratio curves to intersect, providing a 
density measurements, although in principle some departures 
can occur, as each transition is likely formed in a 
slightly different spatial 
region. The curves of the strong \ionm{Fe}{xii}  lines at 192.4, 193.5, 195.1~\AA\
often show discrepancies, related to opacity effects.
The weaker transition at 192.4~\AA\ is less affected and should be used
for the density measurements.
We do not expect that densities measured by different ions
should be exactly the same, but those of the \ionm{Fe}{xii}, \ionm{Fe}{xiii}  and 
\ionm{Si}{x} should be consistent, as these ions are formed within similar
temperatures. 
The emissivity ratios in the off-limb QS regions generally show 
better agreement than those in active regions, as one would expect.

\subsection{Emissivity ratios}

As EIS data are often used to measure electron densities via line ratios,
we have included various checks to make sure that the results have some 
consistency. Generally, one expects densities obtained by ions formed 
at similar temperatures to be similar. Also, that densities obtained from 
lines within the same ion are consistent, although in  principle some differences could occur if the plasma is strongly non-uniformly distributed.
Significant discrepancies within  the \ion{Fe}{12} lines 
and in the densities obtained by  \ion{Fe}{12} and \ion{Fe}{13} 
have been reported in the literature, see e.g. \cite{young_etal:2009_fe_13}.
  Fig.~\ref{fig:rel_den} shows an example: we see  a better consistency in the
  densities obtained from two line ratios for four ions  when using the
present calibration instead of the pre-flight one.

One way to assess at once all the line ratios within an ion is to plot the 
emissivity ratios \citep{delzanna_etal:2004_fe_10}, which are obtained by dividing 
 the observed intensity $I_{\rm ob}$ of a line  by its emissivity
\begin{equation}
F_{ji}= \frac{I_{\rm ob} N_{\rm e} C}{N_j(N_{\rm e} T_{\rm e}) A_{ji}} ,
\end{equation}
 calculated at a fixed electron  temperature $T_{\rm e}$ 
and  plotted  as a function of the density  $N_{\rm e}$
($N_j$ is the population of the upper state, and $A_{ji}$ the transition rate for spontaneous emission). This method  works if the lines have a similar temperature dependence. If the plasma is 
uniformly distributed along the line of sight, the  curves should form a crossing, which indicates the averaged density. 
 The scaling constant $C$ is arbitrarily chosen so that the crossing occurs at a 
 value around 1.

We have applied the  emissivity ratios to selected observations 
of \ion{Fe}{12}, \ion{Fe}{13}, and \ion{Si}{10} over the years. 
As these ions are formed at similar temperatures, we expect the densities 
to be similar. When the emissivities of the lines are independent of density, as in the 
case of the \ion{Fe}{17} lines, the emissivity ratios are plotted 
as a function of temperature.

\begin{figure*}
    \centerline{
    \includegraphics[clip,width=0.45\textwidth]{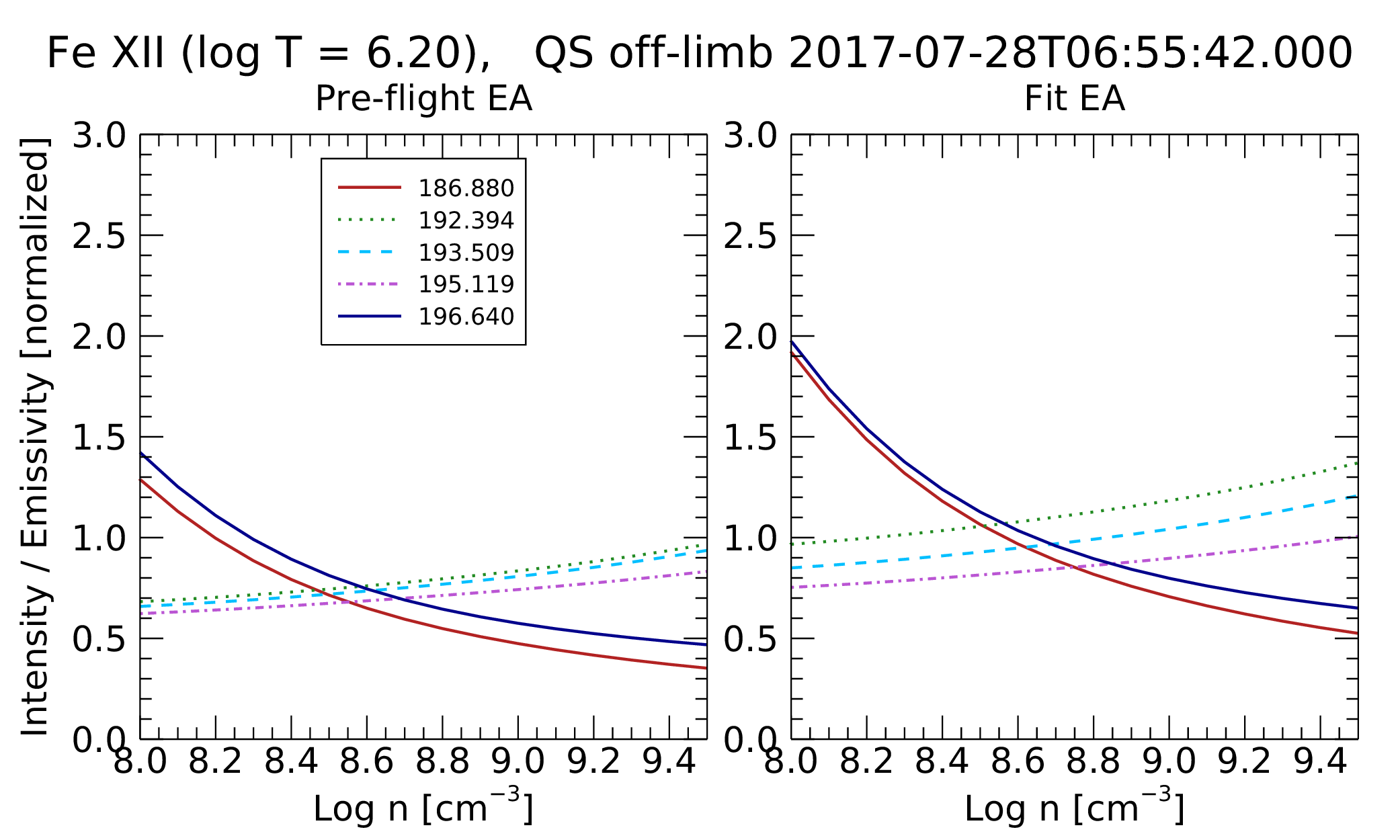}
    \includegraphics[clip,width=0.45\textwidth]{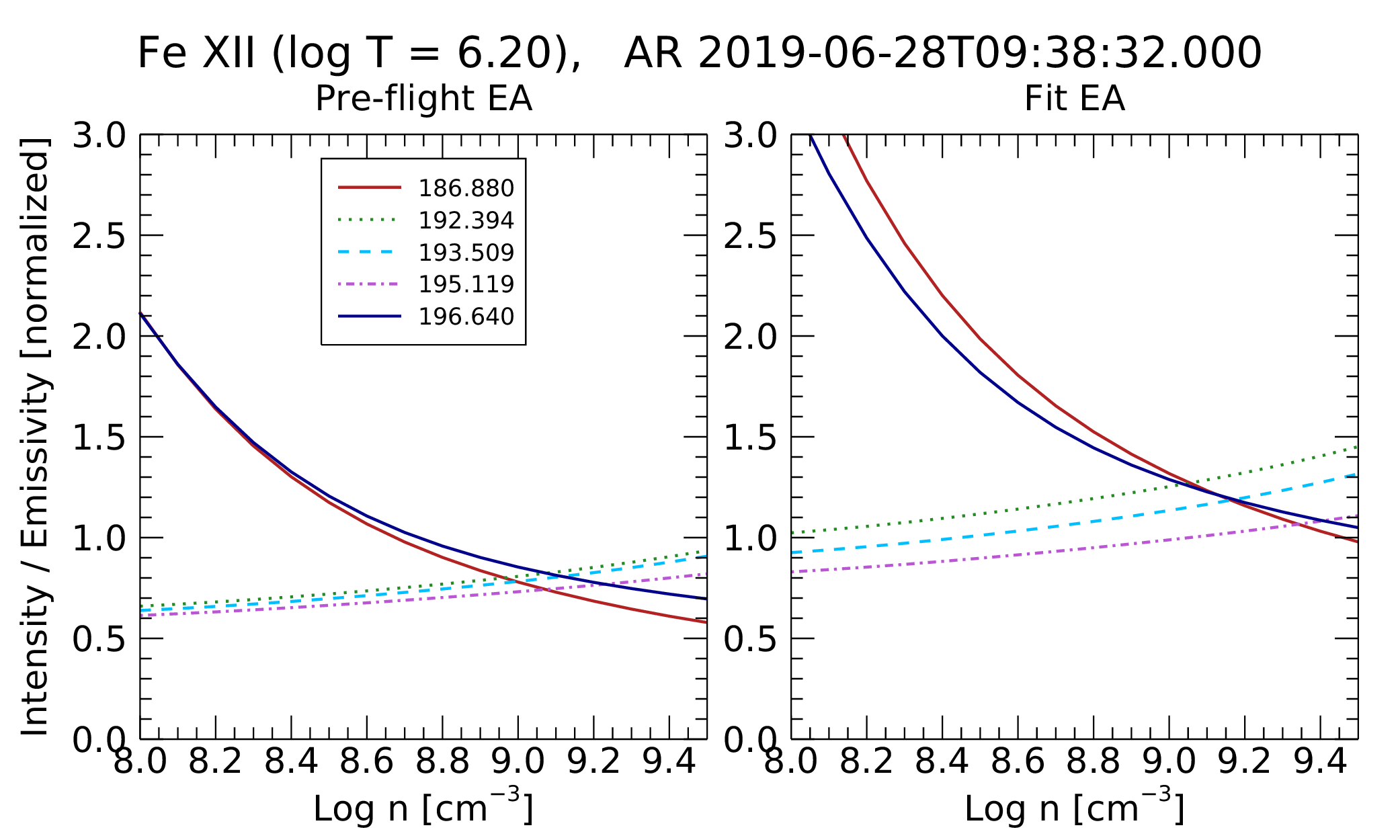}
}
    \centerline{
    \includegraphics[clip,width=0.45\textwidth]{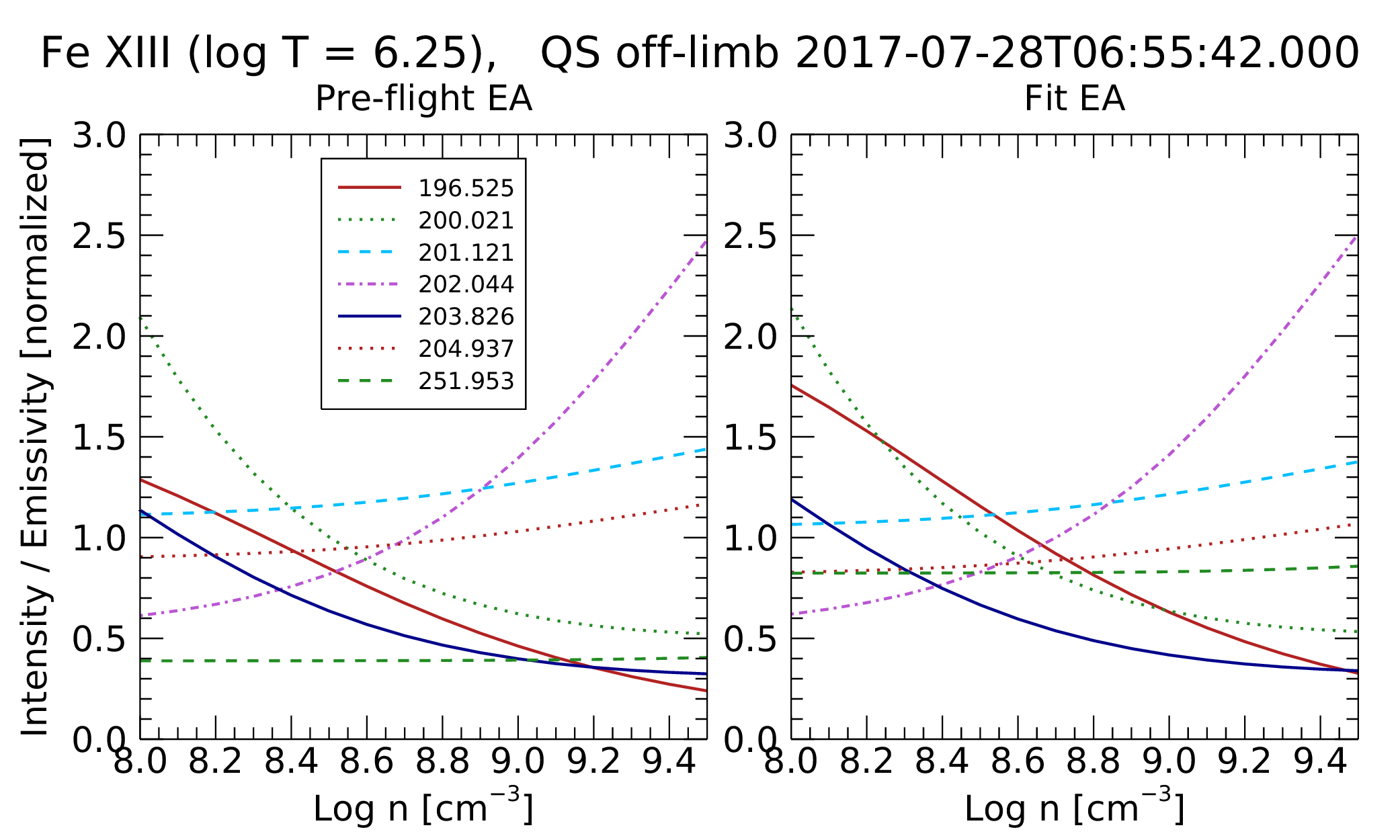}
    \includegraphics[clip,width=0.45\textwidth]{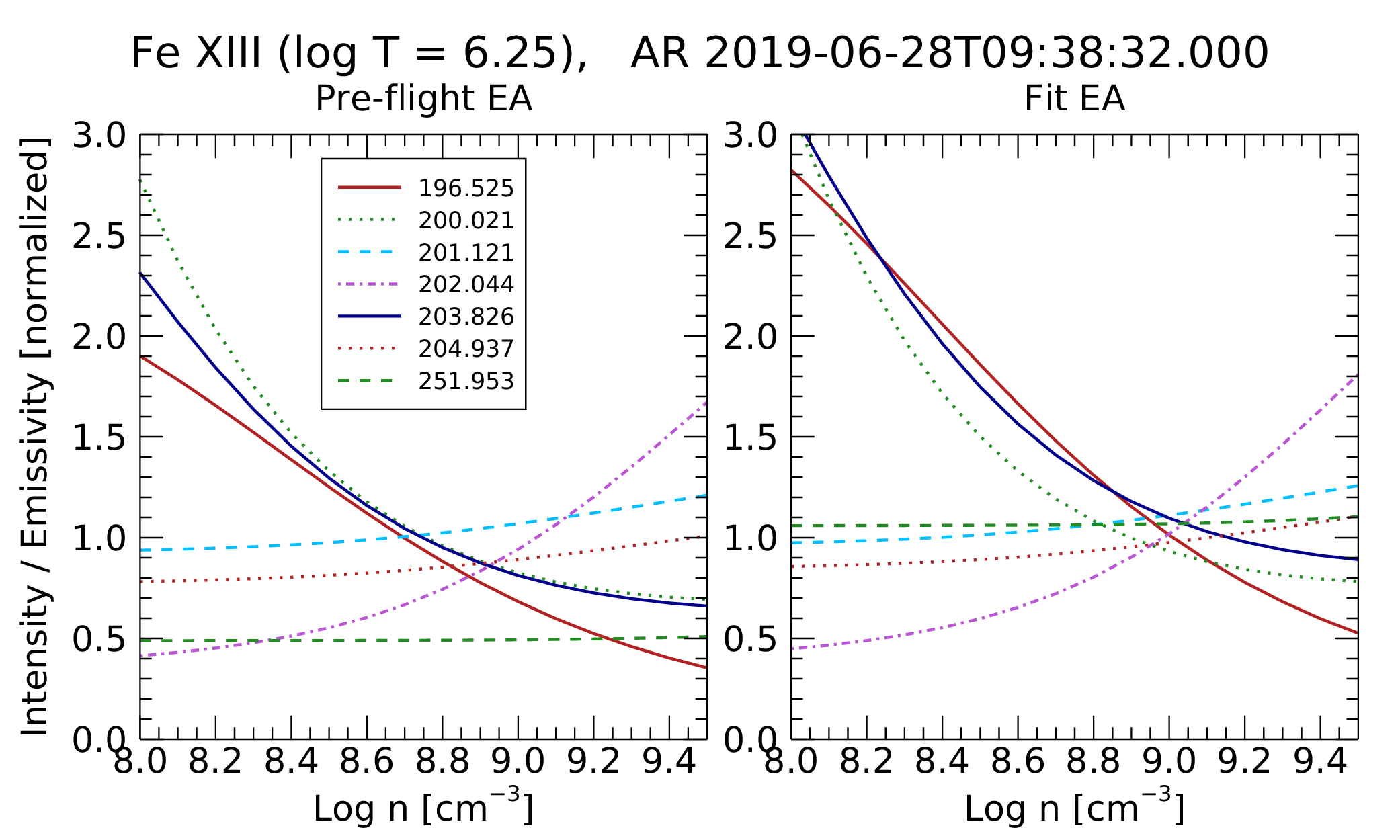}
}
    \centerline{
    \includegraphics[clip,width=0.45\textwidth]{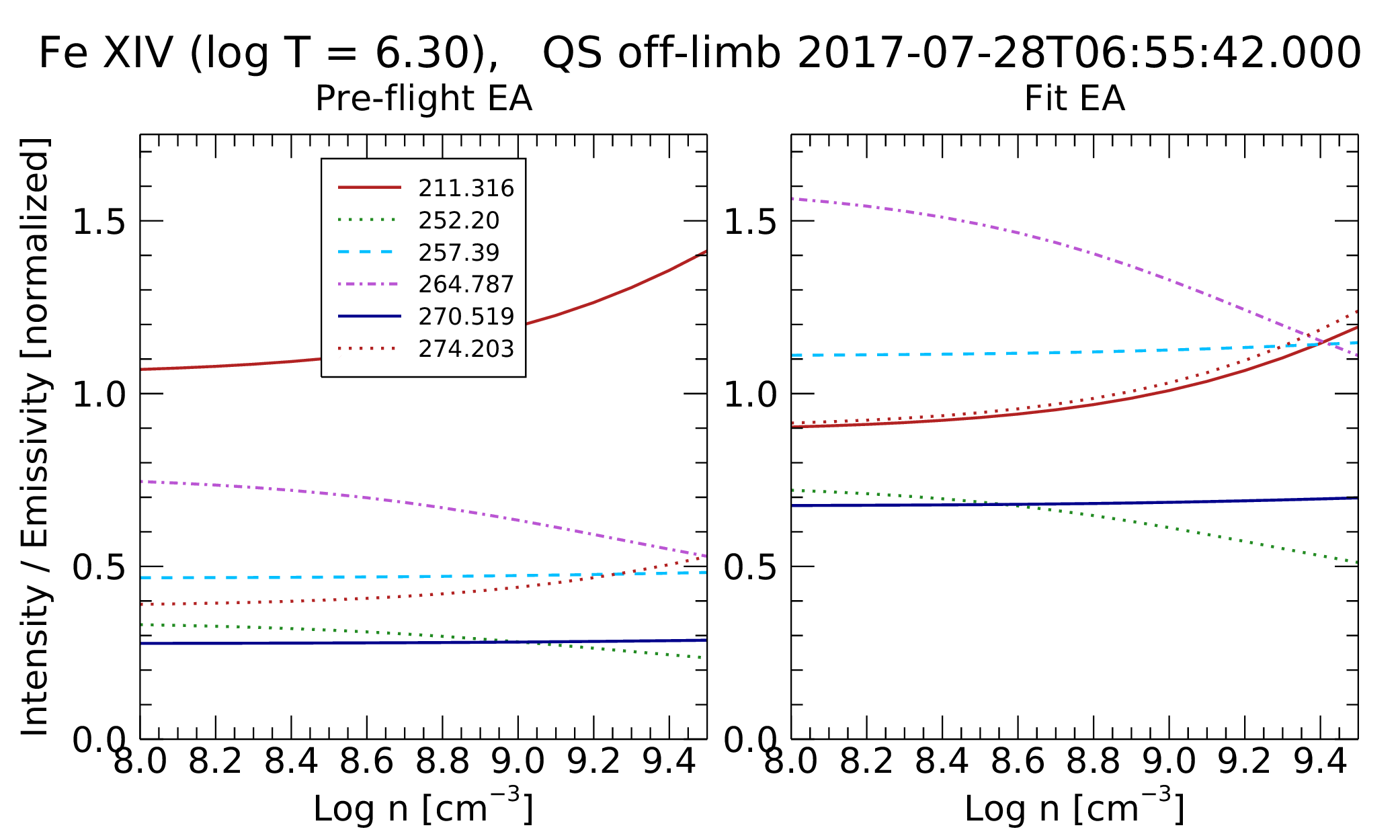}
    \includegraphics[clip,width=0.45\textwidth]{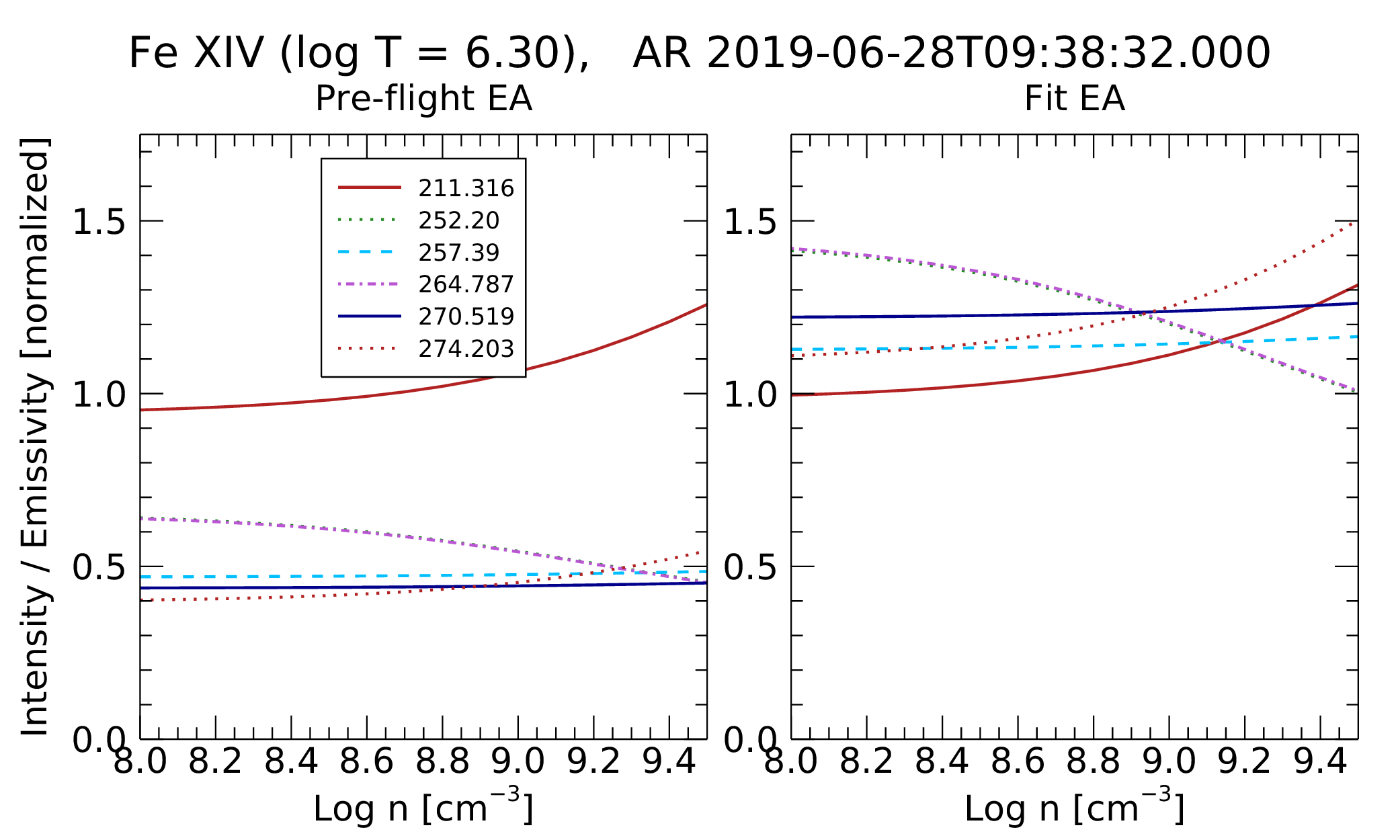}
}
    \centerline{
    \includegraphics[clip,width=0.45\textwidth]{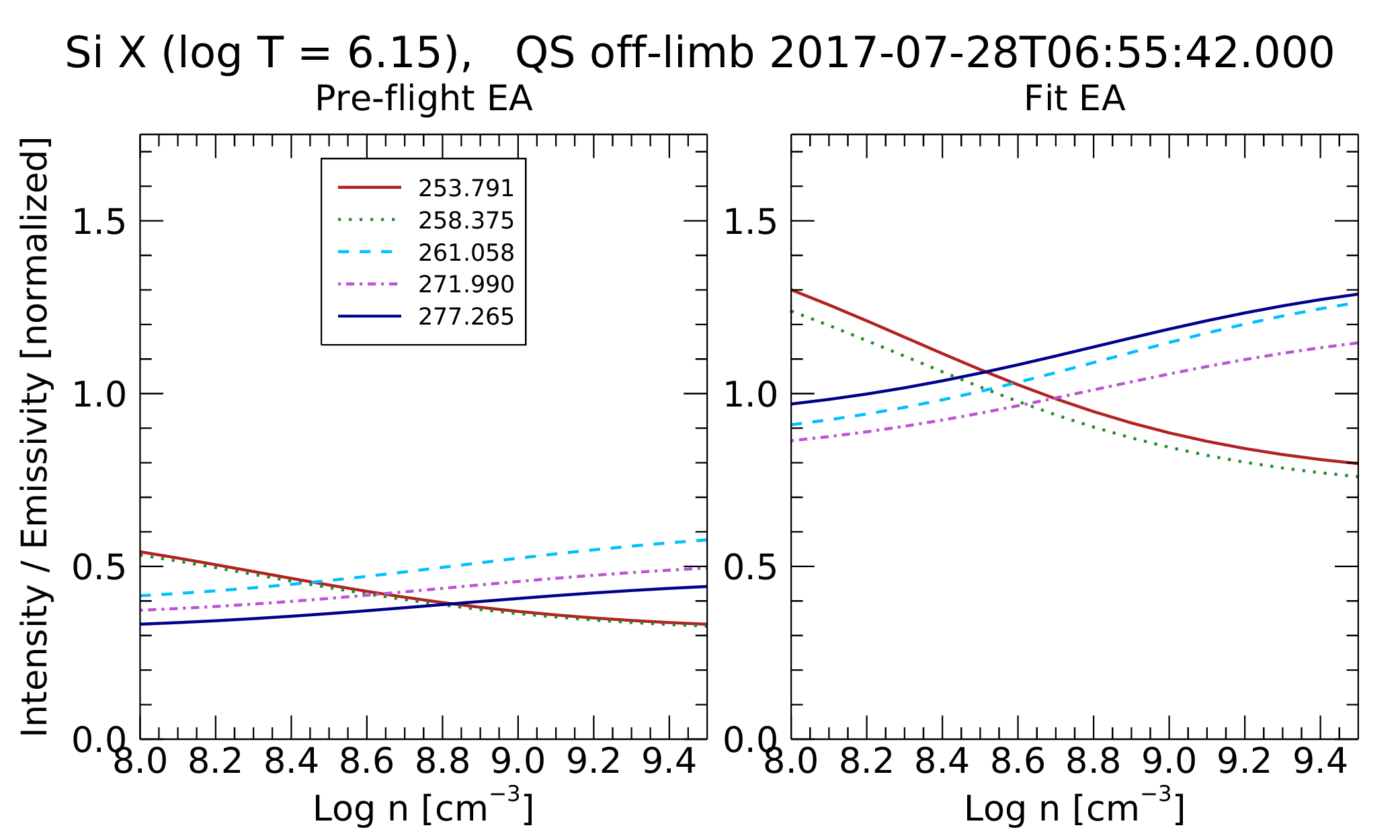}
    \includegraphics[clip,width=0.45\textwidth]{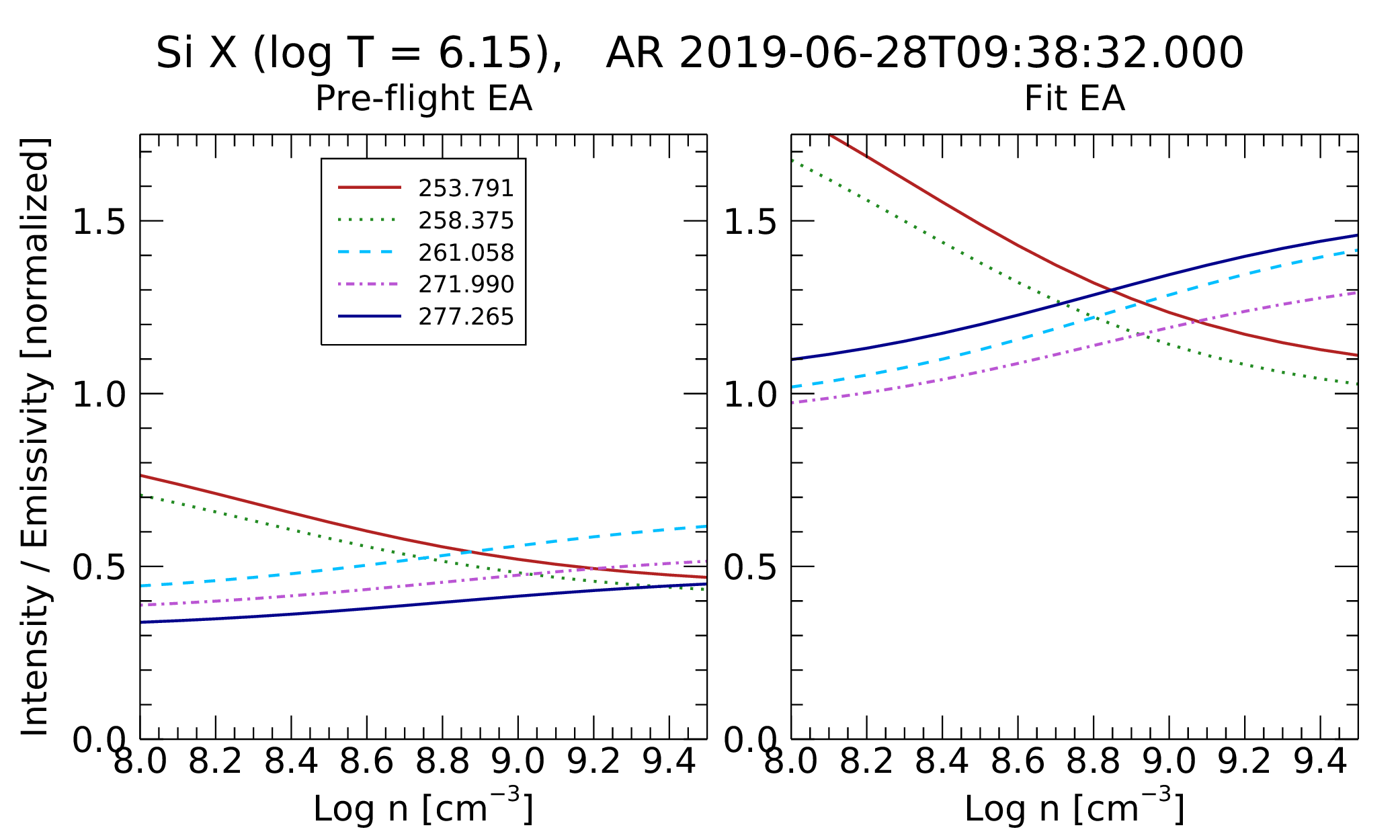}
}
\caption{Normalized emissivity curves for \ion{Fe}{12}, \ion{Fe}{13}, \ion{Fe}{14}, and \ion{Si}{10}. The left two columns show the curves computed using the pre-flight and fit effective areas for a QS off-limb subregion observed on 2017-07-28. The right two columns show the results for a subregion containing an AR observed on 2019-06-28.}
\label{fig:norm_emiss}
\end{figure*}

\begin{figure*}
\includegraphics[width=17cm]{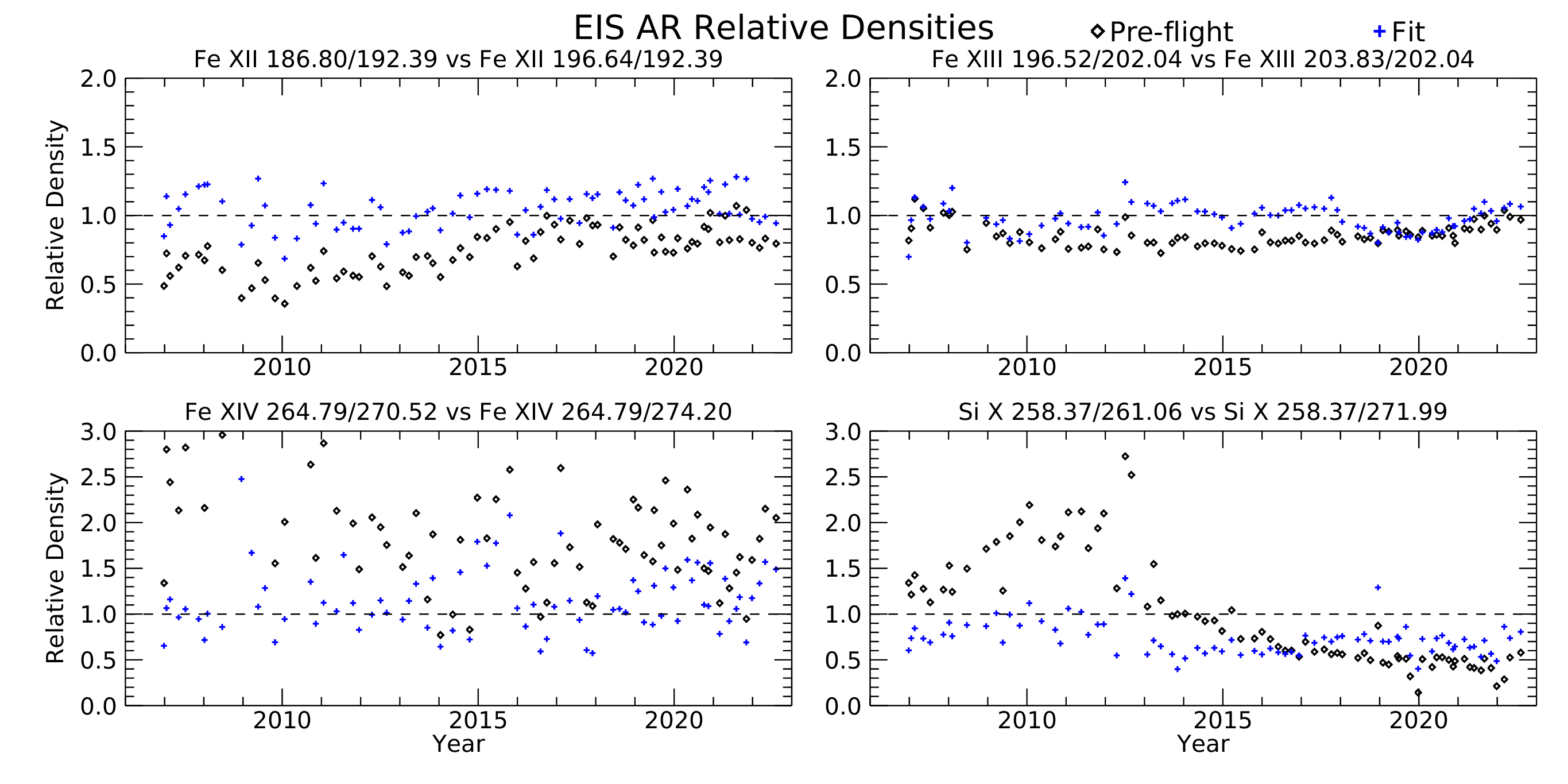}
  \caption{Relative density ratios computed using two density-sensitive line ratios for each of \ion{Fe}{12}, \ion{Fe}{13}, \ion{Fe}{14}, and \ion{Si}{10}. Values computed using the pre-flight calibration are shown as black diamonds and ratios computed from the new fits are shown as blue crosses.}
\label{fig:rel_den}
\end{figure*}

\section{Effective areas}

Fig.~\ref{fig:comp_e} shows a comparison for a few selected dates 
between the present effective areas, shown as a grey band which shows $\pm$ 20\% from the recommended values, and the previous calibrations.
The general shape and the peak location of the effective areas are stable over time.
  The SW channel has degraded in particular at shorter wavelengths, while at longer wavelengths
  actually showed a small (but puzzling) increase. The LW has generally degraded more than the SW, especially at longer wavelengths.

\begin{figure*}
    \centerline{
    \includegraphics[clip,width=0.45\textwidth]{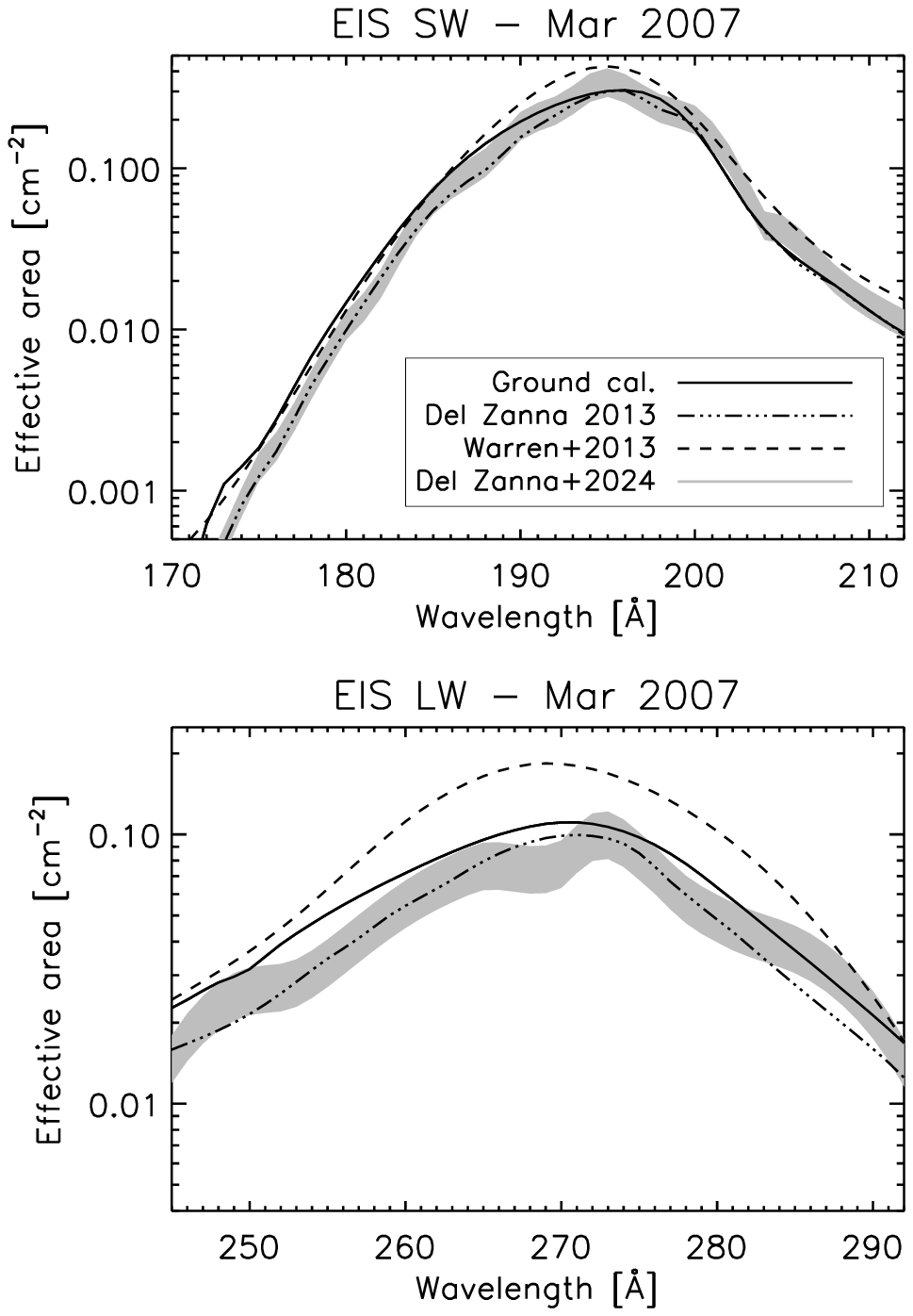}
    \includegraphics[clip,width=0.45\textwidth]{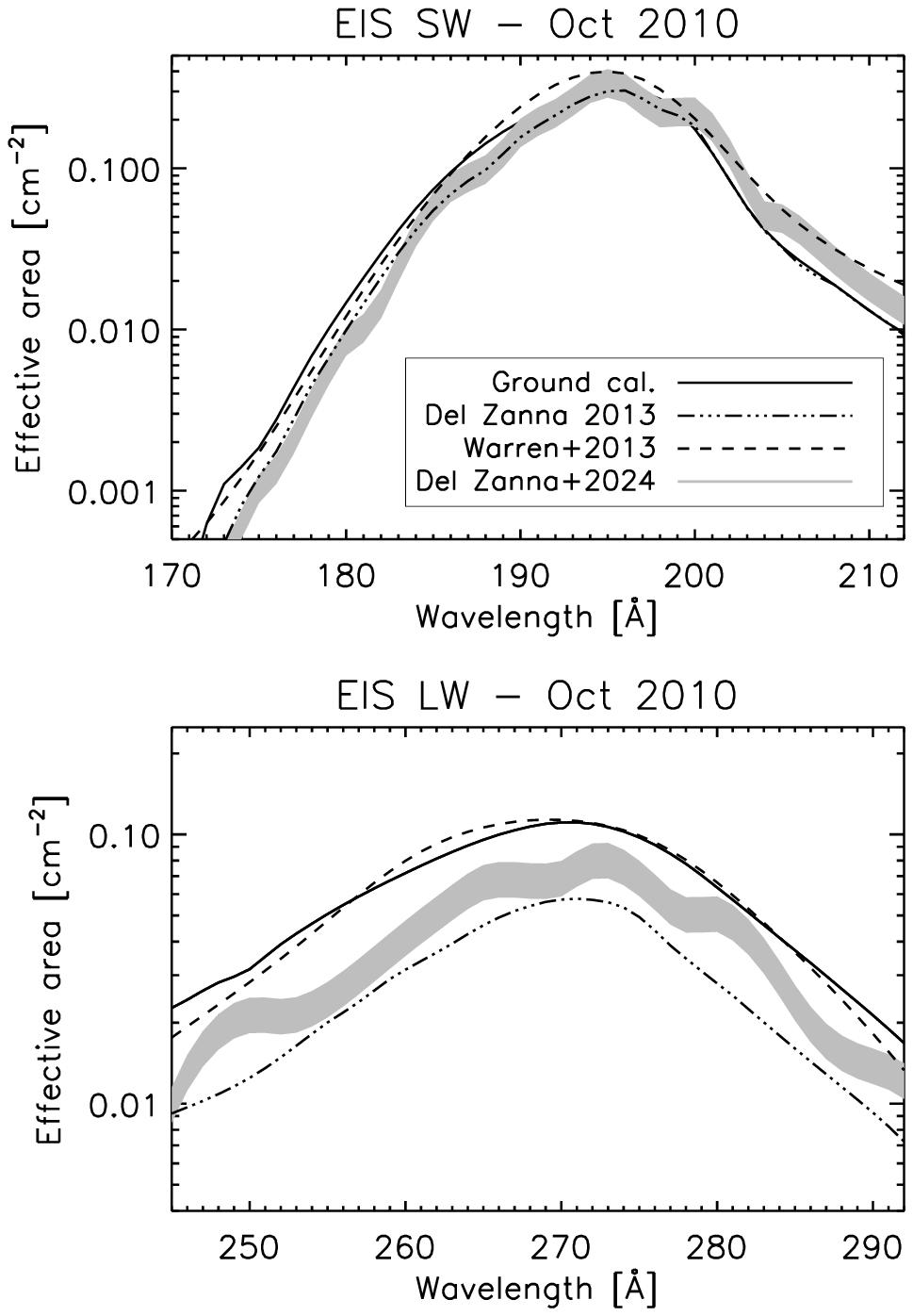}
}
    \centerline{
    \includegraphics[clip,width=0.45\textwidth]{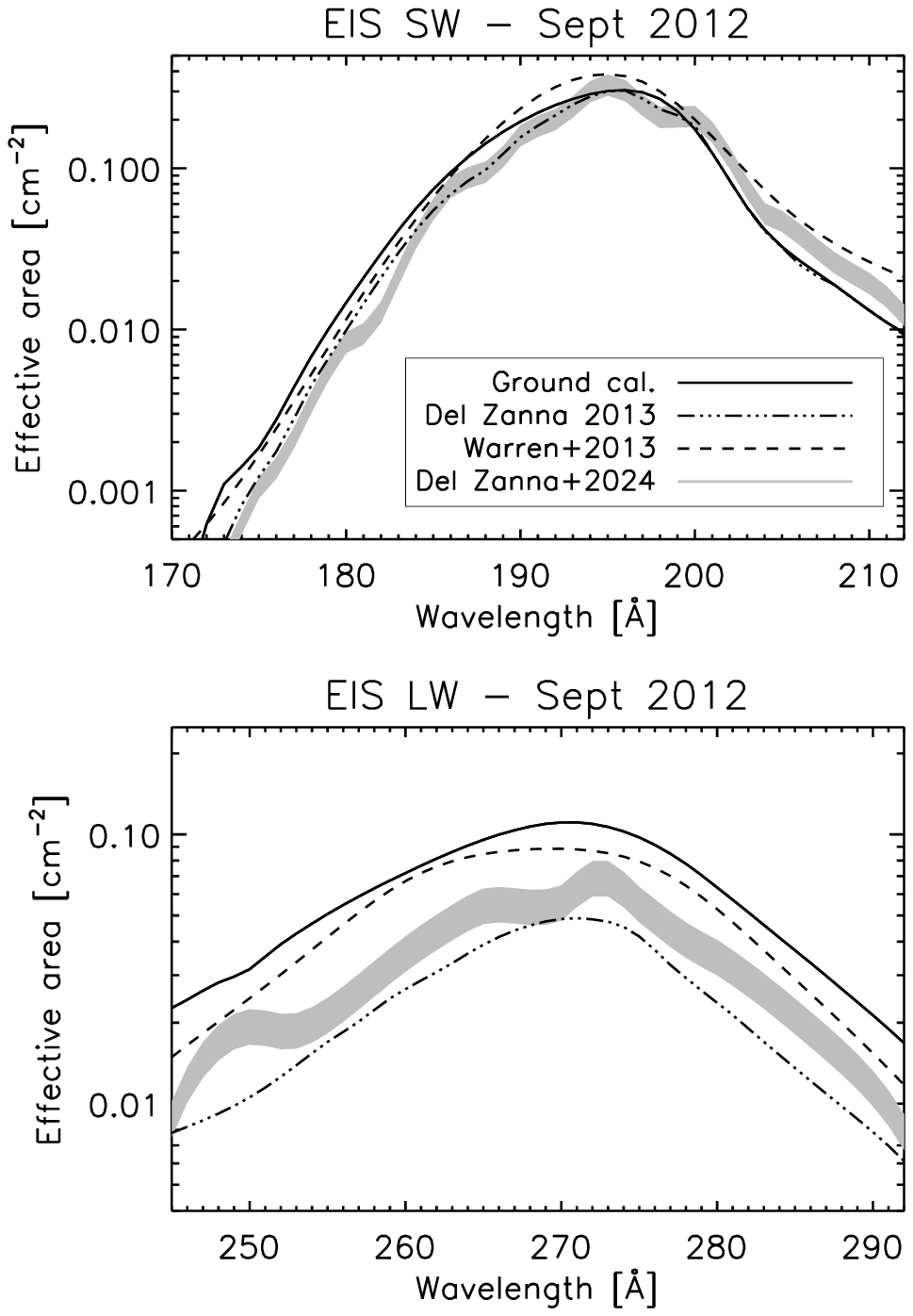}
    \includegraphics[clip,width=0.45\textwidth]{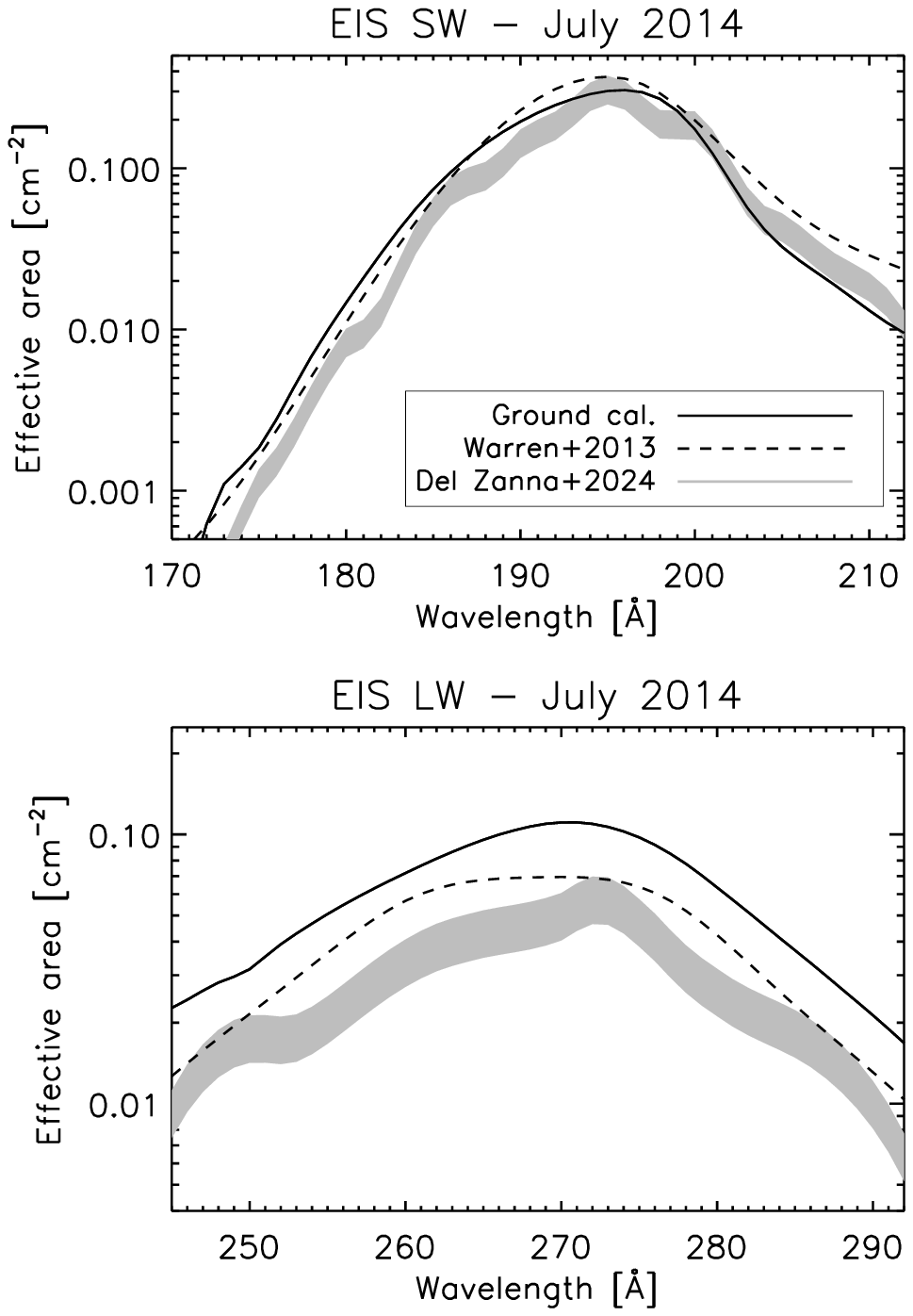}
}
\caption{Effective areas for a sample of dates. The present one is displayed as a grey band which shows $\pm$ 20\% from the recommended 
values. }
\label{fig:comp_e}
\end{figure*}

\section{On-disk AR loop DEM analysis}

As in the QS off-limb case, we have used a constant density
assumption to obtain the $G(T)$ of the lines.
The temperature distributions typically have multiple peaks,
one around log $T$[K]=5.6 produced by the loop legs, and other ones
due to the background/foreground coronal plasma present
in the AR core.

Table~\ref{tab:lines_ar_loop_26_oct_2010} lists the main lines.
The most useful lines are from \ion{Fe}{8}, as they cover both
SW and LW. The stronger lines are well represented within $\pm$10\%.
The  \ion{Fe}{9} lines are also useful to check the SW sensitivity,
especially at the location of the resonance line at
171.0~\AA, as this line is always very weak in any QS observation.
The three LW lines from \ion{Si}{7} are in good agreement with those
from \ion{Fe}{8}, while other lines show larger departures.
The \ion{O}{5} 192.9 vs 248.5~\AA\ ratio is not very reliable as lines
are weak and are both temperature and density sensitive.


\begin{table}
\caption{Same as Table~\ref{tab:lines_11_mar_2007}, but for an
  AR loop  observation on 26 Oct 2010}
\centering
  \scriptsize
   \setlength\tabcolsep{3.pt}
\begin{tabular}{@{}rrrccrlrr@{}}
\hline\hline \noalign{\smallskip}
 $\lambda_{\rm obs}$  & DN & $I_{\rm obs}$   & log $T_{\rm max}$ & log $T_{\rm eff}$  & $R$ & Ion & $\lambda_{\rm exp}$   &  $r$ \\
\noalign{\smallskip}\hline\noalign{\smallskip}
 
 192.91 &     2690 & 170.9 &  5.42 &  5.56 &  1.38 &  \ionm{O}{v} &  192.904 & 0.84 \\ 
 
 248.49 &      260 & 319.8 &  5.42 &  5.58 &  0.87 &  \ionm{O}{v} &  248.461 & 0.91 \\ 
 
 276.58 &     1210 & 296.1 &  5.50 &  5.59 &  0.74 &  \ionm{Mg}{v} &  276.579 & 0.98 \\ 
 
 246.00 &      524 & 799.5 &  5.64 &  5.67 &  0.46 &  \ionm{Si}{vi} &  246.003 & 0.98 \\ 
 
 270.39 &     3630 & 887.6 &  5.65 &  5.68 &  0.36 &  \ionm{Mg}{vi} &  270.400 & 0.12 \\ 
                              &   &   &  &  &  &  \ionm{Mg}{vi} &  270.391 & 0.87 \\ 
 
 268.99 &     1720 & 464.9 &  5.66 &  5.68 &  0.34 &  \ionm{Mg}{vi} &  268.991 & 0.98 \\ 
 
 255.11 &      252 & 130.6 &  5.66 &  5.73 &  0.96 &  \ionm{Fe}{viii} &  255.110 & 0.99 \\ 
 
 255.35 &      391 & 197.3 &  5.67 &  5.73 &  1.03 &  \ionm{Fe}{viii} &  255.350 & 0.96 \\ 
 
 253.96 &      509 & 306.6 &  5.67 &  5.74 &  1.06 &  \ionm{Fe}{viii} &  253.956 & 0.99 \\ 
 
 195.97 &     5720 & 298.8 &  5.31 &  5.75 &  1.12 &  \ionm{Fe}{viii} &  195.972 & 0.92 \\ 
 
 207.12 &      286 & 183.7 &  5.70 &  5.76 &  0.93 &  \ionm{Fe}{viii} &  207.124 & 0.73 \\ 
                              &   &   &  &  &  &  \ionm{Fe}{viii} &  207.124 & 0.25 \\ 
 
 197.36 &     3050 & 172.6 &  5.71 &  5.76 &  0.84 &  \ionm{Fe}{viii} &  197.362 & 0.97 \\ 
 
 194.66 &     8270 & 442.0 &  5.72 &  5.76 &  1.17 &  \ionm{Fe}{viii} &  194.661 & 0.99 \\ 
 
 186.60 &     8150 & 1573.1 &  5.73 &  5.80 &  1.03 &  \ionm{Fe}{viii} &  186.598 & 0.96 \\ 
 
 185.22 &     7630 & 2116.2 &  5.73 &  5.80 &  1.09 &  \ionm{Fe}{viii} &  185.213 & 0.97 \\ 
 
 280.74 &      849 & 304.8 &  5.79 &  5.82 &  0.64 &  \ionm{Mg}{vii} &  280.742 & 0.98 \\ 
 
 276.14 &      613 & 146.1 &  5.79 &  5.83 &  0.66 &  \ionm{Mg}{vii} &  276.154 & 0.98 \\ 
 
 275.37 &     2660 & 612.9 &  5.79 &  5.83 &  1.03 &  \ionm{Si}{vii} &  275.361 & 0.98 \\ 
 
 275.69 &      458 & 106.8 &  5.80 &  5.83 &  1.05 &  \ionm{Si}{vii} &  275.676 & 0.98 \\ 
 
 272.66 &      940 & 183.8 &  5.80 &  5.83 &  1.07 &  \ionm{Si}{vii} &  272.648 & 0.98 \\ 
 
 278.39 &     3270 & 907.6 &  5.79 &  5.83 &  0.53 &  \ionm{Mg}{vii} &  278.404 & 0.97 \\ 
 
 277.00 &     2460 & 615.3 &  5.84 &  5.87 &  0.80 &  \ionm{Mg}{vii} &  277.003 & 0.57 \\ 
                               &  &   &  &  &  &  \ionm{Si}{viii} &  277.058 & 0.39 \\ 
 
 276.84 &      953 & 236.0 &  5.87 &  5.88 &  1.26 &  \ionm{Si}{viii} &  276.850 & 0.47 \\ 
                              &   &   &  &  &  &  \ionm{Si}{vii} &  276.851 & 0.48 \\ 
 
 171.07 &      122 & 8739.7 &  5.93 &  5.91 &  0.92 &  \ionm{Fe}{ix} &  171.073 & 0.99 \\ 
 
 189.94 &     2450 & 250.6 &  5.94 &  5.92 &  0.94 &  \ionm{Fe}{ix} &  189.935 & 0.97 \\ 
 
 191.21 &     1770 & 150.2 &  5.94 &  5.92 &  0.69 &  \ionm{Fe}{ix} &  191.206 & 0.98 \\ 
 
 188.49 &     3240 & 432.2 &  5.94 &  5.93 &  1.00 &  \ionm{Fe}{ix} &  188.493 & 0.91 \\ 
 
 197.85 &     2760 & 162.7 &  5.96 &  5.94 &  1.01 &  \ionm{Fe}{ix} &  197.854 & 0.95 \\ 
 
 190.04 &     2370 & 239.2 &  6.05 &  6.04 &  1.07 &  \ionm{Fe}{x} &  190.037 & 0.68 \\ 
                            &     &   &  &  &  &  \ionm{Fe}{ix} &  190.059 & 0.17 \\ 

\noalign{\smallskip}\hline 
\end{tabular}
\normalsize
 \label{tab:lines_ar_loop_26_oct_2010} 
\end{table}

\section{Off-limb DEM analysis}

Tables~\ref{tab:ol2},\ref{tab:ol3},\ref{tab:ol4},
\ref{tab:ol5},\ref{tab:ol6},\ref{tab:ol7}, and \ref{tab:ol8}
show the summary results of the DEM analysis for the main (strongest) coronal lines, in off-limb 
observations. Note that in the last observation 
hotter active region plasma was present.

\begin{table}
\caption{Same as Table~\ref{tab:lines_11_mar_2007}, but for a 
QS off-limb observation on 16 Oct 2010}
\centering
  \scriptsize
   \setlength\tabcolsep{3.pt}
\begin{tabular}{@{}rrrccrlrr@{}}
\hline\hline \noalign{\smallskip}
 $\lambda_{\rm obs}$  & DN & $I_{\rm obs}$   & log $T_{\rm max}$ & log $T_{\rm eff}$  & $R$ & Ion & $\lambda_{\rm exp}$   &  $r$ \\
\noalign{\smallskip}\hline\noalign{\smallskip}
 194.66 &      238 & 6.4 &  5.72 &  6.02 &  1.27 &  \ionm{Fe}{viii} &  194.661 & 0.94 \\ 
 
 185.22 &      218 & 30.3 &  5.73 &  6.03 &  1.30 &  \ionm{Fe}{viii} &  185.213 & 0.93 \\ 
 
 186.60 &      222 & 23.0 &  5.73 &  6.04 &  1.30 &  \ionm{Fe}{viii} &  186.598 & 0.86 \\ 
 
 189.94 &      382 & 19.6 &  5.94 &  6.10 &  1.24 &  \ionm{Fe}{ix} &  189.935 & 0.97 \\ 
 
 188.49 &      728 & 52.4 &  5.95 &  6.10 &  0.97 &  \ionm{Fe}{ix} &  188.493 & 0.89 \\ 
 
 197.85 &      886 & 26.0 &  5.96 &  6.11 &  1.01 &  \ionm{Fe}{ix} &  197.854 & 0.93 \\ 
 
 257.26 &     1030 & 214.5 &  6.04 &  6.13 &  0.98 &  \ionm{Fe}{x} &  257.259 & 0.22 \\ 
                              &   &   &  &  &  &  \ionm{Fe}{x} &  257.263 & 0.76 \\ 
 
 177.24 &      204 & 646.7 &  6.05 &  6.14 &  1.03 &  \ionm{Fe}{x} &  177.240 & 0.97 \\ 
 
 174.53 &      108 & 1101.1 &  6.05 &  6.14 &  1.07 &  \ionm{Fe}{x} &  174.531 & 0.98 \\ 
 
 184.54 &     1700 & 281.6 &  6.05 &  6.14 &  0.89 &  \ionm{Fe}{x} &  184.537 & 0.95 \\ 
 
 190.04 &     2170 & 109.3 &  6.06 &  6.15 &  1.05 &  \ionm{Fe}{xii} &  190.040 & 0.19 \\ 
                              &   &   &  &  &  &  \ionm{Fe}{x} &  190.037 & 0.68 \\ 
 257.55 &      322 & 65.2 &  6.12 &  6.16 &  1.25 &  \ionm{Fe}{xi} &  257.554 & 0.69 \\ 
                              &   &   &  &  &  &  \ionm{Fe}{xi} &  257.547 & 0.26 \\ 
 
 189.02 &      298 & 18.9 &  6.13 &  6.16 &  0.73 &  \ionm{Fe}{xi} &  188.997 & 0.95 \\ 
 
 256.92 &      492 & 104.7 &  6.12 &  6.16 &  0.97 &  \ionm{Fe}{xi} &  256.919 & 0.91 \\ 
 
 182.17 &      507 & 193.2 &  6.13 &  6.16 &  0.98 &  \ionm{Fe}{xi} &  182.167 & 0.97 \\ 
 
 192.81 &     4550 & 160.2 &  6.13 &  6.16 &  0.89 &  \ionm{Fe}{xi} &  192.813 & 0.96 \\ 
 
 188.30 &     5980 & 448.3 &  6.13 &  6.16 &  0.92 &  \ionm{Fe}{xi} &  188.299 & 0.97 \\ 
 
 181.13 &       95 & 57.3 &  6.13 &  6.16 &  1.13 &  \ionm{Fe}{xi} &  181.130 & 0.97 \\ 
 
 188.22 &     9530 & 726.2 &  6.13 &  6.16 &  0.92 &  \ionm{Fe}{xi} &  188.216 & 0.97 \\ 
 
 180.40 &     1630 & 1340.4 &  6.13 &  6.17 &  1.05 &  \ionm{Fe}{xi} &  180.401 & 0.97 \\ 
 
 180.60 &       69 & 52.7 &  6.13 &  6.17 &  0.88 &  \ionm{Fe}{xi} &  180.594 & 0.96 \\ 
 
 202.71 &      423 & 47.3 &  6.13 &  6.17 &  1.23 &  \ionm{Fe}{xi} &  202.705 & 0.92 \\ 
 
 253.79 &      167 & 51.5 &  6.15 &  6.17 &  0.93 &  \ionm{Si}{x} &  253.790 & 0.97 \\ 
 
 258.37 &     1300 & 248.6 &  6.15 &  6.17 &  0.98 &  \ionm{Si}{x} &  258.374 & 0.97 \\ 
 
 261.06 &      845 & 135.0 &  6.15 &  6.17 &  0.99 &  \ionm{Si}{x} &  261.056 & 0.98 \\ 
 
 271.99 &     1200 & 116.5 &  6.15 &  6.17 &  0.95 &  \ionm{Si}{x} &  271.992 & 0.97 \\ 
 
 277.27 &      687 & 87.2 &  6.15 &  6.17 &  0.98 &  \ionm{Si}{x} &  277.264 & 0.98 \\ 
 
 256.40 &     1030 & 229.8 &  6.16 &  6.18 &  1.18 &  \ionm{Si}{x} &  256.377 & 0.59 \\ 
 &                                &   &  &  &  &  \ionm{Fe}{xiii} &  256.400 & 0.12 \\ 
  &                               &   &  &  &  &  \ionm{Fe}{xii} &  256.410 & 0.18 \\ 
 
 249.39 &      114 & 65.0 &  6.19 &  6.19 &  0.86 &  \ionm{Fe}{xii} &  249.388 & 0.70 \\ 
   &                              &   &  &  &  &  \ionm{Fe}{xii} &  249.384 & 0.23 \\ 
 
 196.64 &     1900 & 51.4 &  6.19 &  6.19 &  1.16 &  \ionm{Fe}{xii} &  196.640 & 0.93 \\ 

 186.88 &     2310 & 225.8 &  6.19 &  6.19 &  0.98 &  \ionm{Fe}{xii} &  186.854 & 0.16 \\ 
                              &   &   &  &  &  &  \ionm{Fe}{xii} &  186.887 & 0.77 \\ 
 
 193.51 &    22900 & 765.5 &  6.19 &  6.19 &  1.02 &  \ionm{Fe}{xii} &  193.509 & 0.95 \\ 
 
 195.12 &    36200 & 852.8 &  6.19 &  6.19 &  1.32 &  \ionm{Fe}{xii} &  195.119 & 0.97 \\ 
 
 192.39 &     9820 & 343.6 &  6.19 &  6.19 &  1.05 &  \ionm{Fe}{xii} &  192.394 & 0.96 \\ 
 
 196.52 &      674 & 18.1 &  6.24 &  6.21 &  1.21 &  \ionm{Fe}{xiii} &  196.525 & 0.92 \\ 
 
 200.02 &     1420 & 59.2 &  6.25 &  6.21 &  0.93 &  \ionm{Fe}{xiii} &  200.021 & 0.95 \\ 
 
 261.73 &      136 & 20.9 &  6.25 &  6.22 &  1.17 &  \ionm{Fe}{xiii} &  261.743 & 0.97 \\ 
 
 203.83 &     1430 & 229.6 &  6.25 &  6.22 &  1.06 &  \ionm{Fe}{xiii} &  203.795 & 0.28 \\ 
                              &   &   &  &  &  &  \ionm{Fe}{xiii} &  203.826 & 0.63 \\ 
 
 246.21 &      142 & 106.4 &  6.25 &  6.22 &  1.02 &  \ionm{Fe}{xiii} &  246.209 & 0.95 \\ 
 
 251.95 &      483 & 193.3 &  6.25 &  6.22 &  1.03 &  \ionm{Fe}{xiii} &  251.952 & 0.97 \\ 
 
 204.94 &      248 & 53.5 &  6.25 &  6.22 &  0.98 &  \ionm{Fe}{xiii} &  204.942 & 0.96 \\ 
 
 209.92 &      271 & 144.5 &  6.25 &  6.22 &  0.94 &  \ionm{Fe}{xiii} &  209.916 & 0.96 \\ 
 
 202.04 &     9310 & 819.2 &  6.25 &  6.22 &  0.95 &  \ionm{Fe}{xiii} &  202.044 & 0.97 \\ 
 
 264.79 &     1300 & 181.7 &  6.29 &  6.23 &  0.97 &  \ionm{Fe}{xiv} &  264.789 & 0.81 \\ 
                                & &   &  &  &  &  \ionm{Fe}{xi} &  264.772 & 0.15 \\ 
 
 274.20 &     1890 & 223.8 &  6.29 &  6.24 &  0.98 &  \ionm{Fe}{xiv} &  274.204 & 0.96 \\ 
 
 252.20 &       81 & 31.6 &  6.29 &  6.24 &  1.16 &  \ionm{Fe}{xiv} &  252.200 & 0.95 \\ 
 
 289.15 &       26 & 19.9 &  6.29 &  6.24 &  0.92 &  \ionm{Fe}{xiv} &  289.151 & 0.96 \\ 
 
 270.52 &      814 & 96.7 &  6.29 &  6.24 &  1.00 &  \ionm{Fe}{xiv} &  270.521 & 0.96 \\ 
 
 257.40 &      221 & 45.3 &  6.29 &  6.24 &  1.53 &  \ionm{Fe}{xiv} &  257.394 & 0.96 \\ 
 
 211.32 &      632 & 392.7 &  6.29 &  6.24 &  1.00 &  \ionm{Fe}{xiv} &  211.317 & 0.96 \\ 
 
 284.16 &     1560 & 508.3 &  6.34 &  6.27 &  1.01 &  \ionm{Fe}{xv} &  284.163 & 0.96 \\ 
\noalign{\smallskip}\hline 
\end{tabular}
\normalsize
 \label{tab:ol2} 
\end{table}

\begin{table}
\caption{Same as Table~\ref{tab:lines_11_mar_2007}, but for a 
QS off-limb observation on 7 Oct 2013}
\centering
  \scriptsize
   \setlength\tabcolsep{3.pt}
\begin{tabular}{@{}rrrccrlrr@{}}
\hline\hline \noalign{\smallskip}
 $\lambda_{\rm obs}$  & DN & $I_{\rm obs}$   & log $T_{\rm max}$ & log $T_{\rm eff}$  & $R$ & Ion & $\lambda_{\rm exp}$   &  $r$ \\
\noalign{\smallskip}\hline\noalign{\smallskip}
 194.66 &      513 & 11.7 &  5.72 &  6.01 &  0.88 &  \ionm{Fe}{viii} &  194.661 & 0.94 \\ 
 
 186.60 &      514 & 38.9 &  5.73 &  6.04 &  1.00 &  \ionm{Fe}{viii} &  186.598 & 0.82 \\ 
 
 185.22 &      365 & 45.3 &  5.73 &  6.05 &  1.17 &  \ionm{Fe}{viii} &  185.213 & 0.86 \\ 
 
  189.94 &      566 & 26.7 &  5.94 &  6.10 &  1.12 &  \ionm{Fe}{ix} &  189.935 & 0.95 \\ 
 
 188.49 &      954 & 60.3 &  5.95 &  6.11 &  1.03 &  \ionm{Fe}{ix} &  188.493 & 0.83 \\ 
 
 197.85 &      981 & 28.0 &  5.96 &  6.11 &  1.10 &  \ionm{Fe}{ix} &  197.854 & 0.91 \\ 
 
 257.26 &     1130 & 268.7 &  6.04 &  6.15 &  0.90 &  \ionm{Fe}{x} &  257.259 & 0.24 \\ 
                                 & &   &  &  &  &  \ionm{Fe}{x} &  257.263 & 0.74 \\

 177.24 &      354 & 824.7 &  6.05 &  6.15 &  0.97 &  \ionm{Fe}{x} &  177.240 & 0.97 \\ 
 
 174.53 &      188 & 1328.5 &  6.05 &  6.15 &  1.07 &  \ionm{Fe}{x} &  174.531 & 0.97 \\ 
 
 184.54 &     2230 & 330.4 &  6.05 &  6.16 &  0.93 &  \ionm{Fe}{x} &  184.537 & 0.94 \\ 
 
 190.04 &     3100 & 142.8 &  6.06 &  6.17 &  1.06 &  \ionm{Fe}{xii} &  190.040 & 0.23 \\

 184.79 &      207 & 28.8 &  6.12 &  6.18 &  1.05 &  \ionm{Fe}{xi} &  184.793 & 0.94 \\ 
 
 257.55 &      479 & 110.9 &  6.12 &  6.18 &  1.02 &  \ionm{Fe}{xi} &  257.554 & 0.68 \\ 
                                 & &   &  &  &  &  \ionm{Fe}{xi} &  257.547 & 0.26 \\ 
 
 256.92 &      725 & 175.9 &  6.12 &  6.18 &  0.78 &  \ionm{Fe}{xi} &  256.919 & 0.89 \\ 
 
 182.17 &      954 & 345.8 &  6.13 &  6.18 &  0.81 &  \ionm{Fe}{xi} &  182.167 & 0.96 \\ 
 
 192.81 &     7840 & 220.4 &  5.42 &  6.18 &  0.90 &  \ionm{Fe}{xi} &  192.813 & 0.96 \\ 
 
 181.13 &      204 & 106.4 &  6.13 &  6.18 &  0.91 &  \ionm{Fe}{xi} &  181.130 & 0.97 \\ 
 
 188.30 &     8990 & 581.8 &  6.13 &  6.18 &  0.98 &  \ionm{Fe}{xi} &  188.299 & 0.97 \\ 
 
 188.22 &    15900 & 1030.8 &  6.13 &  6.18 &  0.91 &  \ionm{Fe}{xi} &  188.216 & 0.97 \\ 
 
 180.40 &     2880 & 1927.1 &  6.13 &  6.18 &  1.01 &  \ionm{Fe}{xi} &  180.401 & 0.97 \\ 
 
 202.71 &      852 & 75.9 &  6.13 &  6.19 &  1.09 &  \ionm{Fe}{xi} &  202.705 & 0.90 \\ 
 
 258.37 &     1990 & 435.5 &  6.15 &  6.19 &  0.88 &  \ionm{Si}{x} &  258.374 & 0.97 \\ 
 
 253.79 &      278 & 88.5 &  6.15 &  6.19 &  0.85 &  \ionm{Si}{x} &  253.790 & 0.97 \\ 
 
 271.99 &     1580 & 192.4 &  6.15 &  6.19 &  0.83 &  \ionm{Si}{x} &  271.992 & 0.97 \\ 
 
 261.06 &     1260 & 229.3 &  6.15 &  6.19 &  0.84 &  \ionm{Si}{x} &  261.056 & 0.98 \\ 
 
 277.27 &      836 & 133.8 &  6.15 &  6.19 &  0.91 &  \ionm{Si}{x} &  277.264 & 0.98 \\ 
 
 256.40 &     1550 & 393.3 &  6.16 &  6.20 &  1.04 &  \ionm{Si}{x} &  256.377 & 0.56 \\ 
                                 & &   &  &  &  &  \ionm{Fe}{xiii} &  256.400 & 0.15 \\ 
                                 & &   &  &  &  &  \ionm{Fe}{xii} &  256.410 & 0.20 \\ 
 
 249.39 &      184 & 91.0 &  6.19 &  6.20 &  1.00 &  \ionm{Fe}{xii} &  249.388 & 0.71 \\ 
                                 & &   &  &  &  &  \ionm{Fe}{xii} &  249.384 & 0.23 \\ 
 
 196.64 &     4250 & 107.1 &  6.19 &  6.20 &  1.04 &  \ionm{Fe}{xii} &  196.640 & 0.94 \\ 
 
 186.88 &     5200 & 367.1 &  6.19 &  6.20 &  1.13 &  \ionm{Fe}{xii} &  186.854 & 0.17 \\ 
                                 & &   &  &  &  &  \ionm{Fe}{xii} &  186.887 & 0.76 \\ 
 
 193.51 &    40100 & 885.9 &  6.19 &  6.20 &  1.40 &  \ionm{Fe}{xii} &  193.509 & 0.95 \\ 
 
 192.39 &    18200 & 550.9 &  6.19 &  6.20 &  1.05 &  \ionm{Fe}{xii} &  192.394 & 0.96 \\ 
 
 195.12 &    63200 & 1284.7 &  6.19 &  6.20 &  1.40 &  \ionm{Fe}{xii} &  195.119 & 0.97 \\ 
 
 196.52 &     1760 & 43.6 &  6.24 &  6.22 &  1.04 &  \ionm{Fe}{xiii} &  196.525 & 0.93 \\ 
 
 200.02 &     3390 & 140.2 &  6.25 &  6.22 &  0.85 &  \ionm{Fe}{xiii} &  200.021 & 0.95 \\ 
 
 261.73 &      241 & 42.2 &  6.25 &  6.22 &  1.02 &  \ionm{Fe}{xiii} &  261.743 & 0.96 \\ 
 
 203.83 &     3600 & 440.5 &  6.25 &  6.22 &  1.21 &  \ionm{Fe}{xiii} &  203.795 & 0.28 \\ 
                                 & &   &  &  &  &  \ionm{Fe}{xiii} &  203.826 & 0.64 \\ 
 
 251.95 &      945 & 354.6 &  6.25 &  6.22 &  1.02 &  \ionm{Fe}{xiii} &  251.952 & 0.97 \\ 
 
 204.94 &      706 & 121.4 &  6.25 &  6.22 &  0.77 &  \ionm{Fe}{xiii} &  204.942 & 0.96 \\ 
 
 209.92 &      691 & 272.0 &  6.25 &  6.22 &  0.81 &  \ionm{Fe}{xiii} &  209.916 & 0.96 \\ 
 
 202.04 &    18000 & 1466.4 &  6.25 &  6.22 &  0.86 &  \ionm{Fe}{xiii} &  202.044 & 0.96 \\ 
 
 264.79 &     2210 & 353.0 &  6.29 &  6.23 &  0.92 &  \ionm{Fe}{xiv} &  264.789 & 0.85 \\ 
                                 & &   &  &  &  &  \ionm{Fe}{xi} &  264.772 & 0.11 \\ 
 
 274.20 &     3270 & 427.2 &  6.29 &  6.24 &  0.95 &  \ionm{Fe}{xiv} &  274.204 & 0.96 \\ 
 
 252.20 &      186 & 68.2 &  6.29 &  6.24 &  1.03 &  \ionm{Fe}{xiv} &  252.200 & 0.95 \\ 

 270.52 &     1460 & 204.2 &  6.29 &  6.24 &  0.89 &  \ionm{Fe}{xiv} &  270.521 & 0.96 \\ 
 
 211.32 &     1670 & 986.8 &  6.29 &  6.24 &  0.74 &  \ionm{Fe}{xiv} &  211.317 & 0.96 \\ 
 
 284.16 &     3350 & 1200.6 &  6.34 &  6.27 &  0.80 &  \ionm{Fe}{xv} &  284.163 & 0.96 \\ 
 
 262.98 &      143 & 23.4 &  6.43 &  6.31 &  0.92 &  \ionm{Fe}{xvi} &  262.976 & 0.63 \\ 
                                 & &   &  &  &  &  \ionm{Fe}{xiii} &  262.984 & 0.32 \\ 
\noalign{\smallskip}\hline 
\end{tabular}
\normalsize
 \label{tab:ol3} 
\end{table}

\begin{table}
\caption{Same as Table~\ref{tab:lines_11_mar_2007}, but for a 
QS off-limb observation on 7 Aug 2014}
\centering
  \scriptsize
   \setlength\tabcolsep{3.pt}
\begin{tabular}{@{}rrrccrlrr@{}}
\hline\hline \noalign{\smallskip}
 $\lambda_{\rm obs}$  & DN & $I_{\rm obs}$   & log $T_{\rm max}$ & log $T_{\rm eff}$  & $R$ & Ion & $\lambda_{\rm exp}$   &  $r$ \\
\noalign{\smallskip}\hline\noalign{\smallskip}
 194.66 &      272 & 13.1 &  5.72 &  6.05 &  1.09 &  \ionm{Fe}{viii} &  194.661 & 0.96 \\ 
 
 185.22 &      265 & 70.9 &  5.73 &  6.06 &  1.01 &  \ionm{Fe}{viii} &  185.213 & 0.93 \\ 
 
 186.60 &      315 & 60.7 &  5.73 &  6.06 &  0.91 &  \ionm{Fe}{viii} &  186.598 & 0.84 \\ 
 
 189.94 &      578 & 60.3 &  5.94 &  6.08 &  0.84 &  \ionm{Fe}{ix} &  189.935 & 0.96 \\ 
 
 188.49 &      678 & 91.4 &  5.95 &  6.08 &  1.11 &  \ionm{Fe}{ix} &  188.493 & 0.87 \\ 
 
 197.85 &     1060 & 64.9 &  5.96 &  6.08 &  0.78 &  \ionm{Fe}{ix} &  197.854 & 0.95 \\ 
 
 257.26 &      928 & 470.9 &  6.04 &  6.09 &  0.72 &  \ionm{Fe}{x} &  257.259 & 0.24 \\ 
                                 & &   &  &  &  &  \ionm{Fe}{x} &  257.263 & 0.74 \\ 
  
 177.24 &      229 & 1152.0 &  6.05 &  6.10 &  0.92 &  \ionm{Fe}{x} &  177.240 & 0.98 \\ 
 
 174.53 &      117 & 1788.6 &  6.05 &  6.10 &  1.04 &  \ionm{Fe}{x} &  174.531 & 0.98 \\ 
 
 184.54 &     1470 & 498.8 &  6.05 &  6.10 &  0.79 &  \ionm{Fe}{x} &  184.537 & 0.97 \\ 
 
 190.04 &     1870 & 191.9 &  6.06 &  6.10 &  0.81 &  \ionm{Fe}{x} &  190.037 & 0.80 \\ 
  
 256.92 &      303 & 157.6 &  6.12 &  6.11 &  0.72 &  \ionm{Fe}{xi} &  256.919 & 0.93 \\ 
 
 181.13 &       69 & 77.8 &  6.13 &  6.11 &  0.99 &  \ionm{Fe}{xi} &  181.130 & 0.96 \\ 
 
 182.17 &      317 & 247.8 &  6.13 &  6.11 &  0.90 &  \ionm{Fe}{xi} &  182.167 & 0.97 \\ 
 
 192.81 &     2670 & 159.7 &  5.42 &  6.11 &  0.98 &  \ionm{Fe}{xi} &  192.813 & 0.96 \\

 188.22 &     5080 & 718.5 &  6.13 &  6.11 &  1.03 &  \ionm{Fe}{xi} &  188.216 & 0.98 \\ 
 
 188.30 &     3560 & 495.2 &  6.13 &  6.11 &  0.92 &  \ionm{Fe}{xi} &  188.299 & 0.98 \\ 
 
 180.40 &     1020 & 1484.1 &  6.13 &  6.11 &  1.04 &  \ionm{Fe}{xi} &  180.401 & 0.98 \\ 
 
 202.71 &      275 & 62.2 &  6.13 &  6.11 &  1.01 &  \ionm{Fe}{xi} &  202.705 & 0.95 \\ 
  
 258.37 &      596 & 279.7 &  6.15 &  6.12 &  0.91 &  \ionm{Si}{x} &  258.374 & 0.97 \\ 
 
 253.79 &       75 & 50.0 &  6.15 &  6.12 &  1.01 &  \ionm{Si}{x} &  253.790 & 0.97 \\ 
 
 256.40 &      511 & 275.1 &  6.16 &  6.12 &  0.89 &  \ionm{Si}{x} &  256.377 & 0.62 \\ 
                                 & &   &  &  &  &  \ionm{Fe}{xii} &  256.410 & 0.15 \\ 
                                 & &   &  &  &  &  \ionm{Fe}{x} &  256.398 & 0.14 \\ 
 
 271.99 &      451 & 117.8 &  6.15 &  6.12 &  0.90 &  \ionm{Si}{x} &  271.992 & 0.98 \\ 
 
 261.06 &      361 & 141.3 &  6.15 &  6.12 &  0.90 &  \ionm{Si}{x} &  261.056 & 0.98 \\ 
 
 277.27 &      240 & 82.2 &  6.15 &  6.12 &  0.99 &  \ionm{Si}{x} &  277.264 & 0.98 \\ 
 
 249.39 &       47 & 49.2 &  6.19 &  6.12 &  0.88 &  \ionm{Fe}{xii} &  249.388 & 0.68 \\ 
                                 & &   &  &  &  &  \ionm{Fe}{xii} &  249.384 & 0.21 \\ 
 
 196.64 &      836 & 45.2 &  6.19 &  6.13 &  1.08 &  \ionm{Fe}{xii} &  196.640 & 0.93 \\ 
 
 186.88 &      872 & 158.9 &  6.19 &  6.13 &  1.12 &  \ionm{Fe}{xii} &  186.854 & 0.17 \\ 
                                 & &   &  &  &  &  \ionm{Fe}{xii} &  186.887 & 0.78 \\ 
 
 193.51 &     7550 & 387.7 &  6.19 &  6.13 &  1.34 &  \ionm{Fe}{xii} &  193.509 & 0.94 \\ 
 
 192.39 &     3240 & 209.5 &  6.19 &  6.13 &  1.13 &  \ionm{Fe}{xii} &  192.394 & 0.96 \\ 
 
 195.12 &    12100 & 528.0 &  6.19 &  6.13 &  1.40 &  \ionm{Fe}{xii} &  195.119 & 0.97 \\ 
 
 264.79 &      187 & 63.9 &  6.29 &  6.14 &  0.90 &  \ionm{Fe}{xiv} &  264.789 & 0.44 \\ 
                                 & &   &  &  &  &  \ionm{Fe}{xi} &  264.772 & 0.53 \\

 200.02 &      394 & 34.6 &  6.25 &  6.14 &  0.73 &  \ionm{Fe}{xiii} &  200.021 & 0.91 \\ 
 
 203.83 &      255 & 78.7 &  6.25 &  6.14 &  1.37 &  \ionm{Fe}{xiii} &  203.795 & 0.28 \\ 
                                 & &   &  &  &  &  \ionm{Fe}{xiii} &  203.826 & 0.64 \\ 
 
 251.95 &       90 & 70.6 &  6.25 &  6.14 &  1.03 &  \ionm{Fe}{xiii} &  251.952 & 0.96 \\ 
 
 209.92 &       42 & 35.3 &  6.25 &  6.14 &  1.22 &  \ionm{Fe}{xiii} &  209.916 & 0.94 \\ 
 
 204.94 &       48 & 18.5 &  6.25 &  6.14 &  1.00 &  \ionm{Fe}{xiii} &  204.942 & 0.95 \\ 
 
 202.04 &     1620 & 286.4 &  6.25 &  6.15 &  0.85 &  \ionm{Fe}{xiii} &  202.044 & 0.96 \\ 
  
 274.20 &      135 & 42.3 &  6.29 &  6.15 &  0.96 &  \ionm{Fe}{xiv} &  274.204 & 0.87 \\ 
 
 211.32 &       60 & 77.2 &  6.29 &  6.16 &  0.86 &  \ionm{Fe}{xiv} &  211.317 & 0.94 \\

 270.52 &       65 & 19.6 &  6.29 &  6.16 &  0.84 &  \ionm{Fe}{xiv} &  270.521 & 0.96 \\ 
 
 284.16 &       45 & 34.9 &  6.34 &  6.19 &  1.14 &  \ionm{Fe}{xv} &  284.163 & 0.96 \\ 

\noalign{\smallskip}\hline 
\end{tabular}
\normalsize
 \label{tab:ol4} 
\end{table}

\begin{table}
\caption{Same as Table~\ref{tab:lines_11_mar_2007}, but for a 
QS off-limb observation on 6 Aug 2016}
\centering
  \scriptsize
   \setlength\tabcolsep{3.pt}
\begin{tabular}{@{}rrrccrlrr@{}}
\hline\hline \noalign{\smallskip}
 $\lambda_{\rm obs}$  & DN & $I_{\rm obs}$   & log $T_{\rm max}$ & log $T_{\rm eff}$  & $R$ & Ion & $\lambda_{\rm exp}$   &  $r$ \\
\noalign{\smallskip}\hline\noalign{\smallskip}
 194.66 &      116 & 6.4 &  5.72 &  6.04 &  0.90 &  \ionm{Fe}{viii} &  194.661 & 0.91 \\ 
 
 185.22 &      110 & 25.5 &  5.73 &  6.06 &  1.17 &  \ionm{Fe}{viii} &  185.213 & 0.83 \\ 
 
 186.60 &      147 & 24.4 &  5.73 &  6.07 &  0.91 &  \ionm{Fe}{viii} &  186.598 & 0.77 \\ 
 
 189.94 &      185 & 19.0 &  5.94 &  6.13 &  1.03 &  \ionm{Fe}{ix} &  189.935 & 0.94 \\ 
 
 188.49 &      401 & 47.3 &  5.95 &  6.14 &  0.88 &  \ionm{Fe}{ix} &  188.493 & 0.80 \\ 
 
 197.85 &      369 & 19.1 &  5.96 &  6.14 &  1.09 &  \ionm{Fe}{ix} &  197.854 & 0.89 \\ 
 
 257.26 &      510 & 224.5 &  6.04 &  6.17 &  0.82 &  \ionm{Fe}{x} &  257.259 & 0.23 \\ 
                                 & &   &  &  &  &  \ionm{Fe}{x} &  257.263 & 0.74 \\

 174.53 &       92 & 1225.3 &  6.05 &  6.17 &  0.91 &  \ionm{Fe}{x} &  174.531 & 0.97 \\ 
 
 177.24 &      166 & 718.4 &  6.05 &  6.17 &  0.87 &  \ionm{Fe}{x} &  177.240 & 0.97 \\ 
 
 184.54 &      956 & 304.6 &  6.05 &  6.17 &  0.79 &  \ionm{Fe}{x} &  184.537 & 0.93 \\ 
 
 190.04 &     1240 & 125.0 &  6.06 &  6.18 &  0.99 &  \ionm{Fe}{xii} &  190.040 & 0.25 \\ 
 
 184.79 &       96 & 27.2 &  6.12 &  6.19 &  0.95 &  \ionm{Fe}{xi} &  184.793 & 0.95 \\ 
 
 257.55 &      228 & 98.7 &  6.12 &  6.19 &  0.98 &  \ionm{Fe}{xi} &  257.554 & 0.68 \\ 
                                 & &   &  &  &  &  \ionm{Fe}{xi} &  257.547 & 0.26 \\ 
 
 189.02 &      199 & 22.3 &  6.13 &  6.19 &  0.94 &  \ionm{Fe}{xi} &  188.997 & 0.95 \\ 
 
 256.92 &      324 & 146.4 &  6.12 &  6.19 &  0.80 &  \ionm{Fe}{xi} &  256.919 & 0.89 \\ 
 
 182.17 &      383 & 258.9 &  6.13 &  6.19 &  0.93 &  \ionm{Fe}{xi} &  182.167 & 0.96 \\ 
 
 192.81 &     3000 & 185.7 &  5.42 &  6.19 &  0.93 &  \ionm{Fe}{xi} &  192.813 & 0.96 \\ 
 
 181.13 &       79 & 77.2 &  6.13 &  6.19 &  1.08 &  \ionm{Fe}{xi} &  181.130 & 0.97 \\ 
 
 188.30 &     4060 & 489.6 &  6.13 &  6.19 &  1.01 &  \ionm{Fe}{xi} &  188.299 & 0.97 \\ 
 
 188.22 &     6520 & 796.2 &  6.13 &  6.19 &  1.02 &  \ionm{Fe}{xi} &  188.216 & 0.97 \\ 
 
 180.40 &     1320 & 1650.5 &  6.13 &  6.19 &  1.03 &  \ionm{Fe}{xi} &  180.401 & 0.97 \\ 
  
 202.71 &      271 & 54.2 &  6.13 &  6.19 &  1.33 &  \ionm{Fe}{xi} &  202.705 & 0.90 \\ 
 
 258.37 &      906 & 370.1 &  6.15 &  6.20 &  0.91 &  \ionm{Si}{x} &  258.374 & 0.97 \\ 
 
 253.79 &      116 & 66.7 &  6.15 &  6.20 &  0.99 &  \ionm{Si}{x} &  253.790 & 0.97 \\ 
 
 271.99 &      646 & 158.1 &  6.15 &  6.20 &  0.89 &  \ionm{Si}{x} &  271.992 & 0.97 \\ 
 
 261.06 &      566 & 192.1 &  6.15 &  6.20 &  0.88 &  \ionm{Si}{x} &  261.056 & 0.98 \\ 
 
 277.27 &      292 & 109.8 &  6.15 &  6.20 &  0.98 &  \ionm{Si}{x} &  277.264 & 0.98 \\

 249.39 &      109 & 97.8 &  6.19 &  6.20 &  0.84 &  \ionm{Fe}{xii} &  249.388 & 0.72 \\ 
                                 & &   &  &  &  &  \ionm{Fe}{xii} &  249.384 & 0.23 \\

 196.64 &     1680 & 87.3 &  6.19 &  6.21 &  1.15 &  \ionm{Fe}{xii} &  196.640 & 0.94 \\

 186.88 &     2280 & 360.3 &  6.19 &  6.21 &  1.05 &  \ionm{Fe}{xii} &  186.854 & 0.17 \\ 
                                 & &   &  &  &  &  \ionm{Fe}{xii} &  186.887 & 0.76 \\ 
 
 193.51 &    17800 & 951.5 &  6.19 &  6.21 &  1.18 &  \ionm{Fe}{xii} &  193.509 & 0.95 \\ 
 
 195.12 &    27900 & 1319.5 &  6.19 &  6.21 &  1.24 &  \ionm{Fe}{xii} &  195.119 & 0.97 \\ 
 
 192.39 &     7770 & 486.5 &  6.19 &  6.21 &  1.08 &  \ionm{Fe}{xii} &  192.394 & 0.96 \\ 
 
 196.52 &      753 & 39.0 &  6.24 &  6.22 &  1.07 &  \ionm{Fe}{xiii} &  196.525 & 0.93 \\ 
 
 200.02 &     1440 & 110.2 &  6.25 &  6.22 &  0.99 &  \ionm{Fe}{xiii} &  200.021 & 0.95 \\ 
 
 261.73 &      108 & 35.1 &  6.25 &  6.22 &  1.12 &  \ionm{Fe}{xiii} &  261.743 & 0.97 \\ 
 
 203.83 &     1580 & 444.0 &  6.25 &  6.22 &  1.10 &  \ionm{Fe}{xiii} &  203.795 & 0.28 \\ 
                                 & &   &  &  &  &  \ionm{Fe}{xiii} &  203.826 & 0.64 \\

 251.95 &      513 & 348.0 &  6.25 &  6.22 &  0.96 &  \ionm{Fe}{xiii} &  251.952 & 0.97 \\ 
 
 204.94 &      266 & 94.3 &  6.25 &  6.22 &  0.91 &  \ionm{Fe}{xiii} &  204.942 & 0.96 \\ 
 
 209.92 &      247 & 193.3 &  6.25 &  6.22 &  1.04 &  \ionm{Fe}{xiii} &  209.916 & 0.96 \\ 
 
 202.04 &     7390 & 1136.1 &  6.25 &  6.22 &  1.02 &  \ionm{Fe}{xiii} &  202.044 & 0.96 \\ 
 
 264.79 &     1070 & 318.3 &  6.29 &  6.23 &  0.91 &  \ionm{Fe}{xiv} &  264.789 & 0.85 \\ 
                                 & &   &  &  &  &  \ionm{Fe}{xi} &  264.772 & 0.11 \\ 
 
 274.20 &     1370 & 373.0 &  6.29 &  6.23 &  0.98 &  \ionm{Fe}{xiv} &  274.204 & 0.96 \\ 
 
 252.20 &       97 & 64.5 &  6.29 &  6.23 &  0.98 &  \ionm{Fe}{xiv} &  252.200 & 0.95 \\

 270.52 &      680 & 180.7 &  6.29 &  6.24 &  0.91 &  \ionm{Fe}{xiv} &  270.521 & 0.96 \\ 
 
 211.32 &      659 & 654.5 &  6.29 &  6.24 &  1.00 &  \ionm{Fe}{xiv} &  211.317 & 0.96 \\ 
 
 284.16 &     1090 & 916.4 &  6.34 &  6.25 &  0.89 &  \ionm{Fe}{xv} &  284.163 & 0.96 \\ 
 
\noalign{\smallskip}\hline 
\end{tabular}
\normalsize
 \label{tab:ol5} 
\end{table}

\begin{table}
\caption{Same as Table~\ref{tab:lines_11_mar_2007}, but for a 
QS off-limb observation on 28  July 2017}
\centering
  \scriptsize
   \setlength\tabcolsep{3.pt}
\begin{tabular}{@{}rrrccrlrr@{}}
\hline\hline \noalign{\smallskip}
 $\lambda_{\rm obs}$  & DN & $I_{\rm obs}$   & log $T_{\rm max}$ & log $T_{\rm eff}$  & $R$ & Ion & $\lambda_{\rm exp}$   &  $r$ \\
\noalign{\smallskip}\hline\noalign{\smallskip}

 194.66 &      278 & 17.5 &  5.72 &  6.05 &  0.86 &  \ionm{Fe}{viii} &  194.661 & 0.95 \\ 
 
 185.22 &      233 & 59.9 &  5.73 &  6.05 &  1.21 &  \ionm{Fe}{viii} &  185.213 & 0.96 \\ 
 
 186.60 &      302 & 55.5 &  5.73 &  6.06 &  1.02 &  \ionm{Fe}{viii} &  186.598 & 0.86 \\

 189.94 &      571 & 62.8 &  5.94 &  6.09 &  0.84 &  \ionm{Fe}{ix} &  189.935 & 0.96 \\ 
 
 188.49 &      850 & 111.0 &  5.95 &  6.09 &  0.95 &  \ionm{Fe}{ix} &  188.493 & 0.89 \\ 
 
 197.85 &      933 & 56.6 &  5.96 &  6.09 &  0.95 &  \ionm{Fe}{ix} &  197.854 & 0.94 \\ 
 
 257.26 &      893 & 425.8 &  6.04 &  6.11 &  0.90 &  \ionm{Fe}{x} &  257.259 & 0.23 \\ 
                                 & &   &  &  &  &  \ionm{Fe}{x} &  257.263 & 0.75 \\ 
 
 177.24 &      241 & 1246.1 &  6.05 &  6.11 &  0.96 &  \ionm{Fe}{x} &  177.240 & 0.98 \\ 
 
 174.53 &      128 & 2006.8 &  6.05 &  6.11 &  1.06 &  \ionm{Fe}{x} &  174.531 & 0.98 \\ 
 
 184.54 &     1570 & 550.6 &  6.05 &  6.11 &  0.81 &  \ionm{Fe}{x} &  184.537 & 0.97 \\ 
 
 190.04 &     2100 & 227.9 &  6.06 &  6.12 &  0.81 &  \ionm{Fe}{xii} &  190.040 & 0.12 \\ 
                                 & &   &  &  &  &  \ionm{Fe}{x} &  190.037 & 0.77 \\ 
 
 192.02 &      518 & 41.4 &  5.74 &  6.12 &  0.91 &  \ionm{Fe}{xi} &  192.021 & 0.84 \\ 
 
 184.79 &      110 & 34.1 &  6.11 &  6.12 &  1.04 &  \ionm{Fe}{xi} &  184.793 & 0.80 \\ 
                                 & &   &  &  &  &  \ionm{Fe}{x} &  184.828 & 0.14 \\ 
 
 257.55 &      254 & 118.4 &  6.12 &  6.13 &  1.02 &  \ionm{Fe}{xi} &  257.554 & 0.70 \\ 
                                 & &   &  &  &  &  \ionm{Fe}{xi} &  257.547 & 0.26 \\ 
 
 256.92 &      379 & 185.2 &  6.12 &  6.13 &  0.79 &  \ionm{Fe}{xi} &  256.919 & 0.92 \\ 
 
 182.17 &      451 & 362.8 &  6.13 &  6.13 &  0.77 &  \ionm{Fe}{xi} &  182.167 & 0.97 \\ 
 
 181.13 &       93 & 107.3 &  6.13 &  6.13 &  0.89 &  \ionm{Fe}{xi} &  181.130 & 0.97 \\ 
 
 188.22 &     6750 & 917.3 &  6.13 &  6.13 &  1.04 &  \ionm{Fe}{xi} &  188.216 & 0.98 \\ 
 
 188.30 &     4190 & 563.3 &  6.13 &  6.13 &  1.04 &  \ionm{Fe}{xi} &  188.299 & 0.98 \\ 
 
 192.81 &     3440 & 244.1 &  5.42 &  6.13 &  0.83 &  \ionm{Fe}{xi} &  192.813 & 0.96 \\

 180.40 &     1310 & 1948.3 &  6.13 &  6.13 &  1.02 &  \ionm{Fe}{xi} &  180.401 & 0.98 \\ 
 
 202.71 &      340 & 76.6 &  6.13 &  6.13 &  1.06 &  \ionm{Fe}{xi} &  202.705 & 0.95 \\

 258.37 &      827 & 364.7 &  6.15 &  6.13 &  0.90 &  \ionm{Si}{x} &  258.374 & 0.97 \\ 
 
 253.79 &      118 & 73.1 &  6.15 &  6.13 &  0.89 &  \ionm{Si}{x} &  253.790 & 0.97 \\ 
 
 271.99 &      556 & 160.0 &  6.15 &  6.13 &  0.90 &  \ionm{Si}{x} &  271.992 & 0.97 \\ 
 
 261.06 &      530 & 194.3 &  6.15 &  6.13 &  0.89 &  \ionm{Si}{x} &  261.056 & 0.98 \\ 
 
 277.27 &      288 & 124.5 &  6.15 &  6.13 &  0.88 &  \ionm{Si}{x} &  277.264 & 0.98 \\ 
 
 256.40 &      697 & 353.8 &  6.16 &  6.13 &  0.94 &  \ionm{Si}{x} &  256.377 & 0.62 \\ 
                                 & &   &  &  &  &  \ionm{Fe}{xii} &  256.410 & 0.16 \\ 
                                 & &   &  &  &  &  \ionm{Fe}{x} &  256.398 & 0.11 \\ 
 
 249.39 &       47 & 46.2 &  6.19 &  6.14 &  1.34 &  \ionm{Fe}{xii} &  249.388 & 0.69 \\ 
                                 & &   &  &  &  &  \ionm{Fe}{xii} &  249.384 & 0.22 \\ 
  
 196.64 &     1220 & 68.3 &  6.19 &  6.14 &  1.00 &  \ionm{Fe}{xii} &  196.640 & 0.93 \\ 
 
 186.88 &     1480 & 260.6 &  6.19 &  6.15 &  0.95 &  \ionm{Fe}{xii} &  186.854 & 0.17 \\ 
                                 & &   &  &  &  &  \ionm{Fe}{xii} &  186.887 & 0.78 \\

 193.51 &    11600 & 707.2 &  6.19 &  6.15 &  1.12 &  \ionm{Fe}{xii} &  193.509 & 0.94 \\ 
 
 192.39 &     5210 & 393.8 &  6.19 &  6.15 &  0.92 &  \ionm{Fe}{xii} &  192.394 & 0.96 \\ 
 
 195.12 &    17700 & 899.4 &  6.19 &  6.15 &  1.26 &  \ionm{Fe}{xii} &  195.119 & 0.97 \\ 
 
 200.02 &      683 & 56.4 &  6.25 &  6.16 &  0.77 &  \ionm{Fe}{xiii} &  200.021 & 0.89 \\ 
 
 196.52 &      379 & 20.9 &  6.24 &  6.16 &  0.81 &  \ionm{Fe}{xiii} &  196.525 & 0.88 \\ 
 
 264.79 &      338 & 108.4 &  6.29 &  6.16 &  0.90 &  \ionm{Fe}{xiv} &  264.789 & 0.57 \\ 
                                 & &   &  &  &  &  \ionm{Fe}{xi} &  264.772 & 0.40 \\

 203.83 &      527 & 168.7 &  6.25 &  6.16 &  1.07 &  \ionm{Fe}{xiii} &  203.795 & 0.28 \\ 
                                 & &   &  &  &  &  \ionm{Fe}{xiii} &  203.826 & 0.64 \\ 
 
 251.95 &      192 & 140.7 &  6.25 &  6.16 &  0.94 &  \ionm{Fe}{xiii} &  251.952 & 0.97 \\ 
 
 209.92 &       92 & 78.2 &  6.25 &  6.16 &  1.07 &  \ionm{Fe}{xiii} &  209.916 & 0.95 \\ 
 
 204.94 &      101 & 44.0 &  6.25 &  6.16 &  0.78 &  \ionm{Fe}{xiii} &  204.942 & 0.95 \\ 
 
 202.04 &     3060 & 558.8 &  6.25 &  6.16 &  0.85 &  \ionm{Fe}{xiii} &  202.044 & 0.97 \\ 
  
 274.20 &      293 & 91.0 &  6.29 &  6.18 &  0.95 &  \ionm{Fe}{xiv} &  274.204 & 0.92 \\

 211.32 &      148 & 159.7 &  6.29 &  6.18 &  0.93 &  \ionm{Fe}{xiv} &  211.317 & 0.95 \\ 
 
 257.40 &       65 & 30.6 &  6.29 &  6.18 &  0.87 &  \ionm{Fe}{xiv} &  257.394 & 0.96 \\ 
 
 270.52 &       99 & 30.4 &  6.29 &  6.18 &  1.22 &  \ionm{Fe}{xiv} &  270.521 & 0.97 \\ 
 
 284.16 &      123 & 119.6 &  6.34 &  6.20 &  0.89 &  \ionm{Fe}{xv} &  284.163 & 0.96 \\
\noalign{\smallskip}\hline 
\end{tabular}
\normalsize
 \label{tab:ol6} 
\end{table}

\begin{table}
\caption{Same as Table~\ref{tab:lines_11_mar_2007}, but for a 
QS off-limb observation on 28  June 2019}
\centering
  \scriptsize
   \setlength\tabcolsep{3.pt}
\begin{tabular}{@{}rrrccrlrr@{}}
\hline\hline \noalign{\smallskip}
 $\lambda_{\rm obs}$  & DN & $I_{\rm obs}$   & log $T_{\rm max}$ & log $T_{\rm eff}$  & $R$ & Ion & $\lambda_{\rm exp}$   &  $r$ \\
\noalign{\smallskip}\hline\noalign{\smallskip}
 194.66 &      417 & 24.4 &  5.72 &  6.07 &  0.80 &  \ionm{Fe}{viii} &  194.661 & 0.95 \\ 
 
 185.22 &      412 & 99.3 &  5.73 &  6.07 &  0.94 &  \ionm{Fe}{viii} &  185.213 & 0.95 \\ 
 
 186.60 &      464 & 74.8 &  5.73 &  6.07 &  0.99 &  \ionm{Fe}{viii} &  186.598 & 0.83 \\ 
  
 189.94 &      785 & 67.3 &  5.94 &  6.09 &  1.11 &  \ionm{Fe}{ix} &  189.935 & 0.97 \\ 
 
 188.49 &     1290 & 170.0 &  5.95 &  6.09 &  0.86 &  \ionm{Fe}{ix} &  188.493 & 0.89 \\ 
 
 197.85 &     1320 & 78.0 &  5.96 &  6.09 &  0.96 &  \ionm{Fe}{ix} &  197.854 & 0.94 \\ 
 
 257.26 &     1510 & 587.6 &  6.04 &  6.11 &  0.90 &  \ionm{Fe}{x} &  257.259 & 0.25 \\ 
                                 & &   &  &  &  &  \ionm{Fe}{x} &  257.263 & 0.73 \\ 
 
 177.24 &      346 & 1672.1 &  6.05 &  6.11 &  1.02 &  \ionm{Fe}{x} &  177.240 & 0.98 \\ 
 
 174.53 &      177 & 2598.2 &  6.05 &  6.11 &  1.16 &  \ionm{Fe}{x} &  174.531 & 0.98 \\ 
 
 184.54 &     2280 & 626.5 &  6.05 &  6.11 &  1.02 &  \ionm{Fe}{x} &  184.537 & 0.96 \\ 
 
 190.04 &     3050 & 256.1 &  6.06 &  6.12 &  1.05 &  \ionm{Fe}{xii} &  190.040 & 0.13 \\ 
                                 & &   &  &  &  &  \ionm{Fe}{x} &  190.037 & 0.75 \\ 
 
 184.79 &      164 & 42.8 &  6.12 &  6.13 &  1.22 &  \ionm{Fe}{xi} &  184.793 & 0.94 \\ 
 
 257.55 &      441 & 169.8 &  6.12 &  6.14 &  1.07 &  \ionm{Fe}{xi} &  257.554 & 0.70 \\ 
                                 & &   &  &  &  &  \ionm{Fe}{xi} &  257.547 & 0.26 \\ 
 
 256.92 &      608 & 241.2 &  6.12 &  6.14 &  0.87 &  \ionm{Fe}{xi} &  256.919 & 0.92 \\ 
 
 189.02 &      427 & 48.1 &  6.13 &  6.14 &  0.86 &  \ionm{Fe}{xi} &  188.997 & 0.96 \\ 
 
 182.17 &      710 & 534.9 &  6.13 &  6.14 &  0.81 &  \ionm{Fe}{xi} &  182.167 & 0.97 \\ 
 
 181.13 &      150 & 163.6 &  6.13 &  6.14 &  0.92 &  \ionm{Fe}{xi} &  181.130 & 0.97 \\ 
 
 188.22 &     9190 & 1271.2 &  6.13 &  6.14 &  1.11 &  \ionm{Fe}{xi} &  188.216 & 0.97 \\ 
 
 188.30 &     6090 & 834.3 &  6.13 &  6.14 &  1.03 &  \ionm{Fe}{xi} &  188.299 & 0.97 \\ 
 
 192.81 &     5040 & 329.8 &  5.42 &  6.14 &  0.91 &  \ionm{Fe}{xi} &  192.813 & 0.96 \\

 180.40 &     1820 & 2525.3 &  6.13 &  6.14 &  1.16 &  \ionm{Fe}{xi} &  180.401 & 0.97 \\ 
 
 202.71 &      424 & 87.9 &  6.13 &  6.14 &  1.38 &  \ionm{Fe}{xi} &  202.705 & 0.93 \\ 
 
 258.37 &     1460 & 545.1 &  6.15 &  6.15 &  0.98 &  \ionm{Si}{x} &  258.374 & 0.97 \\ 
 
 253.79 &      223 & 115.0 &  6.15 &  6.15 &  0.92 &  \ionm{Si}{x} &  253.790 & 0.97 \\ 
 
 271.99 &      852 & 226.4 &  6.15 &  6.15 &  0.95 &  \ionm{Si}{x} &  271.992 & 0.97 \\ 
 
 261.06 &      853 & 274.5 &  6.15 &  6.15 &  0.94 &  \ionm{Si}{x} &  261.056 & 0.98 \\ 
 
 277.27 &      452 & 181.3 &  6.15 &  6.15 &  0.91 &  \ionm{Si}{x} &  277.264 & 0.98 \\

 256.40 &     1050 & 430.4 &  6.16 &  6.15 &  1.20 &  \ionm{Si}{x} &  256.377 & 0.60 \\ 
                                 & &   &  &  &  &  \ionm{Fe}{xii} &  256.410 & 0.17 \\ 
                                 & &   &  &  &  &  \ionm{Fe}{x} &  256.398 & 0.11 \\ 
 
 249.39 &      107 & 92.1 &  6.19 &  6.16 &  1.12 &  \ionm{Fe}{xii} &  249.388 & 0.70 \\ 
                                 & &   &  &  &  &  \ionm{Fe}{xii} &  249.384 & 0.22 \\

 196.64 &     2360 & 129.1 &  6.19 &  6.16 &  0.97 &  \ionm{Fe}{xii} &  196.640 & 0.94 \\ 
 
 186.88 &     3170 & 480.7 &  6.19 &  6.16 &  0.97 &  \ionm{Fe}{xii} &  186.854 & 0.17 \\ 
                                 & &   &  &  &  &  \ionm{Fe}{xii} &  186.887 & 0.77 \\

 193.51 &    18200 & 1032.8 &  6.19 &  6.17 &  1.25 &  \ionm{Fe}{xii} &  193.509 & 0.94 \\ 
 
 192.39 &     8090 & 569.2 &  6.19 &  6.17 &  1.05 &  \ionm{Fe}{xii} &  192.394 & 0.96 \\ 
 
 195.12 &    27800 & 1411.7 &  6.19 &  6.17 &  1.31 &  \ionm{Fe}{xii} &  195.119 & 0.97 \\ 
 
 196.52 &      751 & 40.7 &  6.24 &  6.19 &  0.94 &  \ionm{Fe}{xiii} &  196.525 & 0.91 \\ 
 
 200.02 &     1510 & 109.3 &  6.25 &  6.19 &  0.90 &  \ionm{Fe}{xiii} &  200.021 & 0.94 \\ 
 
 261.73 &       92 & 28.4 &  6.25 &  6.19 &  1.13 &  \ionm{Fe}{xiii} &  261.743 & 0.97 \\

 203.83 &     1600 & 417.7 &  6.25 &  6.19 &  1.04 &  \ionm{Fe}{xiii} &  203.795 & 0.28 \\ 
                                 & &   &  &  &  &  \ionm{Fe}{xiii} &  203.826 & 0.65 \\ 
 
 251.95 &      421 & 269.2 &  6.25 &  6.19 &  1.01 &  \ionm{Fe}{xiii} &  251.952 & 0.96 \\ 
 
 204.94 &      180 & 59.0 &  6.25 &  6.19 &  1.18 &  \ionm{Fe}{xiii} &  204.942 & 0.96 \\ 
 
 209.92 &      135 & 149.7 &  6.25 &  6.19 &  1.04 &  \ionm{Fe}{xiii} &  209.916 & 0.95 \\ 
 
 202.04 &     5070 & 909.0 &  6.25 &  6.19 &  0.98 &  \ionm{Fe}{xiii} &  202.044 & 0.96 \\ 
 
 264.79 &      858 & 239.6 &  6.29 &  6.19 &  0.92 &  \ionm{Fe}{xiv} &  264.789 & 0.71 \\ 
                                 & &   &  &  &  &  \ionm{Fe}{xi} &  264.772 & 0.26 \\ 
 
 274.20 &      835 & 238.7 &  6.29 &  6.21 &  0.96 &  \ionm{Fe}{xiv} &  274.204 & 0.94 \\ 
 
 252.20 &       58 & 36.4 &  6.29 &  6.21 &  1.10 &  \ionm{Fe}{xiv} &  252.200 & 0.94 \\ 
  
 257.40 &      180 & 69.9 &  6.29 &  6.21 &  1.05 &  \ionm{Fe}{xiv} &  257.394 & 0.96 \\ 
 
 211.32 &      370 & 418.2 &  6.29 &  6.21 &  0.97 &  \ionm{Fe}{xiv} &  211.317 & 0.96 \\ 
 
 270.52 &      398 & 112.4 &  6.29 &  6.21 &  0.91 &  \ionm{Fe}{xiv} &  270.521 & 0.96 \\ 
 
 284.16 &      445 & 388.2 &  6.34 &  6.23 &  1.02 &  \ionm{Fe}{xv} &  284.163 & 0.96 \\ 

\noalign{\smallskip}\hline 
\end{tabular}
\normalsize
 \label{tab:ol7} 
\end{table}

\begin{table}
\caption{Same as Table~\ref{tab:lines_11_mar_2007}, but for an off-limb observation on 5 April 2020 (active Sun)}
\centering
  \scriptsize
   \setlength\tabcolsep{3.pt}
\begin{tabular}{@{}rrrccrlrr@{}}
\hline\hline \noalign{\smallskip}
 $\lambda_{\rm obs}$  & DN & $I_{\rm obs}$   & log $T_{\rm max}$ & log $T_{\rm eff}$  & $R$ & Ion & $\lambda_{\rm exp}$   &  $r$ \\
\noalign{\smallskip}\hline\noalign{\smallskip}

 194.66 &     1220 & 71.4 &  5.72 &  5.91 &  0.96 &  \ionm{Fe}{viii} &  194.661 & 0.97 \\ 

 189.94 &     1140 & 97.4 &  5.94 &  6.06 &  1.14 &  \ionm{Fe}{ix} &  189.935 & 0.97 \\ 
 
 188.49 &     1560 & 205.9 &  5.95 &  6.06 &  1.02 &  \ionm{Fe}{ix} &  188.493 & 0.90 \\ 
 
 197.85 &     1540 & 91.2 &  5.96 &  6.08 &  1.05 &  \ionm{Fe}{ix} &  197.854 & 0.91 \\ 
 257.26 &     1400 & 546.5 &  6.04 &  6.12 &  0.88 &  \ionm{Fe}{x} &  257.259 & 0.38 \\ 
                                 & &   &  &  &  &  \ionm{Fe}{x} &  257.263 & 0.60 \\ 
 
 186.60 &     1560 & 252.8 &  5.73 &  6.13 &  1.11 &  \ionm{Ca}{xiv} &  186.610 & 0.17 \\ 
                                 & &   &  &  &  &  \ionm{Fe}{viii} &  186.598 & 0.74 \\ 
 
 174.53 &      217 & 3193.6 &  6.05 &  6.13 &  1.08 &  \ionm{Fe}{x} &  174.531 & 0.98 \\ 
 
 177.24 &      418 & 2020.2 &  6.05 &  6.13 &  0.97 &  \ionm{Fe}{x} &  177.240 & 0.97 \\ 
 
 185.22 &     2190 & 528.3 &  5.73 &  6.14 &  0.80 &  \ionm{Ni}{xvi} &  185.230 & 0.26 \\ 
                                 & &   &  &  &  &  \ionm{Fe}{viii} &  185.213 & 0.71 \\ 
 
 184.54 &     2840 & 781.1 &  6.05 &  6.14 &  1.02 &  \ionm{Fe}{x} &  184.537 & 0.92 \\ 

 190.04 &     3690 & 310.4 &  6.05 &  6.15 &  1.22 &  \ionm{Fe}{xii} &  190.040 & 0.19 \\ 
                                 & &   &  &  &  &  \ionm{Fe}{x} &  190.037 & 0.64 \\ 
 
 184.79 &      498 & 129.9 &  6.12 &  6.17 &  1.30 &  \ionm{Fe}{xi} &  184.793 & 0.95 \\ 
 
 257.55 &      669 & 257.5 &  6.12 &  6.17 &  1.09 &  \ionm{Fe}{xi} &  257.554 & 0.68 \\ 
                                 & &   &  &  &  &  \ionm{Fe}{xi} &  257.547 & 0.25 \\
                                 
 192.81 &     6680 & 437.5 &  5.42 &  6.17 &  0.95 &  \ionm{Fe}{xi} &  192.813 & 0.94 \\ 
 
 189.02 &     1250 & 141.4 &  6.13 &  6.18 &  0.97 &  \ionm{Fe}{xi} &  188.997 & 0.96 \\ 
 
 180.60 &      202 & 262.5 &  6.13 &  6.18 &  0.79 &  \ionm{Fe}{xi} &  180.594 & 0.96 \\ 
 
 181.13 &      263 & 287.3 &  6.13 &  6.18 &  1.01 &  \ionm{Fe}{xi} &  181.130 & 0.96 \\ 
 
 256.92 &      613 & 242.9 &  6.13 &  6.18 &  0.90 &  \ionm{Fe}{xii} &  256.946 & 0.11 \\ 
                                 & &   &  &  &  &  \ionm{Fe}{xi} &  256.919 & 0.86 \\ 
 
 182.17 &     1280 & 964.9 &  6.13 &  6.18 &  0.83 &  \ionm{Fe}{xi} &  182.167 & 0.96 \\ 
 
 188.22 &    13600 & 1881.0 &  6.13 &  6.18 &  1.02 &  \ionm{Fe}{xi} &  188.216 & 0.97 \\ 
 
 188.30 &     8400 & 1150.7 &  6.13 &  6.18 &  0.99 &  \ionm{Fe}{xi} &  188.299 & 0.97 \\ 
 
 180.40 &     2720 & 3787.9 &  6.13 &  6.18 &  1.02 &  \ionm{Fe}{xi} &  180.401 & 0.97 \\

 271.99 &     1250 & 330.8 &  6.15 &  6.20 &  1.00 &  \ionm{Si}{x} &  271.992 & 0.97 \\ 
 
 258.37 &     2900 & 1088.4 &  6.15 &  6.20 &  1.03 &  \ionm{Si}{x} &  258.374 & 0.97 \\ 
 
 261.06 &     1170 & 374.8 &  6.15 &  6.20 &  1.03 &  \ionm{Si}{x} &  261.056 & 0.98 \\ 
 
 253.79 &      381 & 195.6 &  6.15 &  6.20 &  1.14 &  \ionm{Si}{x} &  253.790 & 0.97 \\ 
 
 277.27 &      573 & 230.0 &  6.15 &  6.20 &  1.10 &  \ionm{Si}{x} &  277.264 & 0.98 \\ 
 
 249.39 &      246 & 211.0 &  6.19 &  6.21 &  1.08 &  \ionm{Fe}{xii} &  249.388 & 0.72 \\ 
                                 & &   &  &  &  &  \ionm{Fe}{xii} &  249.384 & 0.20 \\ 
 
 256.40 &     2490 & 1017.8 &  6.17 &  6.22 &  0.95 &  \ionm{Si}{x} &  256.377 & 0.48 \\ 
                                 & &   &  &  &  &  \ionm{Fe}{xiii} &  256.400 & 0.22 \\ 
                                 & &   &  &  &  &  \ionm{Fe}{xii} &  256.410 & 0.19 \\

 196.64 &     7500 & 411.2 &  6.19 &  6.22 &  0.92 &  \ionm{Fe}{xii} &  196.640 & 0.94 \\ 
 
 193.51 &    36700 & 2073.0 &  6.19 &  6.23 &  1.19 &  \ionm{Fe}{xii} &  193.509 & 0.95 \\ 
 
 192.39 &    15800 & 1110.8 &  6.20 &  6.23 &  1.04 &  \ionm{Fe}{xii} &  192.394 & 0.96 \\ 
 
 195.12 &    57900 & 2941.0 &  6.20 &  6.23 &  1.22 &  \ionm{Fe}{xii} &  195.119 & 0.97 \\ 
 
 186.88 &    11000 & 1664.6 &  6.19 &  6.23 &  0.93 &  \ionm{Fe}{xii} &  186.854 & 0.23 \\ 
                                 & &   &  &  &  &  \ionm{Fe}{xii} &  186.887 & 0.69 \\
                                 
 196.52 &     6670 & 360.9 &  6.24 &  6.27 &  0.80 &  \ionm{Fe}{xiii} &  196.525 & 0.95 \\ 
 
 261.73 &      327 & 100.6 &  6.24 &  6.27 &  1.08 &  \ionm{Fe}{xiii} &  261.743 & 0.97 \\ 
 
 246.21 &      632 & 692.5 &  6.25 &  6.27 &  0.84 &  \ionm{Fe}{xiii} &  246.209 & 0.94 \\ 
 
 200.02 &     7700 & 555.3 &  6.25 &  6.27 &  1.08 &  \ionm{Fe}{xiii} &  200.021 & 0.96 \\ 
 
 251.95 &     1780 & 1131.8 &  6.25 &  6.27 &  0.95 &  \ionm{Fe}{xiii} &  251.952 & 0.97 \\ 
 
 203.83 &     9910 & 2591.9 &  6.25 &  6.27 &  1.09 &  \ionm{Fe}{xiii} &  203.795 & 0.27 \\ 
                                 & &   &  &  &  &  \ionm{Fe}{xiii} &  203.826 & 0.69 \\ 
 
 204.94 &      684 & 224.1 &  6.25 &  6.27 &  1.11 &  \ionm{Fe}{xiii} &  204.942 & 0.96 \\ 
 
 209.92 &      270 & 300.5 &  6.25 &  6.27 &  1.11 &  \ionm{Fe}{xiii} &  209.916 & 0.95 \\ 
 
 202.04 &    11600 & 2080.3 &  6.25 &  6.27 &  0.91 &  \ionm{Fe}{xiii} &  202.044 & 0.96 \\ 
 
 264.79 &     7330 & 2033.9 &  6.29 &  6.31 &  0.89 &  \ionm{Fe}{xiv} &  264.789 & 0.92 \\ 
 
 274.20 &     6900 & 1978.6 &  6.29 &  6.31 &  0.93 &  \ionm{Fe}{xiv} &  274.204 & 0.96 \\ 
 
 252.20 &      771 & 476.8 &  6.29 &  6.32 &  0.88 &  \ionm{Fe}{xiv} &  252.200 & 0.96 \\

 270.52 &     3910 & 1102.0 &  6.29 &  6.32 &  0.83 &  \ionm{Fe}{xiv} &  270.521 & 0.97 \\ 
 
 211.32 &     3020 & 3410.9 &  6.30 &  6.32 &  0.99 &  \ionm{Fe}{xiv} &  211.317 & 0.97 \\ 
 
 257.40 &     1540 & 595.2 &  6.29 &  6.32 &  1.12 &  \ionm{Fe}{xiv} &  257.394 & 0.96 \\ 
 
 284.16 &    12800 & 11211.0 &  6.34 &  6.39 &  1.00 &  \ionm{Fe}{xv} &  284.163 & 0.98 \\
\noalign{\smallskip}\hline 
\end{tabular}
\normalsize
 \label{tab:ol8} 
\end{table}

\section{The 2021 June 16 QS off-limb observation}

{
We analysed the last suitable off-limb observation, taken on 
2021 June 16 with the EIS study 
``dhb\_atlas\_120m\_30", which had 120s exposures. 
We averaged a spectrum over a region and applied our latest effective 
areas. We carried out a DEM analysis adopting photospheric 
abundances. Table~\ref{tab:ol_last} shows the results for all the lines we measured 
in the spectra, except a few very weak ones. 
We assumed a constant electron density of 3 $\times$ 10$^8$ cm$^{-3}$ which 
produces good results for all the density-sensitive lines from 
\ionm{Si}{x},  \ionm{Fe}{x}, \ionm{Fe}{xi}, \ionm{Fe}{xii}, \ionm{Fe}{xiii}, \ionm{Fe}{xiv}.
All of the Sulphur and Argon lines are relatively well predicted, except the 
\ionm{S}{x} 264.23~\AA\ anomalous line which is weaker by about a factor of 2.5 
compared to expectation. 

Figure~\ref{fig:si_s_ratio} shows the ratio of the two nearby 
Si and S lines as measured in the quiet Sun on-disk datasets. The ratio
is remarkably constant over time, except after the beginning of 2021 when the 
count rates in the \ionm{S}{x} 264.23~\AA\ start to decrease considerably. 

}

 \begin{figure*}[!ht]
\centerline{\includegraphics[width=0.9\linewidth]{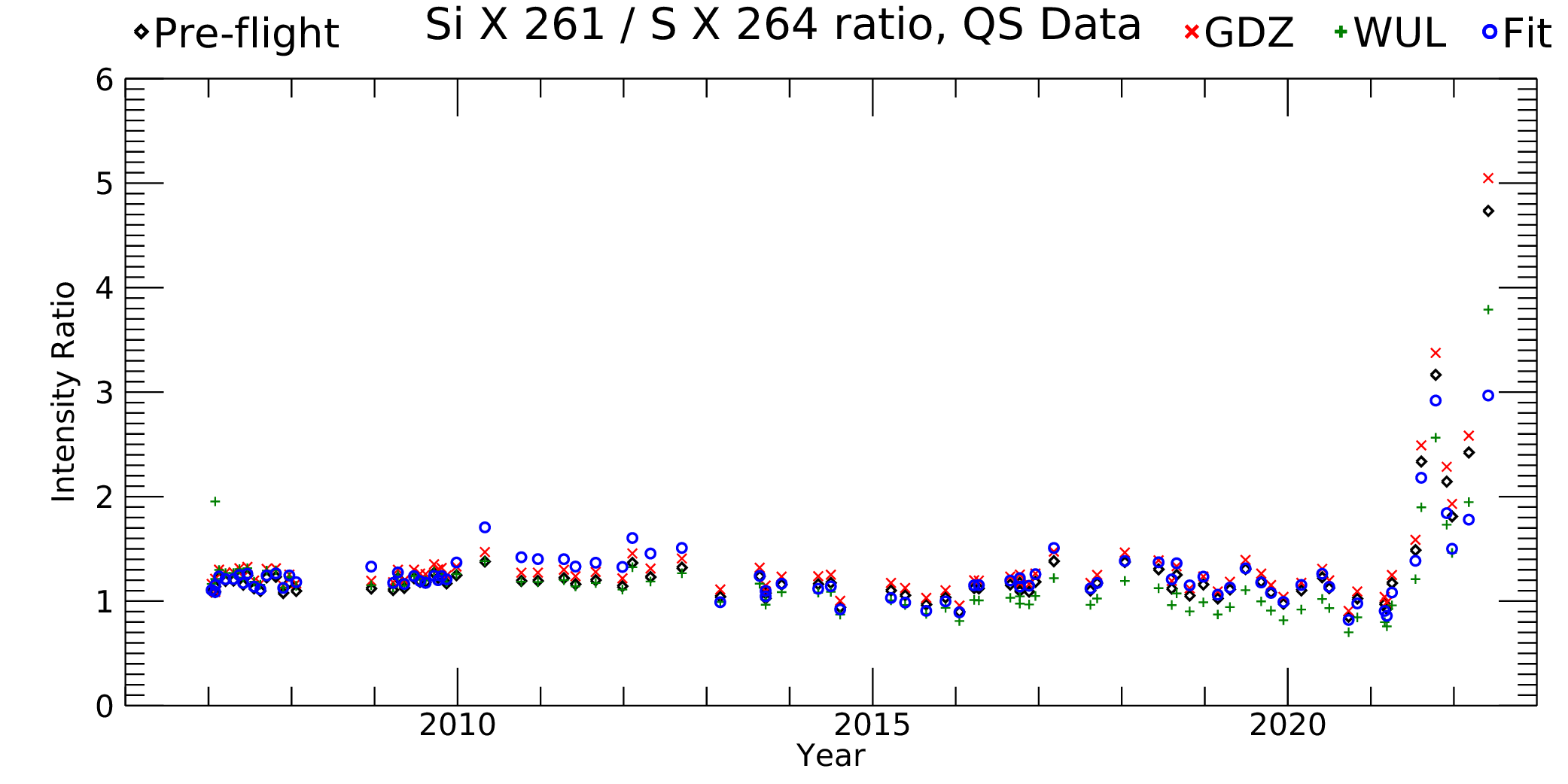}}
 \caption{Ratio of two relatively strong Si and S lines in the LW channel, 
 from the on-disk quiet Sun observations, and applying the various calibrations.
 Note the constancy of the ratio until early 2021 when the count rates in the \ionm{S}{x} 264.23~\AA\ start to decrease considerably. }
\label{fig:si_s_ratio}
\end{figure*}

\begin{table}
  \caption{Same as Table~\ref{tab:lines_11_mar_2007}, but for the last quiet Sun off-limb observation analysed, taken on
   16 June 2021.}
\centering
  \scriptsize
   \setlength\tabcolsep{3.pt}
\begin{tabular}{@{}rrrccrlrr@{}}
\hline\hline \noalign{\smallskip}
 $\lambda_{\rm obs}$  & DN & $I_{\rm obs}$   & log $T_{\rm max}$ & log $T_{\rm eff}$  & $R$ & Ion & $\lambda_{\rm exp}$   &  $r$ \\
\noalign{\smallskip}\hline\noalign{\smallskip}

 275.36 &       76 & 17.6 &  5.79 &  6.04 &  0.99 &  \ionm{Si}{vii} &  275.361 & 0.99 \\ 
 
 185.22 &      212 & 28.0 &  5.73 &  6.04 &  0.96 &  \ionm{Fe}{viii} &  185.213 & 0.95 \\ 

 186.62 &      237 & 25.1 &  5.72 &  6.05 &  0.81 &  \ionm{Fe}{viii} &  186.598 & 0.87 \\

 194.66 &      199 & 6.7 &  5.72 &  6.05 &  0.90 &  \ionm{Fe}{viii} &  194.661 & 0.88 \\ 
 
 277.02 &      114 & 31.8 &  5.83 &  6.07 &  0.76 &  \ionm{Mg}{vii} &  277.001 & 0.15 \\ 
                              &   &   &  &  &  &  \ionm{Si}{viii} &  277.055 & 0.77 \\

 189.95 &      311 & 16.1 &  5.92 &  6.07 &  1.27 &  \ionm{Fe}{ix} &  189.935 & 0.97 \\ 
 
 188.50 &      575 & 45.3 &  5.93 &  6.07 &  0.95 &  \ionm{Mn}{ix} &  188.480 & 0.13 \\ 
                             &    &   &  &  &  &  \ionm{Fe}{ix} &  188.493 & 0.84 \\

 197.87 &      557 & 24.3 &  5.94 &  6.08 &  0.87 &  \ionm{Fe}{ix} &  197.854 & 0.93 \\ 
 
 257.26 &      667 & 169.0 &  6.01 &  6.09 &  0.85 &  \ionm{Fe}{x} &  257.259 & 0.23 \\ 
                              &   &   &  &  &  &  \ionm{Fe}{x} &  257.261 & 0.75 \\

 184.55 &     1090 & 175.0 &  6.02 &  6.10 &  0.93 &  \ionm{Fe}{x} &  184.537 & 0.97 \\ 
 
 177.25 &      194 & 576.0 &  6.03 &  6.10 &  0.76 &  \ionm{Fe}{x} &  177.240 & 0.98 \\ 
 
 174.54 &      119 & 1090.0 &  6.03 &  6.10 &  0.71 &  \ionm{Fe}{x} &  174.531 & 0.98 \\ 
 
 193.72 &      473 & 17.1 &  6.01 &  6.10 &  1.15 &  \ionm{Fe}{x} &  193.715 & 0.91 \\ 
 
 190.05 &     1320 & 66.6 &  6.02 &  6.10 &  0.92 &  \ionm{Fe}{x} &  190.037 & 0.85 \\ 
 
 202.62 &      135 & 18.4 &  5.95 &  6.10 &  1.04 &  \ionm{S}{viii} &  202.610 & 0.58 \\ 
                             &    &   &  &  &  &  \ionm{Fe}{xi} &  202.609 & 0.38 \\ 
 
 198.57 &      767 & 34.5 &  5.96 &  6.11 &  1.37 &  \ionm{S}{viii} &  198.553 & 0.51 \\ 
                             &    &   &  &  &  &  \ionm{Fe}{xi} &  198.538 & 0.45 \\

 184.12 &       47 & 8.9 &  5.51 &  6.12 &  1.04 &  \ionm{O}{vi} &  184.117 & 0.98 \\

 257.54 &      196 & 48.6 &  6.10 &  6.12 &  1.33 &  \ionm{Fe}{xi} &  257.554 & 0.70 \\ 
                            &     &   &  &  &  &  \ionm{Fe}{xi} &  257.547 & 0.25 \\ 
 
 190.40 &      323 & 15.7 &  6.10 &  6.12 &  1.19 &  \ionm{Fe}{xi} &  190.382 & 0.96 \\ 
 
 189.03 &      141 & 9.7 &  6.10 &  6.12 &  1.24 &  \ionm{Fe}{xi} &  188.997 & 0.95 \\ 
 
 256.92 &      329 & 85.6 &  6.10 &  6.12 &  0.91 &  \ionm{Fe}{xi} &  256.919 & 0.92 \\

 202.44 &      323 & 40.0 &  6.10 &  6.12 &  0.99 &  \ionm{Fe}{xi} &  202.424 & 0.94 \\ 
 
 182.18 &      309 & 168.0 &  6.10 &  6.12 &  0.89 &  \ionm{Fe}{xi} &  182.167 & 0.96 \\ 
 
 180.41 &     1250 & 1100.0 &  6.10 &  6.12 &  1.00 &  \ionm{Fe}{xi} &  180.401 & 0.93 \\

 188.31 &     3460 & 287.0 &  6.11 &  6.13 &  1.08 &  \ionm{Fe}{xi} &  188.299 & 0.97 \\ 
 
 188.23 &     5690 & 483.0 &  6.11 &  6.13 &  1.04 &  \ionm{Fe}{xi} &  188.216 & 0.98 \\ 
 
 192.83 &     2500 & 101.0 &  6.11 &  6.13 &  1.05 &  \ionm{Fe}{xi} &  192.813 & 0.97 \\

 201.75 &      235 & 20.7 &  6.11 &  6.13 &  1.49 &  \ionm{Fe}{xi} &  201.734 & 0.90 \\ 
 
 201.57 &      242 & 19.3 &  6.02 &  6.13 &  0.79 &  \ionm{Fe}{xiii} &  201.552 & 0.42 \\ 
                              &   &   &  &  &  &  \ionm{Fe}{x} &  201.565 & 0.53 \\ 
 
 189.13 &      333 & 22.1 &  6.13 &  6.13 &  1.20 &  \ionm{Fe}{xii} &  189.120 & 0.25 \\ 
                             &    &   &  &  &  &  \ionm{Fe}{xi} &  189.123 & 0.70 \\ 
 
 258.37 &      700 & 164.0 &  6.15 &  6.14 &  1.09 &  \ionm{Si}{x} &  258.374 & 0.97 \\ 
 
 253.78 &       87 & 29.7 &  6.15 &  6.14 &  1.18 &  \ionm{Si}{x} &  253.790 & 0.97 \\ 
 
 271.98 &      484 & 74.7 &  6.15 &  6.14 &  1.04 &  \ionm{Si}{x} &  271.992 & 0.98 \\ 
 
 261.05 &      453 & 92.5 &  6.15 &  6.14 &  1.01 &  \ionm{Si}{x} &  261.056 & 0.98 \\ 
 
 277.26 &      217 & 61.9 &  6.15 &  6.14 &  0.96 &  \ionm{Si}{x} &  277.264 & 0.98 \\ 
 
 256.38 &      664 & 180.0 &  6.16 &  6.14 &  0.99 &  \ionm{Si}{x} &  256.377 & 0.62 \\ 
                              &   &   &  &  &  &  \ionm{Fe}{xii} &  256.410 & 0.17 \\ 
 
 249.38 &       67 & 32.5 &  6.17 &  6.14 &  0.83 &  \ionm{Fe}{xii} &  249.388 & 0.91 \\ 
 
 259.49 &      241 & 52.8 &  6.18 &  6.15 &  1.10 &  \ionm{S}{x} &  259.496 & 0.98 \\ 
 
 264.23 &      172 & 33.1 &  6.18 &  6.15 &  2.55 & ** \ionm{S}{x} &  264.230 & 0.98 \\ 
 
 188.81 &      163 & 11.8 &  6.28 &  6.15 &  0.83 &  \ionm{Ar}{xi} &  188.820 & 0.75 \\ 
                              &   &   &  &  &  &  \ionm{Fe}{ix} &  188.817 & 0.20 \\ 
 
 257.14 &      101 & 25.9 &  6.19 &  6.15 &  1.23 &  \ionm{S}{x} &  257.147 & 0.93 \\ 
 
 196.66 &      815 & 29.5 &  6.17 &  6.15 &  1.33 &  \ionm{Fe}{xii} &  196.640 & 0.94 \\ 
 
\noalign{\smallskip}\hline 
\end{tabular}
\normalsize
 \label{tab:ol_last} 
\end{table}

\addtocounter{table}{-1}
\begin{table}
  \caption{Contd.}
\centering
  \scriptsize
   \setlength\tabcolsep{3.pt}
\begin{tabular}{@{}rrrccrlrr@{}}
\hline\hline \noalign{\smallskip}
 $\lambda_{\rm obs}$  & DN & $I_{\rm obs}$   & log $T_{\rm max}$ & log $T_{\rm eff}$  & $R$ & Ion & $\lambda_{\rm exp}$   &  $r$ \\
\noalign{\smallskip}\hline\noalign{\smallskip}

 191.05 &      186 & 8.4 &  6.17 &  6.15 &  1.07 &  \ionm{Fe}{xii} &  191.049 & 0.93 \\ 
 
186.89 &     1100 & 113.0 &  6.17 &  6.15 &  1.24 &  \ionm{Fe}{xii} &  186.854 & 0.17 \\ 
                            &     &   &  &  &  &  \ionm{Fe}{xii} &  186.887 & 0.80 \\ 
 
203.74 &      230 & 47.3 &  6.17 &  6.15 &  0.82 &  \ionm{Fe}{xii} &  203.728 & 0.97 \\ 
 
 193.52 &    13100 & 487.0 &  6.18 &  6.15 &  0.94 &  \ionm{Fe}{xii} &  193.509 & 0.94 \\ 
 
 192.41 &     5410 & 226.0 &  6.18 &  6.15 &  0.93 &  \ionm{Fe}{xii} &  192.394 & 0.97 \\ 
 
 195.13 &    20700 & 679.0 &  6.18 &  6.15 &  0.97 &  \ionm{Fe}{xii} &  195.119 & 0.97 \\ 
 
 188.68 &      197 & 14.8 &  6.28 &  6.16 &  0.94 &  \ionm{S}{xi} &  188.675 & 0.83 \\ 
                               &  &   &  &  &  &  \ionm{Fe}{ix} &  188.680 & 0.13 \\ 
 
 201.13 &     1460 & 96.4 &  6.23 &  6.16 &  1.24 &  \ionm{Fe}{xiii} &  201.126 & 0.73 \\ 
                            &     &   &  &  &  &  \ionm{Fe}{xi} &  201.112 & 0.21 \\ 
 
 196.53 &      271 & 9.7 &  6.23 &  6.17 &  1.46 &  \ionm{Fe}{xiii} &  196.525 & 0.81 \\ 
                             &    &   &  &  &  &  \ionm{Ni}{xii} &  196.558 & 0.11 \\ 
 
 191.26 &      499 & 22.3 &  6.28 &  6.17 &  0.96 &  \ionm{S}{xi} &  191.266 & 0.93 \\ 
 
 200.03 &      523 & 24.5 &  6.23 &  6.17 &  1.28 &  \ionm{Fe}{xiii} &  200.021 & 0.95 \\ 
 
 264.78 &      427 & 82.0 &  6.29 &  6.17 &  0.92 &  \ionm{Fe}{xiv} &  264.788 & 0.70 \\ 
                            &     &   &  &  &  &  \ionm{Fe}{xi} &  264.772 & 0.27 \\ 
 
 204.95 &      109 & 26.8 &  6.24 &  6.17 &  1.00 &  \ionm{Fe}{xiii} &  204.942 & 0.94 \\ 
 
 197.44 &      444 & 18.2 &  6.24 &  6.17 &  1.27 &  \ionm{Fe}{xiii} &  197.431 & 0.94 \\

 203.84 &      441 & 93.2 &  6.23 &  6.17 &  0.98 &  \ionm{Fe}{xiii} &  203.826 & 0.97 \\ 
 
 246.20 &      113 & 70.5 &  6.23 &  6.17 &  0.77 &  \ionm{Fe}{xiii} &  246.209 & 0.97 \\ 
 
 251.94 &      279 & 111.0 &  6.23 &  6.17 &  0.91 &  \ionm{Fe}{xiii} &  251.952 & 0.97 \\ 
 
 209.93 &      104 & 71.2 &  6.24 &  6.17 &  0.90 &  \ionm{Fe}{xiii} &  209.916 & 0.96 \\ 
 
 202.06 &     4320 & 455.0 &  6.24 &  6.17 &  0.80 &  \ionm{Fe}{xiii} &  202.044 & 0.97 \\ 
 
 274.20 &      476 & 90.0 &  6.29 &  6.19 &  0.90 &  \ionm{Fe}{xiv} &  274.203 & 0.94 \\

 257.38 &      113 & 28.5 &  6.29 &  6.19 &  0.88 &  \ionm{Fe}{xiv} &  257.394 & 0.96 \\ 
 
 211.33 &      194 & 149.0 &  6.29 &  6.19 &  0.94 &  \ionm{Fe}{xiv} &  211.317 & 0.96 \\ 
 
 270.51 &      194 & 34.6 &  6.29 &  6.19 &  1.01 &  \ionm{Fe}{xiv} &  270.520 & 0.97 \\ 
 
 284.17 &      168 & 87.5 &  6.34 &  6.20 &  1.18 &  \ionm{Fe}{xv} &  284.163 & 0.97 \\ 
 
\noalign{\smallskip}\hline 
\end{tabular}
\normalsize
\end{table}

 
\end{document}